\documentclass[prc,preprint,showpacs,tightenlines,floatfix,amsfonts,nofootinbib]{revtex4}

\usepackage{graphicx}
\usepackage{bm}
\usepackage{amssymb}
\usepackage{latexsym}
\usepackage{amsmath}

\begin{document}

%
%
%
\newcommand{\rhoc}{\rho_{\mathrm{c}}}
\newcommand{\rhop}{\rho_{\mathrm{p}}}
\newcommand{\rhopcn}[1]{\rho_{\mathrm{p},#1}^{c}}
\newcommand{\rhopsn}[1]{\rho_{\mathrm{p},#1}^{s}}
\newcommand{\rhopn}[1]{\rho_{\mathrm{p},#1}}
\newcommand{\rhoczero}{\rho_{\mathrm{c}}^{0}}
\newcommand{\rhopzero}{\rho_{\mathrm{p}}^{0}}
\newcommand{\drhoc}{\delta \rho_{\mathrm{c}}}
\newcommand{\drhop}{\delta \rho_{\mathrm{p}}}
\newcommand{\T}[1]{T_{#1}}
\newcommand{\Tzero}[1]{T_{#1}^{0}}
\newcommand{\dT}[1]{\delta T_{#1}}
\newcommand{\e}[1]{\epsilon_{#1}}
\newcommand{\eT}[2]{\epsilon_{#1}^{T#2}}
\newcommand{\n}[1]{\nu_{#1}}
\newcommand{\ave}[1]{< #1 >}
\newcommand{\nus}[1]{\nu_{#1}^{s}}
\newcommand{\nuh}[1]{\nu_{#1}^{h}}
\newcommand{\psis}[1]{\psi^{S}_{#1}}
\newcommand{\psih}[1]{\psi^{J}_{#1}}
\newcommand{\psie}{\psi^{\mathrm{EP}}}
\newcommand{\psir}{\psi^{\mathrm{RP}}}
\newcommand{\Ns}{N^{s}}
\newcommand{\Nh}{N^{h}}
\newcommand{\phis}{\phi^{s}}
\newcommand{\phih}{\phi^{h}}
\newcommand{\dphi}{\Delta \phi}
\newcommand{\dpsis}[1]{\Delta \psi_{#1}^{S}}
\newcommand{\dpsih}[1]{\Delta \psi_{#1}^{J}}
\newcommand{\nuhre}[1]{\tilde{\nu}^{h}_{#1}}
\newcommand{\nuhreprime}[1]{\tilde{\nu}^{h\prime}_{#1}}
\renewcommand{\vec}{\boldsymbol}

\title{Event-by-event azimuthal anisotropy of jet quenching \\ in relativistic heavy ion collisions}

\author{Xilin Zhang$^{1,3}$} \email{zhangx4@ohio.edu}
\author{Jinfeng Liao$^{1,2}$}\email{liaoji@indiana.edu}
\affiliation{$^1$ Physics Department and Center for Exploration of Energy and Matter, Indiana University, 2401 N Milo B. Sampson Lane, Bloomington, IN 47408, USA.\\
$^2$ RIKEN BNL Research Center, Bldg. 510A, Brookhaven National Laboratory, Upton, NY 11973, USA.\\
$^3$ Institute of Nuclear and Particle Physics and Department of
Physics and Astronomy, Ohio University, Athens, OH 45701, USA.}


%
\date{April 21, 2013\\[20pt]}

\begin{abstract}
\begin{description}
\item[Background:] Strong jet quenching has been observed in heavy ion collisions at both the Relativistic Heavy Ion Collider (RHIC) and the Large Hadron Collider (LHC) that can be understood through substantial jet energy loss in the created hot QCD matter. Yet the azimuthal anisotropy of jet quenching has not been thoroughly studied in the presence of strong fluctuations in the initial condition.
\item[Purpose:] We present with full details a systematic quantification of the hard probe azimuthal response to the geometry and fluctuations of the hot QCD matter at both RHIC and LHC. We also examine the hard-soft dihadron correlation arising from the hard and soft sectors' responses to the common fluctuating initial condition.
\item[Methods:] An even-by-event MC simulation is employed. Different geometrical jet-quenching models are tested. The azimuthal anisotropy of jet quenching is extracted and decomposed as harmonic responses (for n=1-6) to the corresponding harmonics in the initial state.
\item[Results:] We show that such jet response harmonics are sensitive to the jet quenching models as well as to the initial composition of bulk matter. Their  centrality dependence puts a strong constraint on the path-length and medium-density dependence of jet energy loss. The computed hard-soft dihadron correlation shows a strong peak on the near side in RHIC central collisions. The triggered correlation in noncentral collisions is also presented.  
\item[Conclusions:] Only the jet-quenching model with near-$T_c$ enhancement survives the second-harmonic test by the RHIC and LHC. Other harmonics in this model are consistent with the available data. We also demonstrate that the experimentally observed ``hard ridge'' can be explained in our calculation and that its trigger-azimuthal-angle and associate-$p_t$ dependence could also be qualitatively understood.  
\end{description}
\end{abstract}


\maketitle

\section{Introduction}

In high energy nucleus-nucleus ($AA$) collisions, the hot deconfined QCD matter, so-called quark-gluon plasma (QGP), is believed to be created at both the Relativistic Heavy Ion Collider (RHIC) and the Large Hadron Collider (LHC). Also created in such collisions are jets from the initial hard scatterings, which penetrate the soft matter and  eventually convert into hadrons. Along its path through the medium, the jet keeps interacting with the medium and losses energy. The quenching of jets due to the energy loss can be used to characterize the jet-medium interaction as well as the medium  properties, which provides an imaging tool of QGP often called jet tomography (see, e.g., reviews in Refs.~\citep{Gyulassy:2003mc}).  

A traditional single hadron observable for quantifying the jet quenching is the so-called nuclear modification factor $R_{AA}$, i.e., the ratio between the hadron production in $AA$ collisions and that  in $NN$ collisions (multiplied by the expected binary collision number). In the high $p_t$ region that is dominated by jet physics, the $R_{AA}$ has been measured to be substantially less than 1  in central collisions  at RHIC as well as at LHC, indicating a strong jet energy loss in the hot, color-opaque medium. With both RHIC and LHC $R_{AA}$ measurements available now, there has been a lot of interest recently in quantifying the evolution of medium opaqueness with collision beam energy, and a few model studies indicate a sizable reduction of the opaqueness from RHIC to LHC \citep{Liao:2011kr, Horowitz:2011gd, Betz:2012qq}.

A very important aspect of jet quenching study is the azimuthal anisotropy in the high $p_t$ hadron production, which can be quantified by the azimuthal angle $\phi$ dependent nuclear modification factor $R_{AA}(\phi)$ \citep{Gyulassy:2000gk,Wang:2000fq,Shuryak:2001me,Drees:2003zh,Jia:2011pi,Rodriguez:2010di,Renk:2011aa,Renk:2011qi,xzjl2012}. 
Such anisotropy originates from the geometric features of the underlying matter distribution as well as the distribution of initial hard scattering spots. In off-central $AA$ collisions, the overlap zone has an almond-like geometry in the transverse plane, which is then inherited by the created medium at early time. As a result, the jets with different transverse orientations (at mid-rapidity) would see different medium ``thickness'' and lose different amounts of energy, which leads to the azimuthal anisotropy in the distribution of final-state high $p_t$ hadrons. The dominant component of this anisotropy is the second  harmonic (often called the elliptic component) in the Fourier decomposition of $R_{AA}(\phi)\sim 1 + \nuh{2}\cos[2(\phi-\psie)]$. 
 This $\nuh{2}$ should be differentiated from the commonly known elliptical flow $\nus{2}$ of the soft hadrons, because of their distinctive origins: the former is from jet quenching while the latter is from the bulk medium's collective expansion. At RHIC, most model calculations predicted a $\nuh{2}$ that is much smaller than the measured data. A first resolution of such discrepancy was proposed  in Ref.~\citep{Liao:2008dk}  (referred to as the NTcE model hereafter)  with a radical insight that the jet-medium interaction bears a nontrivial dependence on matter density and is strongly enhanced in the near-$T_c$ matter via a nonperturbative mechanism, as motivated by the ``magnetic scenario'' for sQGP \citep{Liao:2006ry, Liao:2008vj}. The NTcE model successfully described the $R_{AA}$ and $\nuh{2}$ as well as their centrality dependence at RHIC.

More recently it has been realized that anisotropy arises not only from the (average) geometric shape but also from strong fluctuations in the initial condition~\citep{Alver:2010gr}.  The aforementioned almond-shape picture for $\nuh{2}$ and $\nus{2}$ assumes an averaged smooth geometry for the medium. The reality, however, turns out to be more complicated and interesting: event by event, the nucleon positions in the two colliding nuclei fluctuate, leading to the so-called initial state fluctuation (ISF), i.e., the fluctuation of the medium's initial (entropy and energy) density distributions, which therefore bears different anisotropy in each event and leaves imprints in many observables. Detailed investigations over the past few years have revealed that such initial state fluctuations are quite strong,
and in fact even in perfectly central collisions the final hadron distribution shows nontrivial anisotropy. 
For example in the bulk matter evolution, such event-by-event anisotropy is translated by hydrodynamic expansion into different harmonic flows in the finally observed soft-hadron production. 
 These have been observed in the RHIC and LHC experiments \citep{Lacey:2011av,Sorensen:2011fb,GrosseOetringhaus:2011kv,Jia:2011hfa,Li:2011mp} and also demonstrated in various hydrodynamic calculations \citep{Alver:2010gr,Alver:2010dn,Luzum11plb,Bhalerao2011,Luzum:2011jpg,Sorensen:2008bf,Teaney:2010vd,ZhiQiu2011PRC84,Qiu:2011iv,Staig:2010pn,Takahashi:2009na,Schenke:2010rr,Xu:2010du,Qin:2010pf,Ma:2010dv}. It has also been shown that the so-called ``soft ridge'' --- a strong peak on the near side in azimuthal angle and extending long in rapidity in the dihadron correlation --- can be explained by convoluting such harmonic flow components in the single hadron distributions; see, e.g., \citep{Alver:2010gr,Luzum11plb,Sorensen:2008bf,Takahashi:2009na,Xu:2010du,WLQian2012plb}.

Since studies of the soft hadron distributions show convincing evidence for strong initial state fluctuations and event-by-event anisotropy beyond average geometry, and since the jet energy loss is a sensitive tomographic tool for anisotropy, one is naturally led to the questions of how such event-by-event anisotropy can be manifested in jet quenching observables such as the $R_{AA}(\phi)$ and what we can learn about jet energy loss from such observables. These are the central issues we aim to address, at least partially, in this study.  Following the same argument about $\nuh{2}$ induced by the almond-like average geometry, one expects event-by-event azimuthal anisotropy for the high $p_t$ hadrons   can also arise from the fluctuating geometric distributions of {\em both the soft matter density and the initial jet-spot profile (the binary collision density)}.  
Some earlier discussions and data from RHIC and LHC can be found in Refs.~\citep{Jia:2010ee, Betz:2011tu, Rodriguez:2010di, Adare:2011tg, Sorensen:2011fb, Abelev:2012di,Chatrchyan:2012wg, Aad:2012bu}. In Ref.~\citep{xzjl2012} we made the first attempt to study the connection between ISF and the jet azimuthal anisotropy ($\nuh{n}$) and quantify the high $p_t$ harmonics as the hard-probe response to the ISF, for the central collisions at RHIC with the fluctuations implemented via cumulant expansions. In the present paper we will present a systematic quantification,  with full details and on an even-by-event basis, of how the hard probe responds to the geometry and fluctuations of the hot QCD matter created in heavy ion collisions from RHIC to LHC, substantiating previous studies reported in  \citep{Liao:2008dk, Liao:2011kr, xzjl2012, xzjl2012jqst}.  We emphasize that this is an important shift for jet quenching physics, i.e., from studying an average anisotropy pattern  $R_{AA}(\phi)\sim 1 + \nuh{2}\cos[2(\phi-\psie)]$ toward the event-by-event extraction of full anisotropy information $R_{AA}(\phi)\sim 1 + \sum_{n=1,2,3,...} \nuh{n}\cos[n(\phi-\psih{n})]$. 
In particular as alluded above, we want to gain a coherent picture about the jet quenching and its azimuthal anisotropy from RHIC to LHC, which is already partially investigated in Ref.~\citep{xzjl2012jqst}. The hope here is that by using geometric data at both collision energies, a strong constraint can be put on different jet quenching models, especially on how the energy loss depends on the path length and the medium density (temperature). In addition, understanding the jet-response anisotropy provides a sensitive tool for probing the initial conditions, which is complimentary to what can be learned from studying the bulk expansion responses. The analysis in the present paper will further address a number of important questions, including the event-by-event determinations of eccentricities $\e{n}$ and participant angles $\psi_{n}$ in both RHIC and LHC experiments; the jet azimuthal anisotropy in terms of harmonics arising from geometry and fluctuations; comparison of these jet-response harmonics in different jet-energy-loss models; the transfer of eccentricities to jet-response harmonics and the angular correlations between $\psih{n}$ and $\psi_{n}$; examination of the separate contributions to the jet anisotropy from the initial jet-spot-profile fluctuation and from the medium-density-profile fluctuation; and the sensitivity of the jet-response anisotropy to the composition of the matter density (in terms of participant and collision profiles). 

In addition we will also study the azimuthal correlation between hard and soft hadrons~\citep{Zhang:2007ja,Renk:2006pk} (hard-soft correlation), much in parallel to the study of dihadron correlation in the kinematic region dominated by soft hadrons. In particular we examine the contribution  from  their common correlations to the fluctuating initial conditions, as also initiated in Ref.~\citep{xzjl2012}. Our improved calculation confirms what we previously found for the central collisions at RHIC: the hard-soft correlation shows a strong peak at near side
 (``hard-ridge'') and double-hump structure on the away side due to the concurrent harmonics in both soft and hard particle azimuthal distributions. We think this is a possible explanation of the ``hard-ridge''  \citep{STAR_ridge,Shuryak:2007fu,Majumder:2006wi,Dumitru:2008wn}.
 Recently, the STAR sollaboration has extracted the trigger-azimuthal-angle dependence of the dihadron correlation  \citep{star2004prl,AFeng2008,star2010dc}. In Ref.~\citep{Luzum11plb},  the trigger-angle (relative to $\psie$) and $p_{t}$ dependence of all the harmonic components extracted from the dihadron correlation were studied based on the  hydrodynamic calculations. This approach should be good when the associated hadron $p_t$ is around $0.15-3$ GeV and trigger $p_t$ is around $3-4$ GeV). However for even higher $p_t$ trigger (e.g., $4-6$ GeV), the picture should be different because the jets become the dominant source and the correlation is of the hard-soft type. This hard-soft correlation will be studied based on our computation of jet response to ISF. We will also rederive the formula for the triggered correlation \citep{Bielckiova2004} by lifting the often used (and not fully correct) assumption that different event-plane angles $\psih{n}$ (for hard hadron) and $\psis{n}$ (for soft hadron) are  totally correlated with $\psie$ (i.e., $\psis{2}$).    

Before getting into the main body of this study, we would like to emphasize two important discussions on the near-$T_c$ enhancement model for jet quenching, which have been included as Appendices C and D. The first issue is about the possible relation between the strong near-$T_c$ peak seen in the QCD trace anomaly (a measure of nonconformal behavior) and the proposed near-$T_c$ enhancement in jet quenching, which arose during a discussion \citep{Miklos}\citep{Ficnar:2012yu}. In Appendix C we will show that indeed the two phenomena could be simultaneously and consistently understood from the scenario of a near-$T_c$ plasma of magnetic monopoles. The second issue concerns a quantitative estimate of the evolution of medium opaqueness from RHIC to LHC. The NTcE model with strong near-$T_c$ enhancement of jet-medium interaction  naturally predicts a less opaque medium at LHC, as the RHIC fireball has a larger fraction of its space-time evolution in the near-$T_c$ region. In Appendix D, this is quantified to be a reduction of about $30\%$ for the average jet-medium interaction, i.e., 
$ <\kappa>_{\rm RHIC} : <\kappa>_{\rm LHC} \approx 1 : 0.72 $---consistent with a number of recent jet quenching analyses \citep{{Horowitz:2011gd,Betz:2012qq,Zakharov:2011ws,Lacey:2012bg,Majumder:2012sh}}. These are two important points and we choose to leave them in the Appendices primarily because they are somewhat less attached to the stream of main discussions in this paper. 

The rest of the paper is organized as follows. In Sec.~\ref{sec:MC}, we collect the details of our Monte Carlo (MC) simulation. The initial state is modeled by using the MC Glauber model. Three different geometric jet-energy-loss models are presented. The structure of the simulation code can be found in Appendix~\ref{app:flowchart}. Sec.~\ref{sec:jetanisotropy} summarizes our results for the jet anisotropy at RHIC and LHC, and 
the comparison of different models. 
In Sec.~\ref{sec:HScorrelation}, we discuss the (un)triggered hard-soft correlations. The re-derived formula for the triggered dihadron correlation can be found in Appendix~\ref{app:correlation}. A summary and some discussions will be given in Sec.~\ref{sec:summary}.

\section{MC simulation} \label{sec:MC}

In this section, we discuss the details of the MC simulation. The structure of the code can be found in Appendix~\ref{app:flowchart}.

\subsection{Glauber model}

In the $AA$ collisions event by event, the positions of the nucleons fluctuate, as the result of the ``measurement'' of their positions (in the quantum language). This leads to the fluctuation of matter density, i.e., ISF. In the simulation, it is realized in a simplified way: sampling every nucleon position in the nuclei according to its density as measured in the low-energy scattering process \citep{miller2007}. The short range correlation between nucleons is included via requiring a smallest distance between them, which is set to $0.4$ fm in most simulations (e.g., \citep{miller2007,ZhiQiu2011PRC84}) including this work. The Glauber model is applied to deal with multiple $NN$ collisions \citep{miller2007}: binary collision happens only if the transverse distance between two nucleons is smaller than $\sqrt{\sigma_{\mathrm{inel}}/\pi}$. 
Here $\sigma_{\mathrm{inel}}$ is the total inelastic $NN$ scattering cross section at the corresponding center of mass energy. The trajectory of each nucleon is always along the beam direction. 
The parameters in the simulation of the initial state are summarized in the following. First, the density in each colliding nuclei is parameterized in terms of the Wood-Saxon form \citep{Heinz1108.5379}: 
\begin{eqnarray}
\rho(\vec{r})=\frac{\rho_{0}}{\exp[\frac{(r-R_{A})}{d_{A}}]+1} \ .
\end{eqnarray}
The values of $\rho_{0}$, $R_{A}$, and $d_{A}$ for $Au$ and $Pb$ nuclei as used in RHIC and LHC can be found in Table~\ref{tab:parameters} \citep{Heinz1108.5379, Hirano11prcra, ZQiucommun} \footnote{The finite size of nucleons is taken into account when fixing the parameters in $\rho(\vec{r})$ as the sampling probability.}.
\begin{table}
 \centering
   \begin{tabular}{|c|c|c|c|c|c|c|c|} \hline
           & $\rho_{0} \ (\mathrm{fm}^{-3})$
           & $R_{A} \ (\mathrm{fm})$ 
           & $d_{A} \ (\mathrm{fm})$
           & $\sqrt{B} \ (\mathrm{fm})$
           & $\sigma_{\mathrm{inel}} \ (\mathrm{fm}^{2})$ 
           & $S_{0} \ (\mathrm{fm}^{-3})$ 
           & $n$   \\  \hline
$\mathrm{RHIC} \ (0.2 \ \mathrm{TeV})$
         & $0.170$
           & $6.42$
           & $0.45$
           & $0.544$
           & $4.2$
           & $116$
           & $8.1$  \\   \hline
$\mathrm{LHC} \ (2.76 \ \mathrm{TeV})$ 
           & $0.166$
           & $6.67$
           & $0.44$
           & $0.660$
           & $6.2$
           & $291$
           & $6.0$  \\   \hline    
$\mathrm{LHC} \ (5.50 \ \mathrm{TeV})$ 
           & $0.166$
           & $6.67$
           & $0.44$
           & $0.680$
           & $6.6$
           & $ 364$
           & $5.4$  \\   \hline                          
   \end{tabular}
   \caption{Parameters in the simulations for both RHIC and LHC. The explanation and the references for them can be found in the text.} \label{tab:parameters}
\end{table}
For the density in a nucleon, a Gaussian distribution is assumed in the transverse plane \citep{Heinz1108.5379}:  
\begin{eqnarray}
\T{p}(\vec{r}^{\perp})=\frac{\exp(-\frac{|\vec{r}^{\perp}|^{2}}{2B})}{2\pi B} \ ,
\end{eqnarray}
where the $B$ value can be found in Table~\ref{tab:parameters} for three different energies. Second, the total cross sections $\sigma_{\mathrm{inel}}$ for $NN$ collisions at these energies are also listed in Table~\ref{tab:parameters}. After the binary collisions are sampled by using the Glauber model based on $\sigma_{\mathrm{inel}}$, we sum up contributions of all the ``wounded'' nucleons to get the participant density in the transverse plane:
\begin{eqnarray}
\rhop(\vec{r}^{\perp})=\sum_{i=1}^{N_{\mathrm{p}}} \T{p}(|\vec{r}^{\perp}-\vec{r}^{\perp}_{i}|) \ . 
\end{eqnarray}
Here $N_{\mathrm{p}}$ is the number of wounded nucleons. A similar procedure can be applied to calculate the collision density $\rhoc(\vec{r}^{\perp})$. 

To make a connection between the initial state and the state of equilibrated medium, we assume that the entropy density $S$ at $\tau\equiv\tau_{0}=0.6 \ \mathrm{fm}/c$ (around the equilibration time) is proportional to $\rhop$ at RHIC ($0.2$ TeV) \citep{Liao:2008dk, Liao:2011kr, xzjl2012}, and a two-component profile,  $(1-\delta)\times\rhop/2+\delta\times\rhoc$, at LHC ($\delta=0.118$ for both $2.76$ and $5.5$ TeV) \citep{Hirano11prcra, ZQiucommun, Qiu:2011iv}. The maximum entropy density in the central collisions, $S_{0}$, is listed in  Table~\ref{tab:parameters} for different experiments \citep{ZQiucommun,Qiu:2011iv,CShen2011prc}, which are calibrated to the observed multiplicities in the hydrodynamic calculation. Based on this, we compute the proportionality coefficient in the relation between $\rhop$ and $\rhoc$, and $S(\tau_{0})$. Before $\tau_{0}$, we assume effectively for jet quenching, the entropy density $S$ grows linearly with time, while after that, the longitudinal expansion is applied and $S$ decreases as $1/\tau$. See Ref.~\citep{Jia:2010ee} for a detailed discussion about the effects of different preequilibrium models on jet quenching. Unfortunately, the medium's transverse dynamics is simplified without its hydrodynamical evolution, i.e., the shape is frozen, which will be discussed later. For the jets, their productions at the early stage of the collision should distribute according to  $\rhoc$ and are isotropic in their momentum space. 
In our simulation for each event, we  integrate over all jet spots weighted by the binary collisional density from the same event, which is equivalent to using a large number of jets in each event. In the following section, we focus on the jet energy loss when traveling through the medium.

\subsection{Jet quenching models} \label{subsec:models}


Three different geometric models are applied to compute the jet energy loss, which have distinctive geometric features (e.g., the path-length dependence) and matter-density dependence that are most crucial for describing geometric data \citep{Shuryak:2001me,Liao:2008dk,Jia:2010ee,Jia:2011pi,Betz:2011tu,Liao:2011kr,xzjl2012}. Suppose the jet is produced with initial energy $E_{i}$. After traveling a path $\vec{P}$ in the medium, the ratio between its final energy $E_f$ and the initial energy $E_i$, i.e., the suppression factor $f_{\vec P}$, is given by
\begin{equation} \label{Eq_fP}
f_{\vec P} = \exp\left\{ - \int_{\vec P}\, \kappa[s(l)]\, s(l)\, l^m dl  \right\} \ . 
\end{equation}
In this expression the $s(l)$ is the local entropy density at a given point on the jet path, and the $\kappa(s)$ is the local jet quenching strength which as a property of matter should in principle depend on the local density $s(l)$. There can be different choices of the parameter $m$ for path-length dependence (e.g., LPM-motivated quadratic or AdS/CFT-motivated cubic) and of the jet-medium interaction $\kappa(s)$. In this study, we compare the following models \citep{Liao:2008dk,Liao:2011kr}: the near-$T_c$ enhancement (NTcE) model, $\mathrm{L}^2$ model, and $\mathrm{L}^3$ model. 
The NTcE model assumes $m=1$ and introduces a strong jet quenching component in the vicinity of $T_c$ (with density $s_c$ and span of $s_w$) via
\begin{eqnarray} \label{Eq_kappa}
\kappa(s)=\kappa_0 [1+ \xi\, exp(-(s-s_c)^2/s_w^2)] \ , 
\end{eqnarray}
with $\xi=6$, $s_c=7/fm^3$, and $s_w=2/fm^3$. (see \citep{Liao:2008dk} for  details.)
In the $\mathrm{L}^2$ ($\mathrm{L}^3$) model, $\kappa(s)=\kappa_0$ is constant and $m=1$ ($m=2$). In each model, the parameter $\kappa_0$ will be fixed by $R_{AA}\approx 0.18$ in the $0-5\%$ collisions at RHIC $\sqrt{s}=0.2 \ \rm TeV$, and then will be applied at LHC $\sqrt{s}=2.76, \ 5.5 \ \rm TeV$.

To calculate the azimuthal angle dependence of the nuclear modification factor, we can apply the following formula: 
\begin{equation}
R_{AA} (\phi) = <\, (f_{\vec P})^{n-2}  \,>_{\vec P(\phi)} \ . \label{eqn:Raa}
\end{equation}
A detailed derivation of Eq.~(\ref{eqn:Raa}) can be found in Ref.~\cite{Adler:2006bw}. \footnote{This formula shows that the quenching effect increases when the overall jet energy loss increases. Meanwhile, for a given energy loss fraction, a softer p-p collision spectrum (i.e., with a larger $n$) leads to a larger quenching.}
Here $< \, \, >_{\vec{P}(\phi)}$ means averaging over all jet paths with the  propagation orientation fixed at angle $\phi$ relative to the reaction plane, including all the possible initial jet production points [distributed according to $\rhoc(\vec{r}^{\perp})$]. The exponent $n$ is the spectrum index of the measured high $p_{t}$ spectrum in a reference p-p collision. It should be emphasized here that the index $n$ depends on the p-p collision energy. In Table~\ref{tab:parameters}, we show $n$ at the three different energies \citep{Adler:2006bw, ALICE2011pts, Francois}. To calculate the over all modification factors, $R_{AA}$, we need to average over all the jet paths with different orientation angles that are equally distributed. 

Moreover, in these models, assuming the fractional energy loss leads to  $R_{AA} (\phi)$ being independent of $p_t$.  This at RHIC energy may be justified by the approximate ``flatness'' seen in the $R_{AA} \ \mathrm{vs} \ p_{t}$ data. At LHC energies, we expect the $p_{t}$ dependence of the azimuthal anisotropy, i.e., $R_{AA} (\phi)/R_{AA}$, to be weak, which should be dictated by the length and density dependence of the jet energy loss in the transverse plane. The results from such modeling  apply only to the high-$p_t$ region, e.g., $p_t>6\, \rm GeV$ at RHIC $0.2$ TeV, and $p_t>8\, \rm GeV$ at LHC $2.76$ TeV. \footnote{At LHC $5.5$ TeV, the threshold maybe somewhat higher.} 

In the simulation of each event, we apply the Fourier-decomposition to $R_{AA} (\phi)/R_{AA}$: 
\begin{equation}
R_{AA}(\phi)=R_{AA}\left(1+2\sum_{n} \nuh{n}\cos[n(\phi-\psih{n})]\right) \ . \label{eqn:Raadecomp}
\end{equation}
In each event the azimuthal anisotropy can be represented by the collection of $\nuh{n}$ and $\psih{n}$. The importance of different harmonics decreases with increasing $n$ except for the second harmonics, as will be shown later. So we focus on the first six harmonics in this work.

\section{The ISF and jet azimuthal anisotropy} \label{sec:jetanisotropy}

\subsection{The ISF}

To quantify the ISF in each $AA$ collision, the following eccentricities $\e{n}$ are defined, as motivated by the cumulant expansion method in Ref.~\citep{Teaney:2010vd}: 
\begin{eqnarray}
\e{n\geqslant 2}&\equiv&-\frac{\ave{r^{n}\cos[n(\phi-\psi_{n})]}}{\ave{r^{n}}} \ ,  \\
\psi_{n\geqslant 2}&=&\frac{1}{n}\arctan\left(\frac{\ave{r^{n}\sin(n\phi)}}{\ave{r^{n}\cos(n\phi)}}\right)+\frac{\pi}{n} \ , \label{eqn:psin} \\
\e{1}&\equiv&-\frac{\ave{r^{3}\cos[(\phi-\psi_{1})]}}{\ave{r^{3}}} \ , \\
\psi_{1}&=&\arctan\left(\frac{\ave{r^{3}\sin(\phi)}}{\ave{r^{3}\cos(\phi)}}\right)+\pi \  . \label{eqn:psi1}
\end{eqnarray} 
The $\psi_{n}$ is the reaction plane angle of $nth$ harmonics.  $\ave{\cdots}$ means averaging over the entropy density at $\tau_{0}$. 

\begin{figure}
\centering
\includegraphics[scale=0.6, angle=-90]{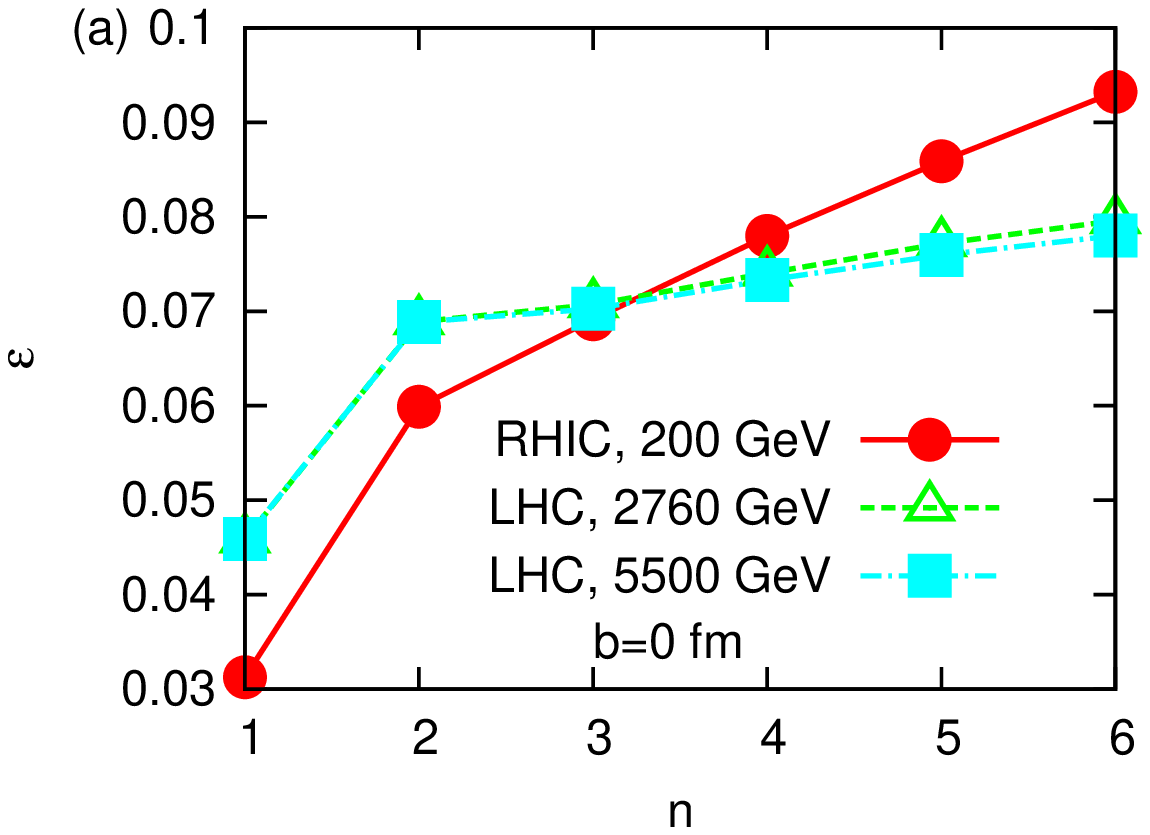}
\includegraphics[scale=0.6, angle=-90]{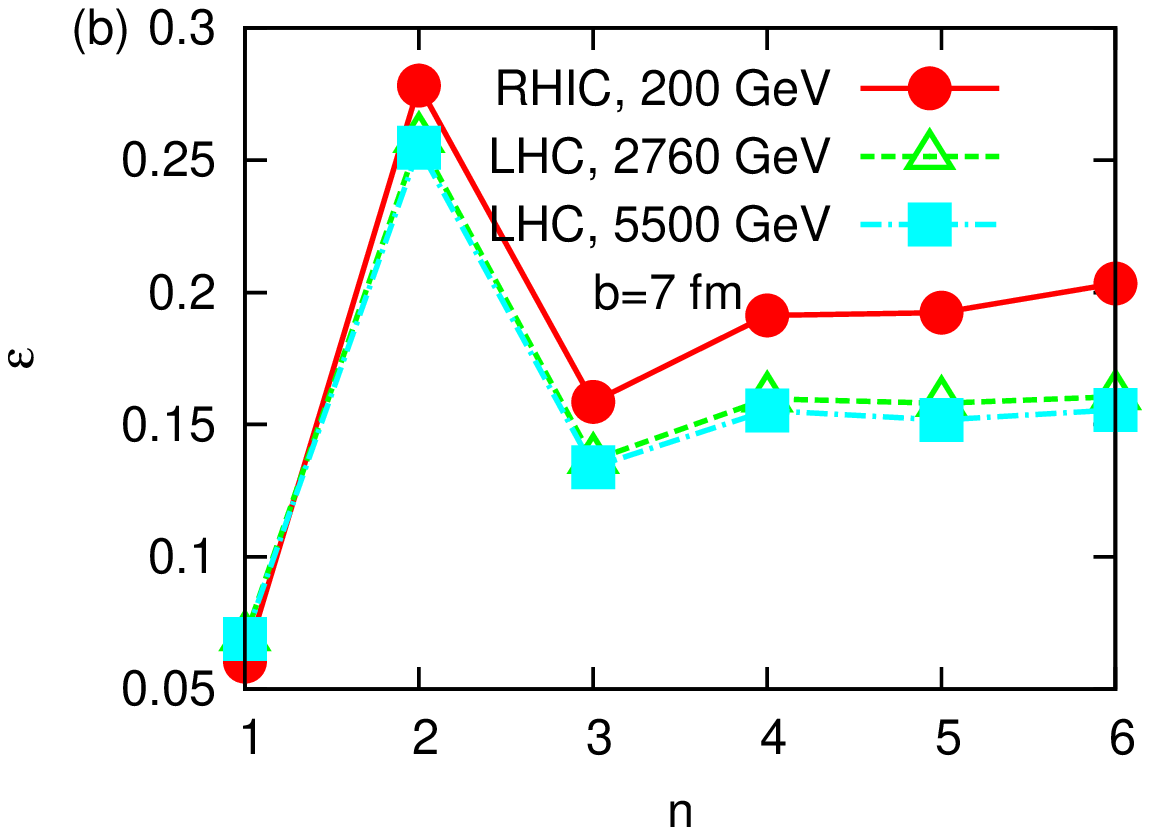}
\caption{(Color online). The spectrum of $\e{n}$ at $b=0, \ 7$ fm for RHIC with $\sqrt{s}=200$ GeV and for LHC with $\sqrt{s}=2760, \ 5500$ GeV.}
\label{fig:epsvsn}
\end{figure}  

\begin{figure}
\centering
\includegraphics[scale=0.6, angle=-90]{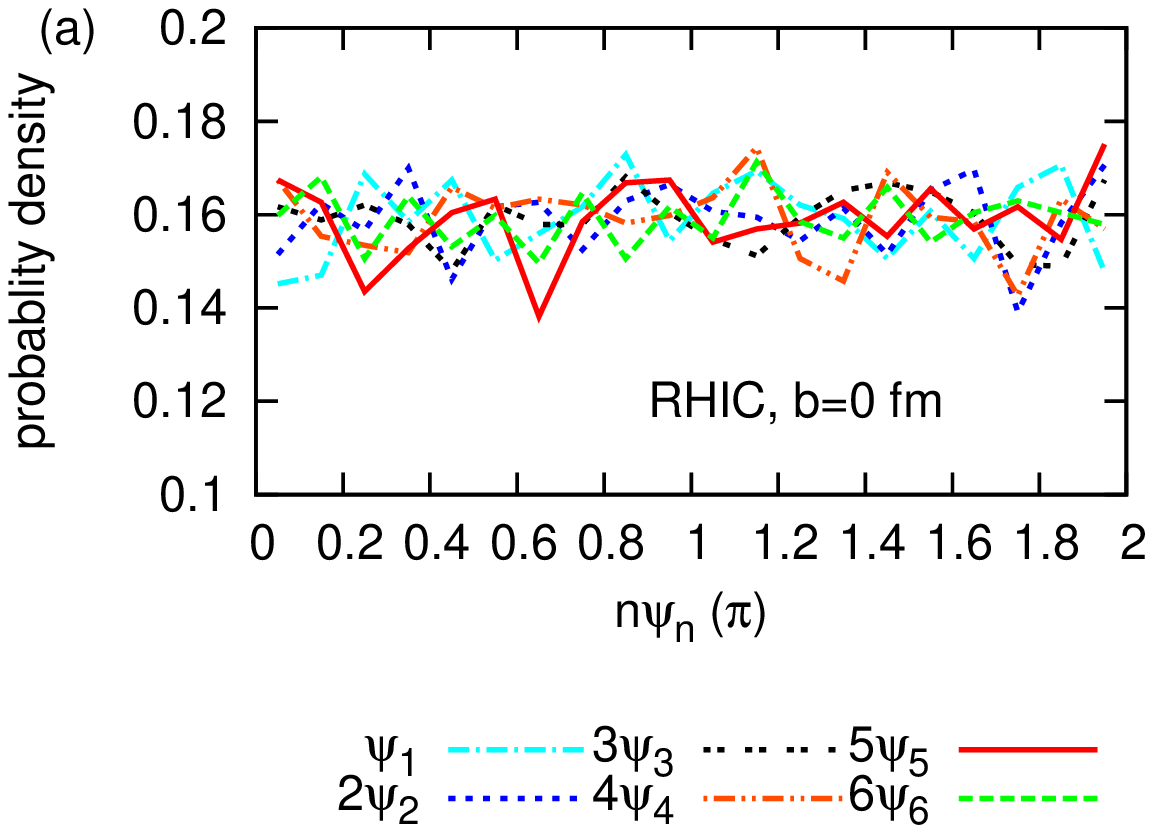}
\includegraphics[scale=0.6, angle=-90]{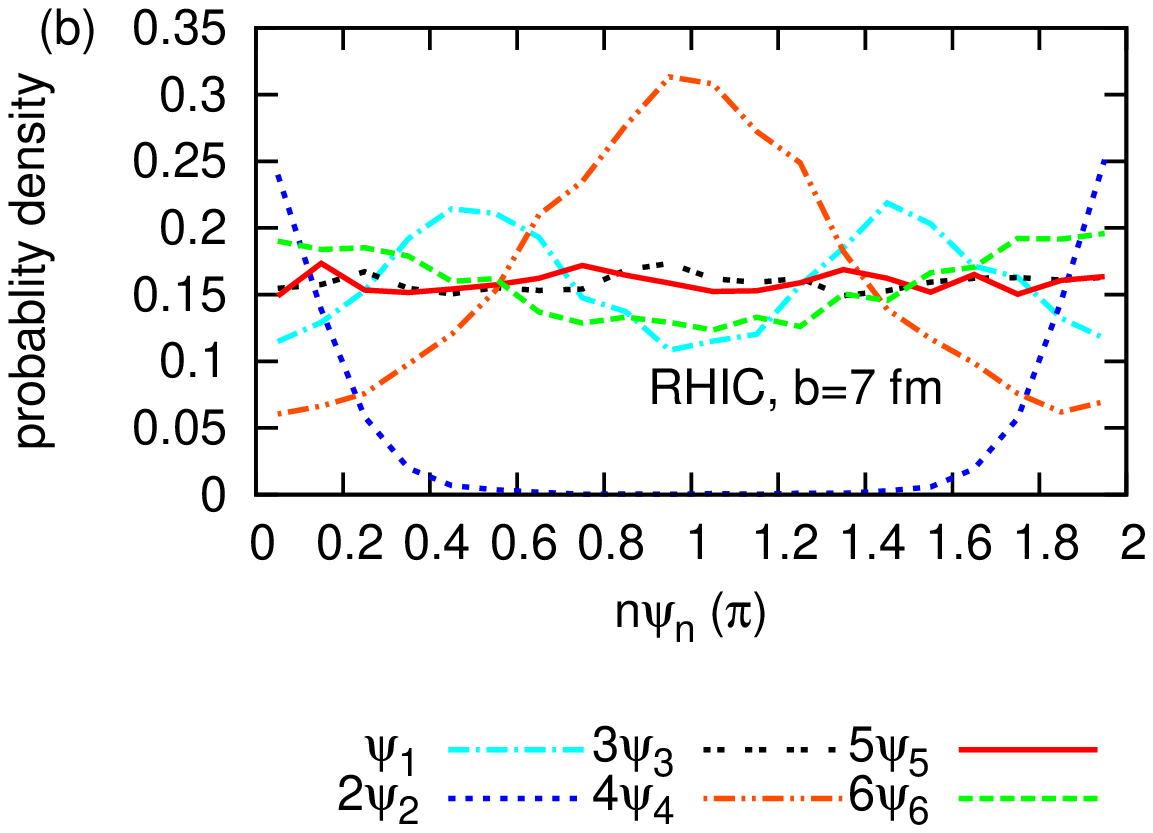}
\caption{(Color online). The distribution of $\psi_{n}$ for $b=0, \ 7$ fm at RHIC ($\sqrt{s}=200$ GeV). For $b=7$ fm, (2$\psi_{2}$)'s probability density is scaled down by a factor $0.3$. Similar results for LHC ($\sqrt{s}=2760, \ 5500$ GeV) can be found in Figure~\ref{fig:psidissupp}.}
\label{fig:psidis}
\end{figure}  

\begin{figure}
\centering
\includegraphics[scale=0.6, angle=-90]{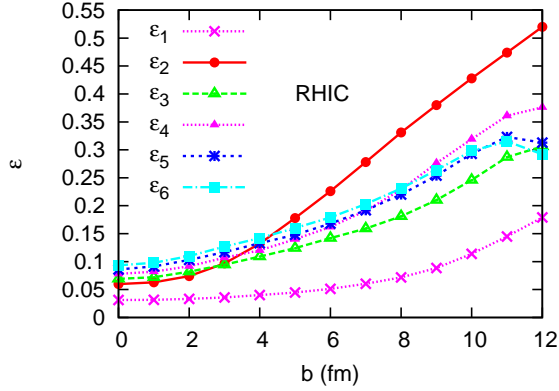}
\caption{(Color online). The $b$ dependence of $\e{n}$ for RHIC ($\sqrt{s}=200$ GeV). Similar results for LHC ($\sqrt{s}=2760, \ 5500$ GeV) can be found in Figure~\ref{fig:epsvsbsupp}.}
\label{fig:epsvsb}
\end{figure}

\begin{figure}
\centering
\includegraphics[scale=0.6, angle=-90]{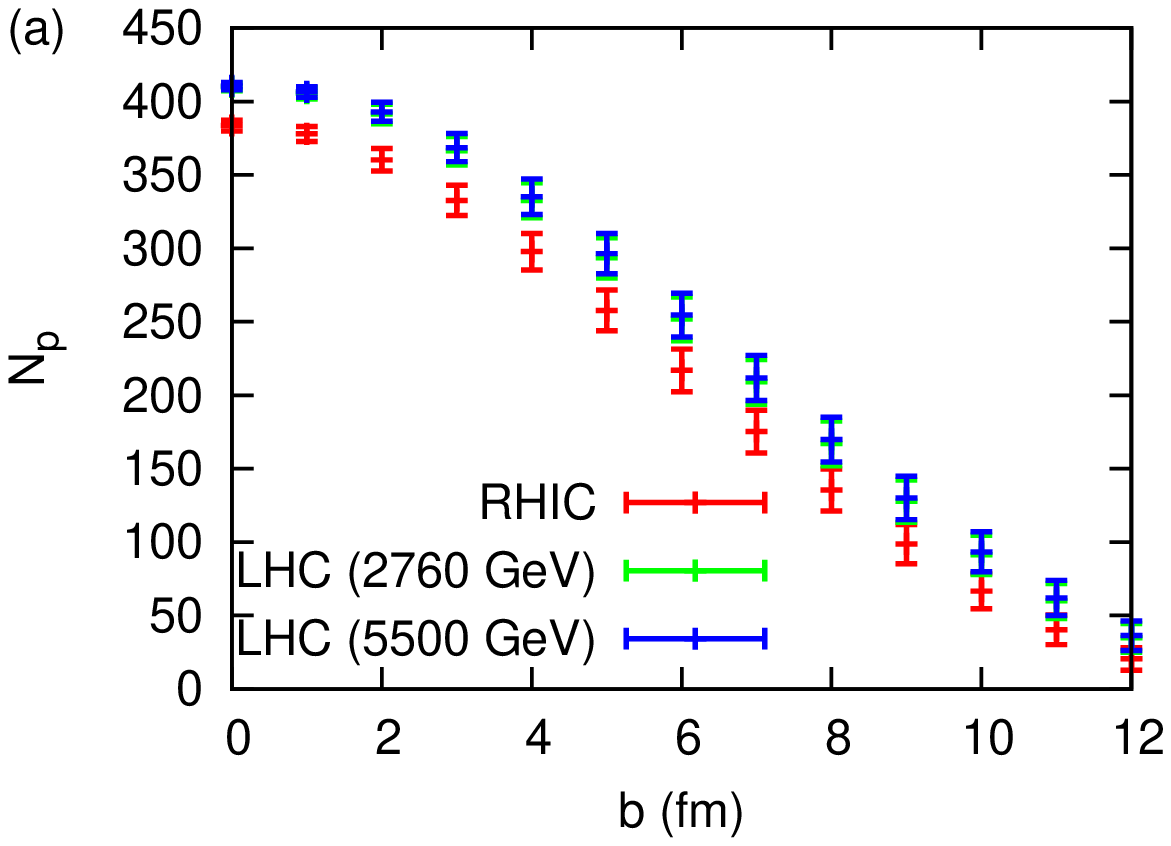}
\includegraphics[scale=0.6, angle=-90]{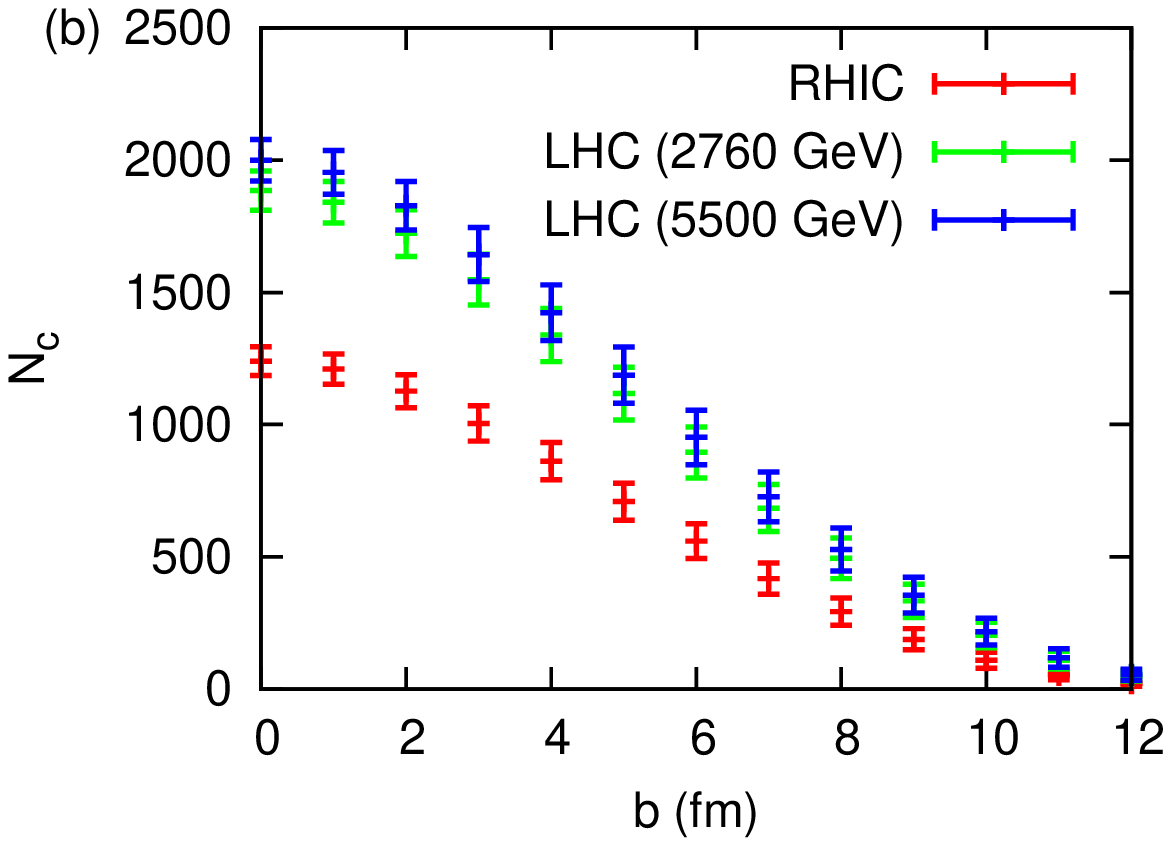}
\caption{(Color online). The $b$ dependence of $N_{p}$ and $N_{c}$  for both RHIC ($\sqrt{s}=200$ GeV) and LHC ($\sqrt{s}=2760, \ 5500$ GeV).}
\label{fig:NpNcvsb}
\end{figure}

Figure~\ref{fig:epsvsn} shows the $\e{n}$ spectrum for $b=0$ fm (most central collision) and $b=7$ fm (peripheral collision) at three different collision energies. For all the experiments, in the central collisions, $\e{n}$ are close  while in the peripheral collisions, $\e{2}$ dominates due to the almond-like geometry. Moreover, $\e{n}$ at LHC $2.76$ TeV and LHC $5.5$ TeV are almost the same. In Figs.~\ref{fig:psidis} and~\ref{fig:psidissupp}, we plot the distributions of $\psi_{n}$ relative to the reaction plane for $b=0$ and $7$ fm in three cases \citep{Teaney:2010vd}: in the central collisions, $\psi_{n}$ do not have specific orientation; in the peripheral collisions, $\psi_{1}$ and $\psi_{4}$ tend to lie in the $\pi/2$ and $\pi/4$ directions; $\psi_{2}$ is strongly correlated with the reaction plane (in the plot, $\psi_{2}$'s probability density is scaled down by a factor $0.3$); other odd $\psi_{n}$ distribute randomly. We also check that the correlation between any two angles vanishes in the central collisions. Figs.~\ref{fig:epsvsb} and~\ref{fig:epsvsbsupp} list the $b$ dependence of the eccentricities, which shows the similarity in three experiments and also the emergence of the second harmonic dominance in the peripheral collisions. In addition, Figure~\ref{fig:NpNcvsb} shows the relations between $N_{p}$ ($N_{c}$) and $b$. The error bars indicates their r.m.s values, which are small except in the very peripheral collisions. In the following, we will show the results against $b$. Figure~\ref{fig:NpNcvsb} can be used to compute the corresponding $N_{p}$ dependence.

\subsection{The jet azimuthal anisotropy}

\begin{figure}
\centering
\includegraphics[scale=0.6, angle=-90]{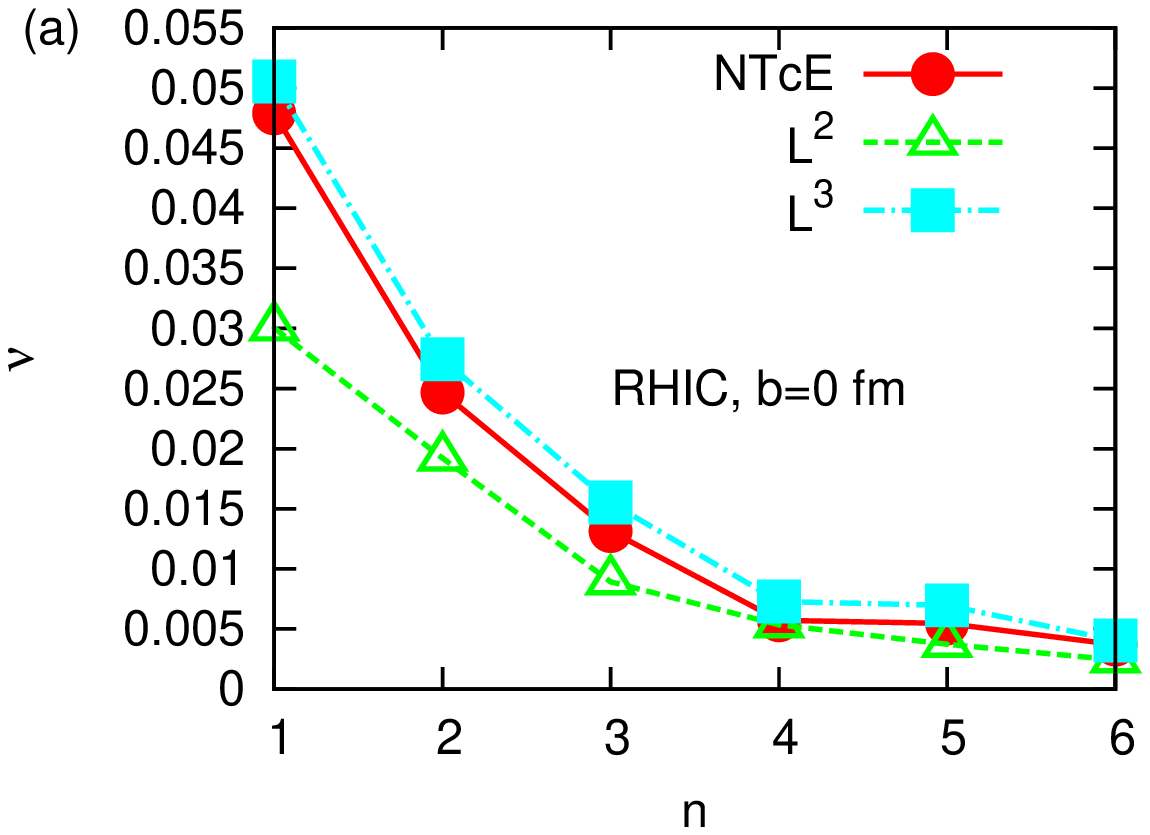}
\includegraphics[scale=0.6, angle=-90]{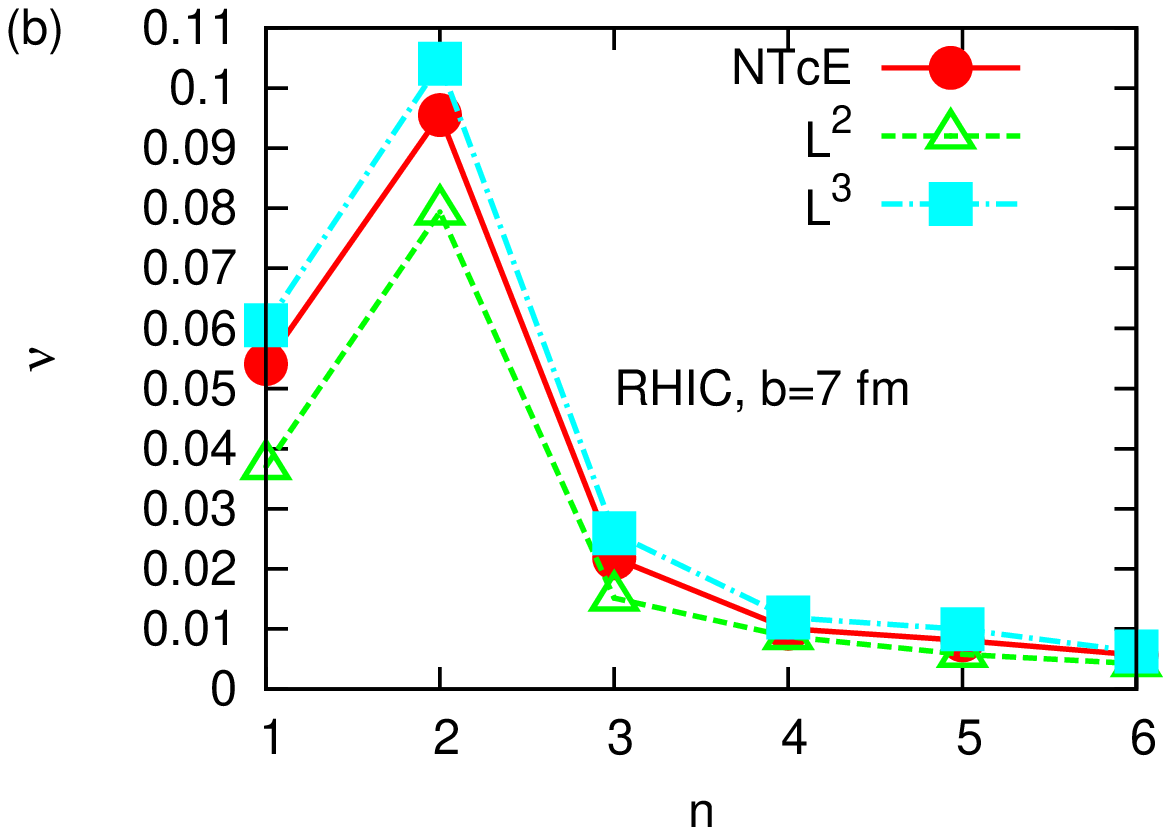}
\includegraphics[scale=0.6, angle=-90]{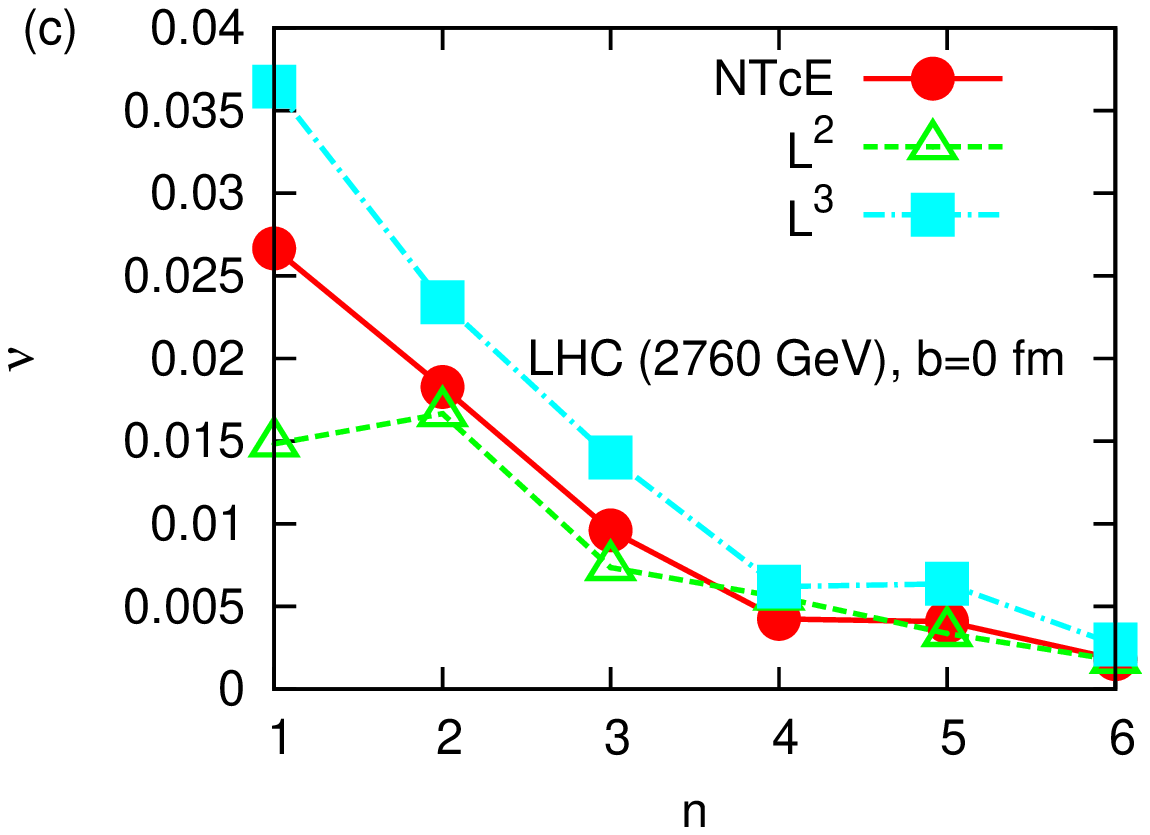}
\includegraphics[scale=0.6, angle=-90]{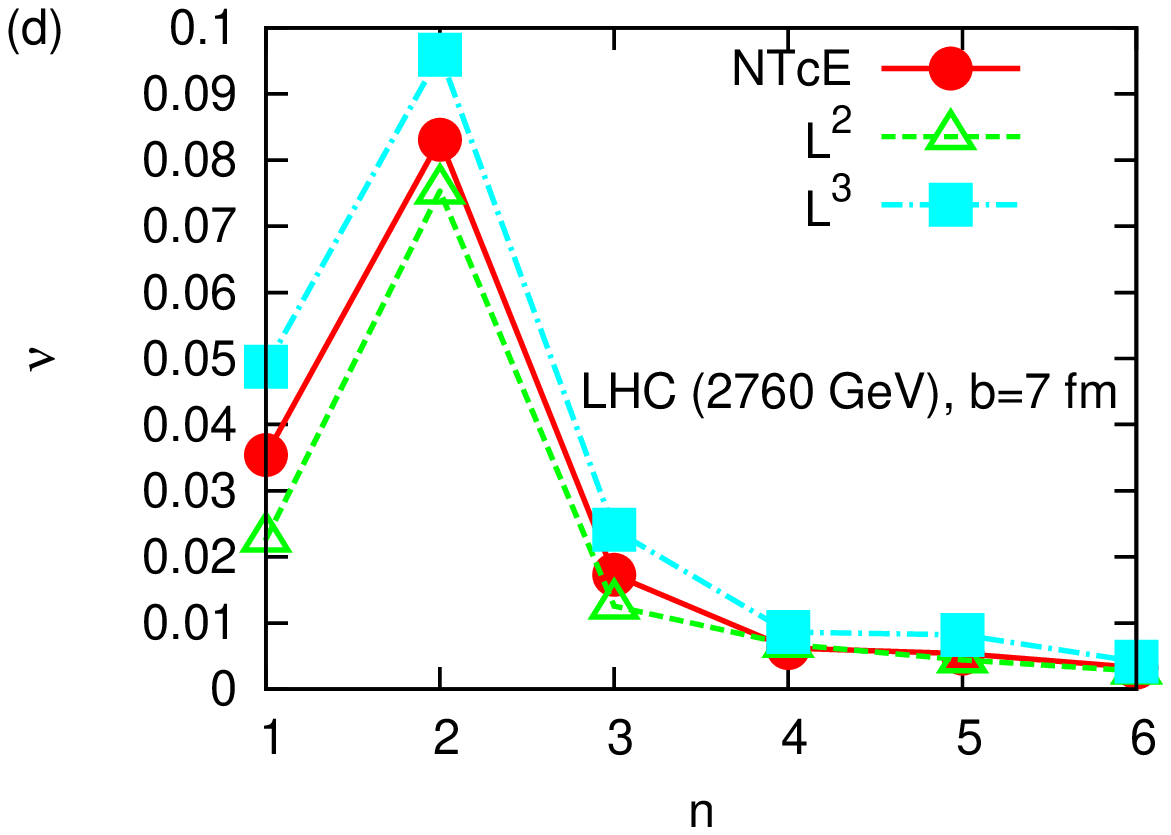}
\caption{(Color online). The spectrum of $\nuh{n}$ for RHIC ($\sqrt{s}=200$ GeV) and LHC ($\sqrt{s}=2760$ GeV) at $b=0, \ 7$ fm,  based on three different models. Similar results for LHC ($\sqrt{s}=5500$ GeV) can be found in Figure~\ref{fig:nuspectrumRHICLHCsupp}.}
\label{fig:nuspectrumRHICLHC}
\end{figure}

\begin{figure}
\centering
\includegraphics[scale=0.6, angle=-90]{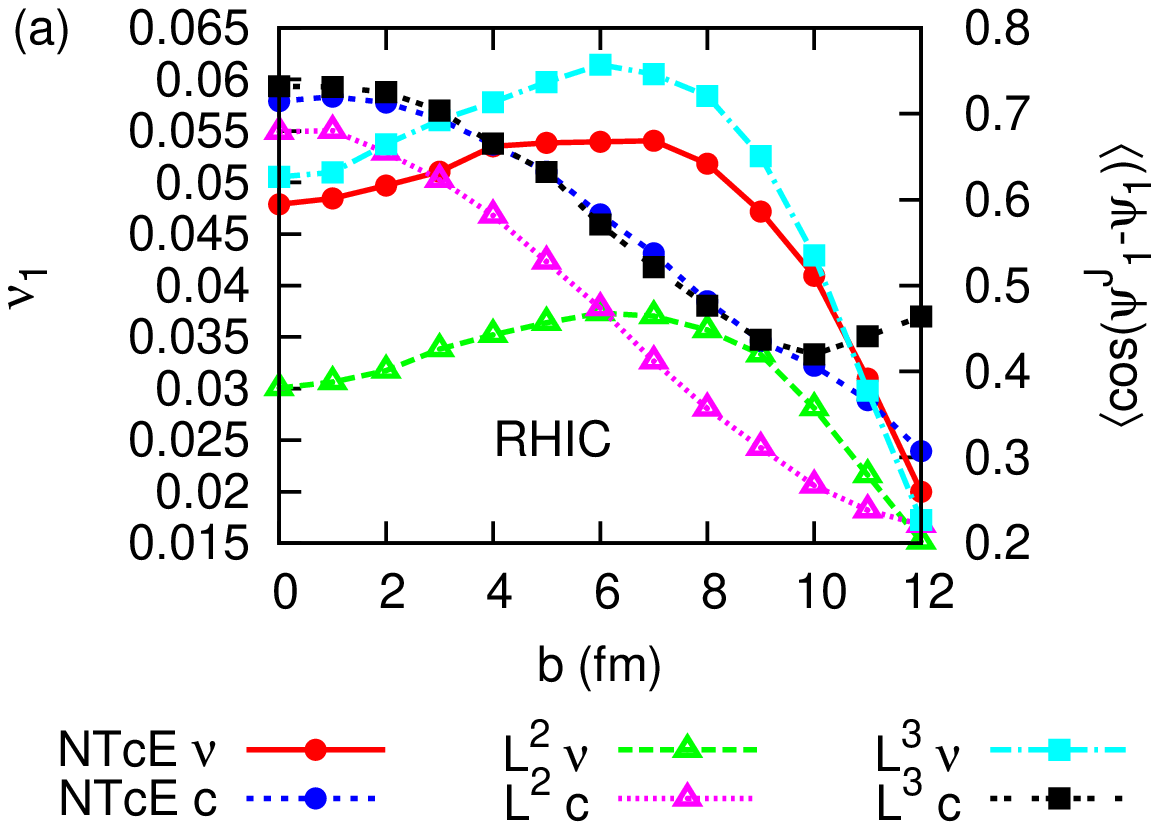}
\includegraphics[scale=0.6, angle=-90]{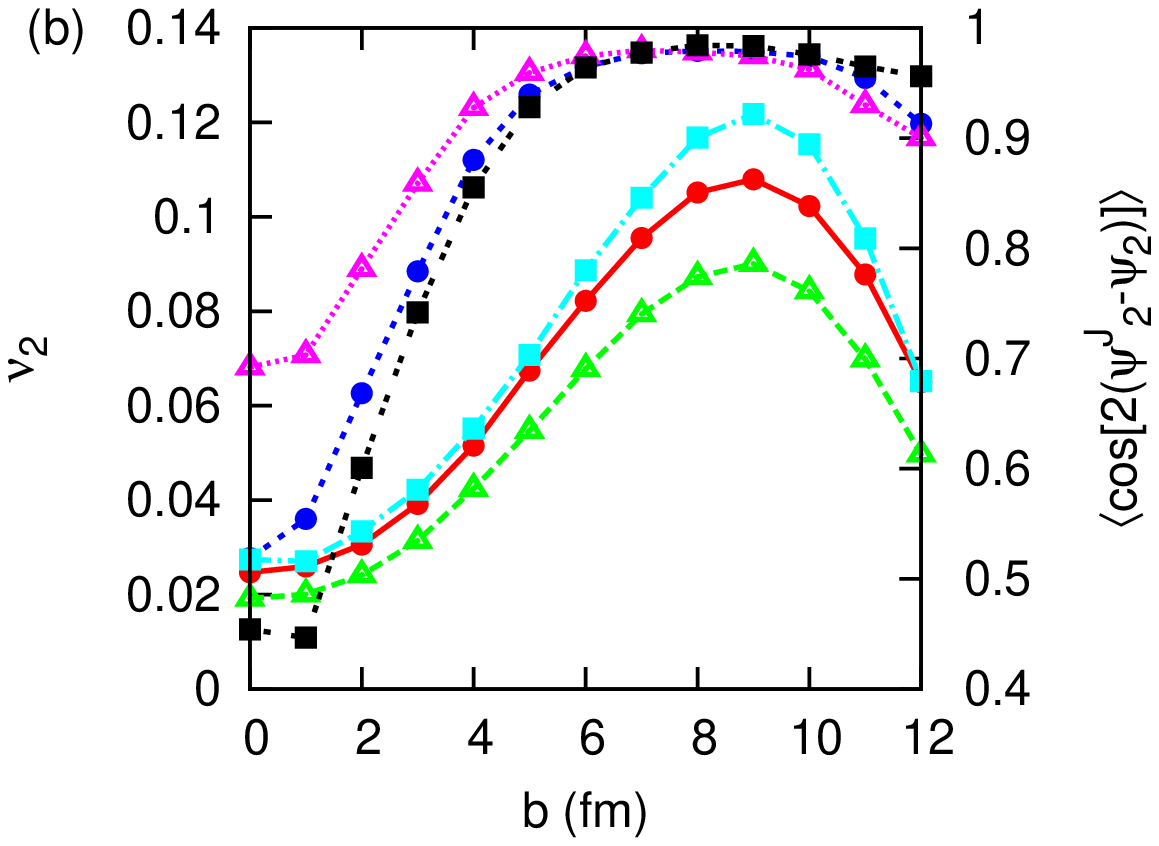}
\includegraphics[scale=0.6, angle=-90]{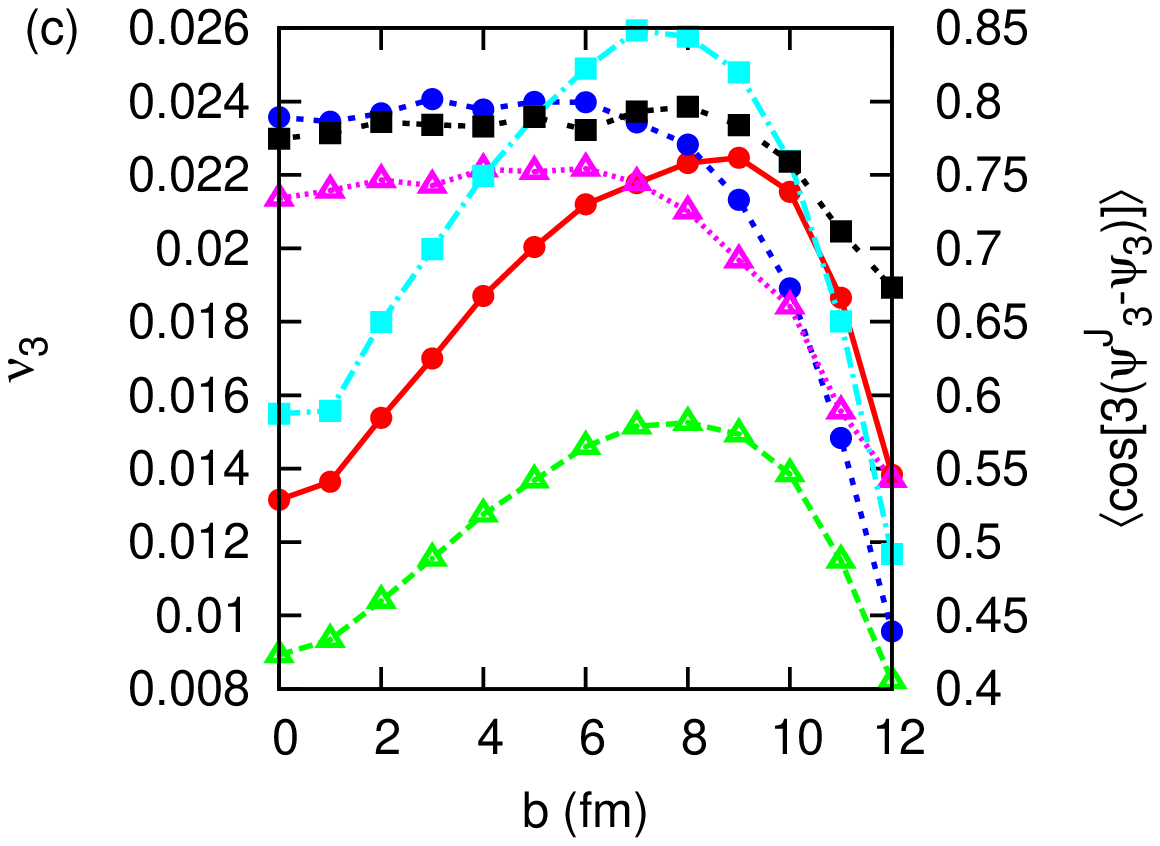}
\includegraphics[scale=0.6, angle=-90]{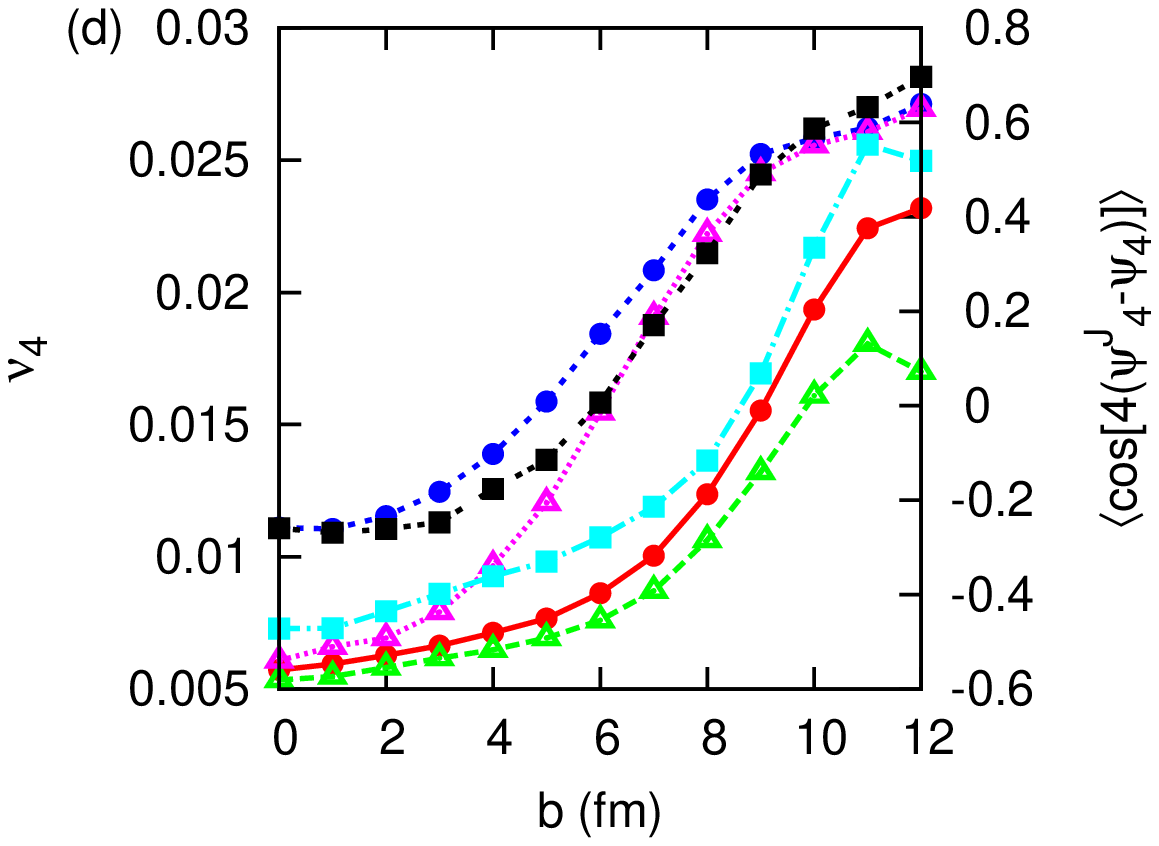}
\includegraphics[scale=0.6, angle=-90]{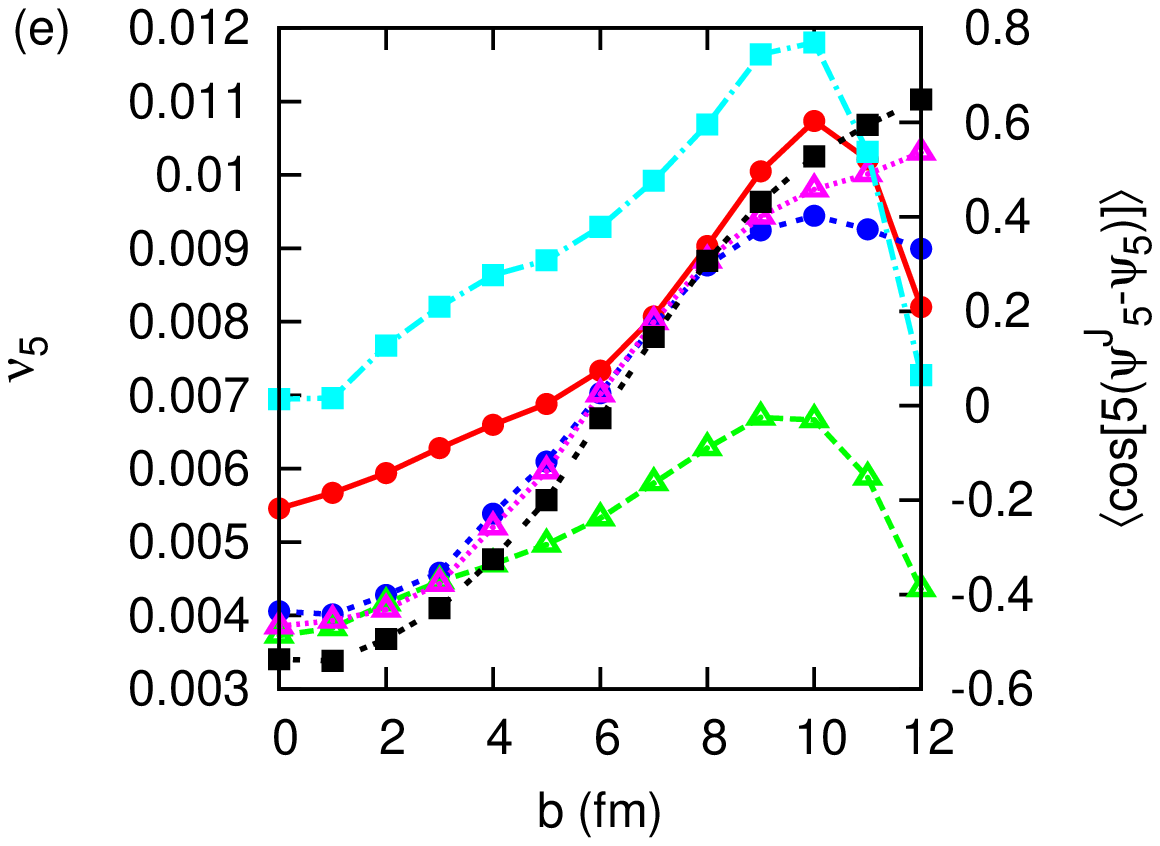}
\includegraphics[scale=0.6, angle=-90]{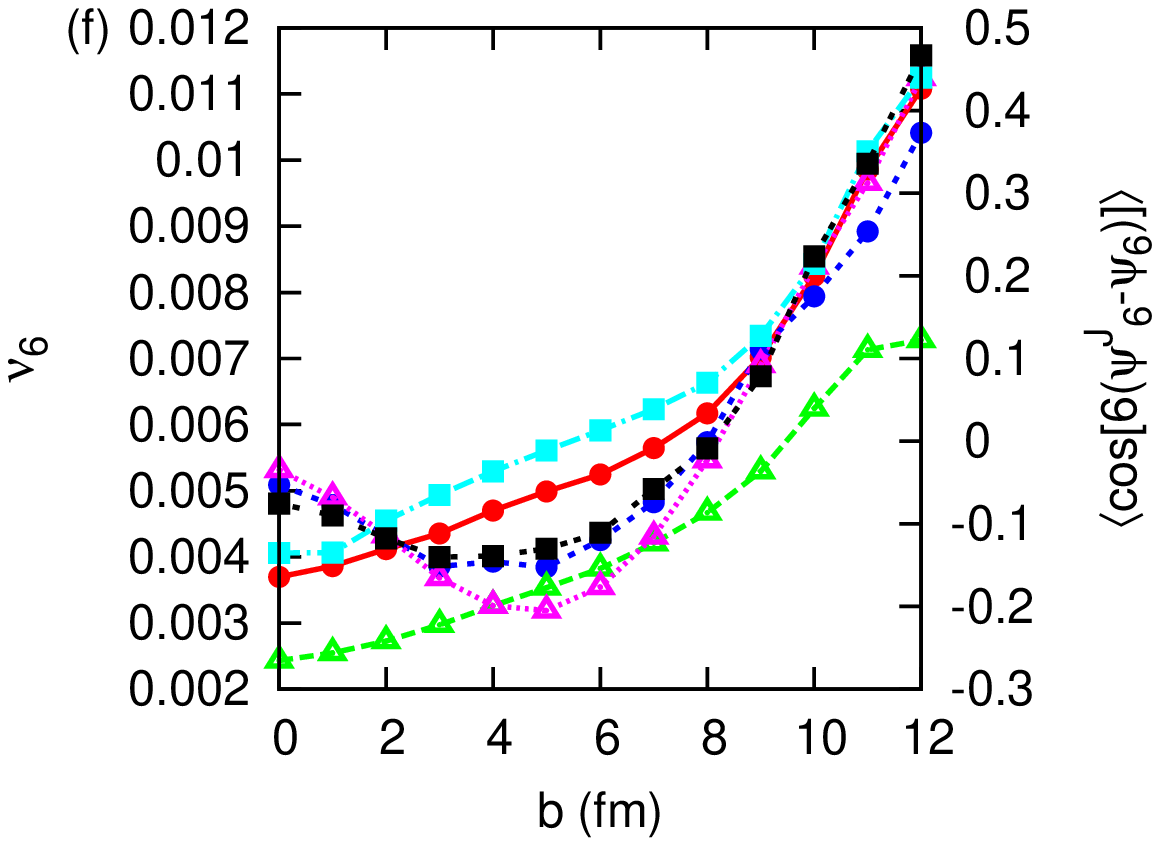}
\caption{(Color online). The $b$ dependence of $\nuh{n}$ and $\ave{\cos[n(\psih{n}-\psi_{n})]}$ for RHIC ($\sqrt{s}=200$ GeV) based on three different models.}
\label{fig:nuvsbRHIC}
\end{figure}  

\begin{figure}
\centering
\includegraphics[scale=0.6, angle=-90]{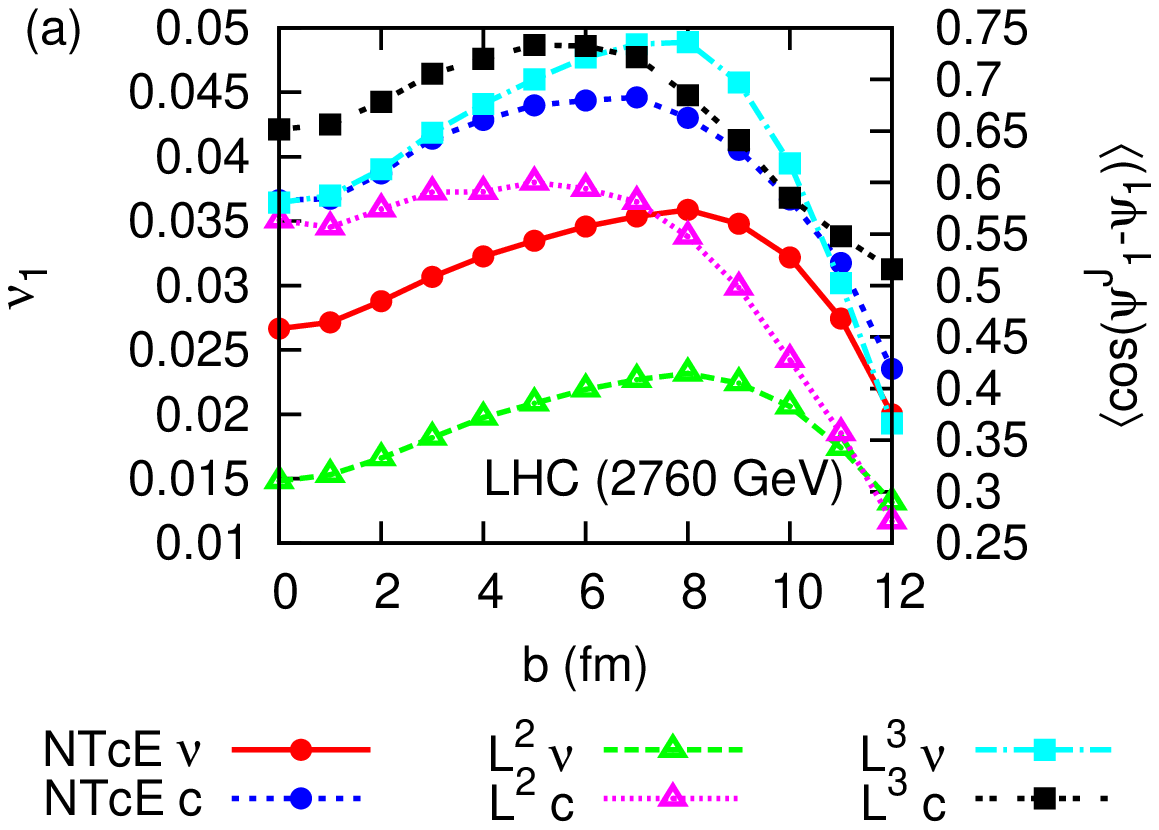}
\includegraphics[scale=0.6, angle=-90]{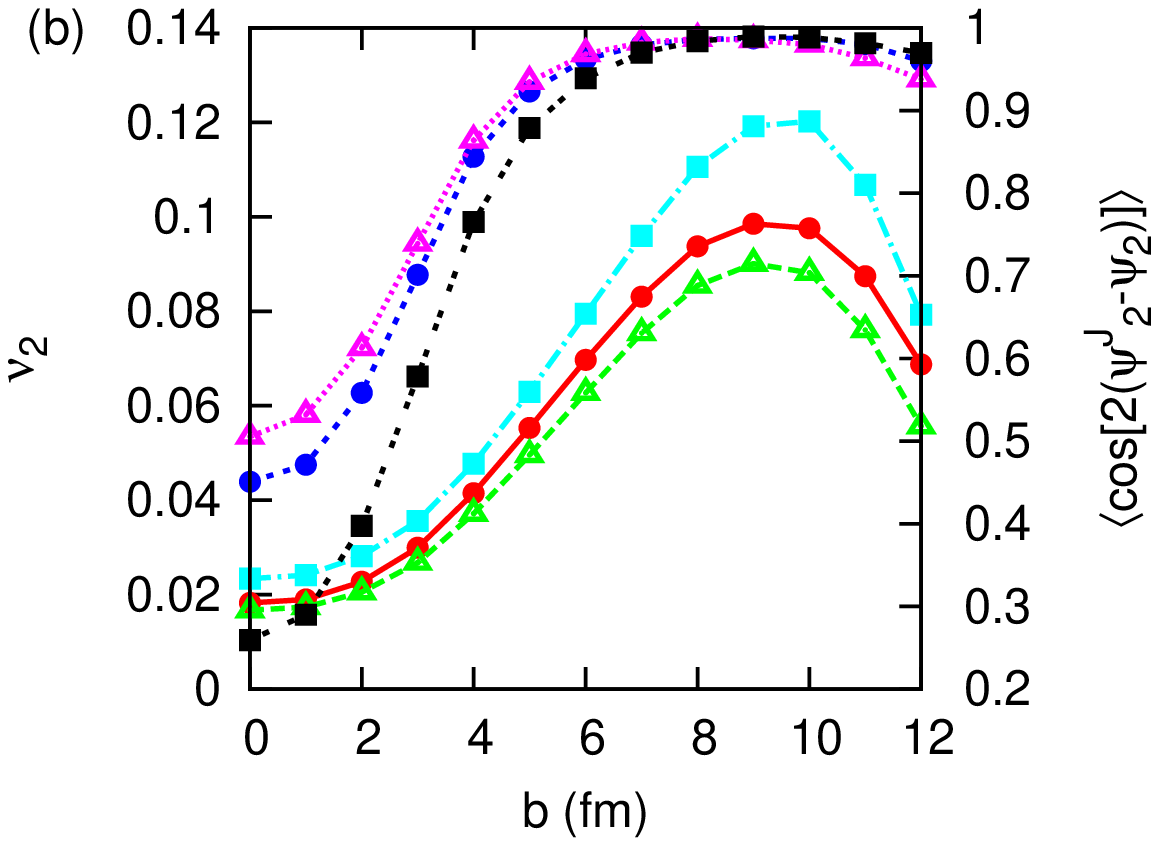}
\includegraphics[scale=0.6, angle=-90]{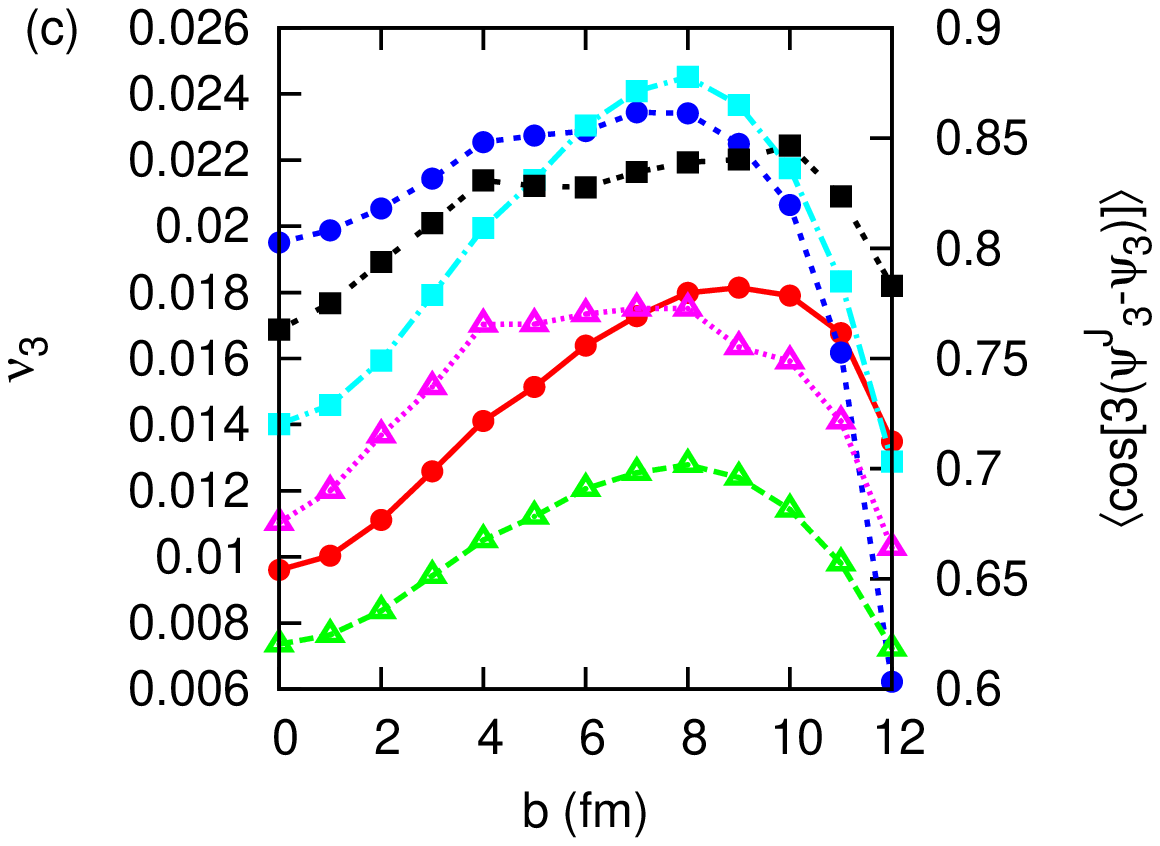}
\includegraphics[scale=0.6, angle=-90]{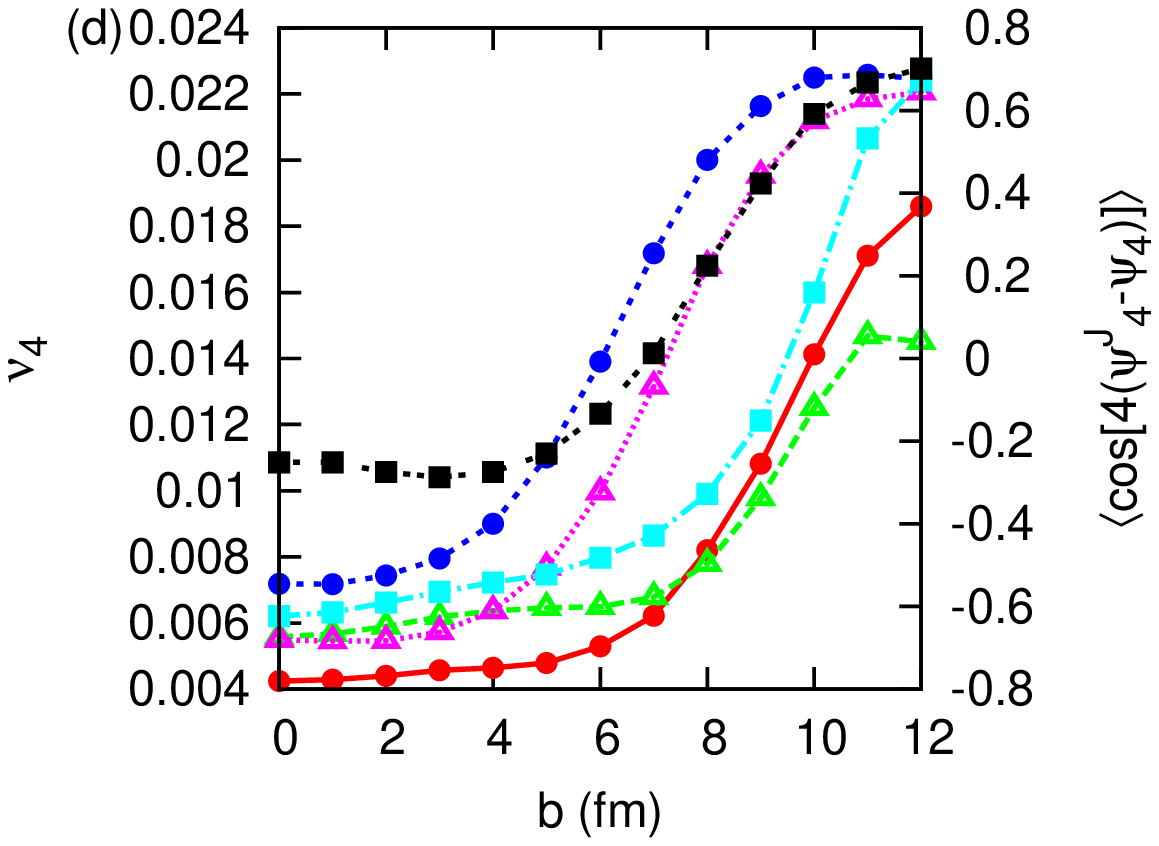}
\includegraphics[scale=0.6, angle=-90]{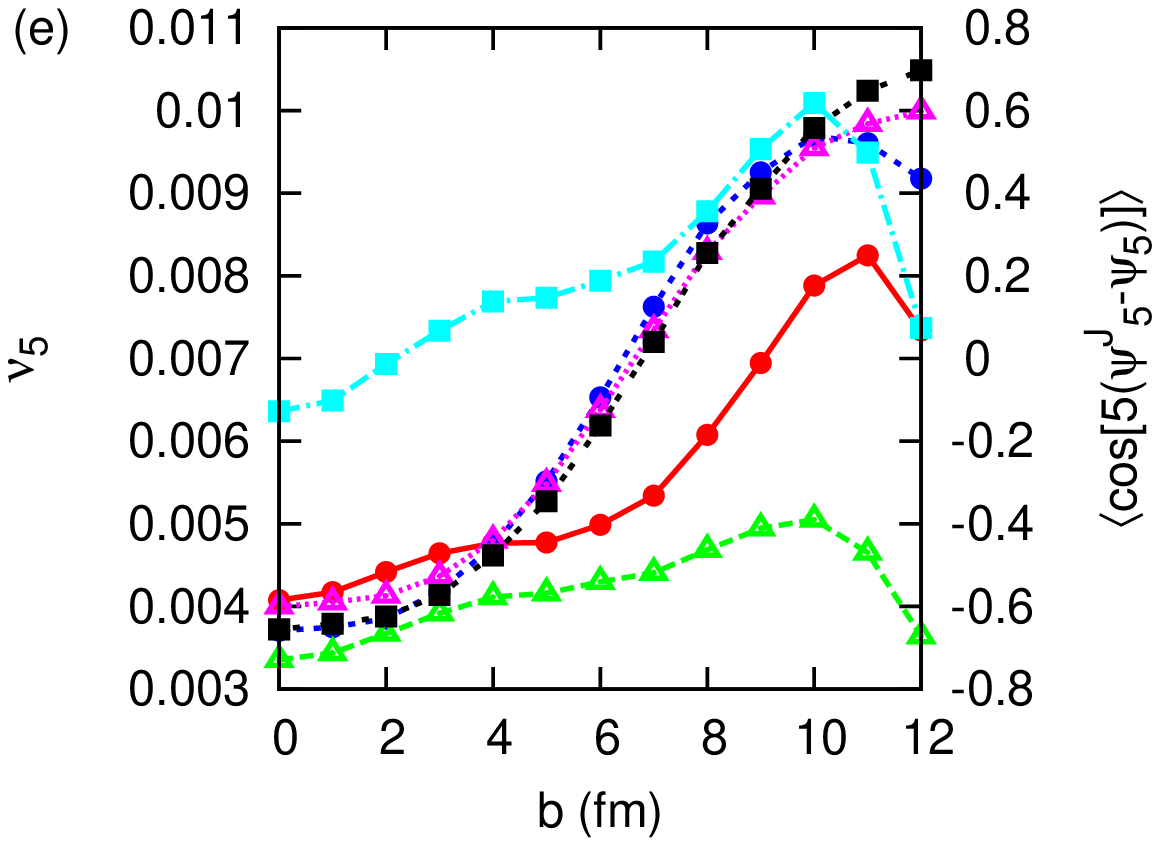}
\includegraphics[scale=0.6, angle=-90]{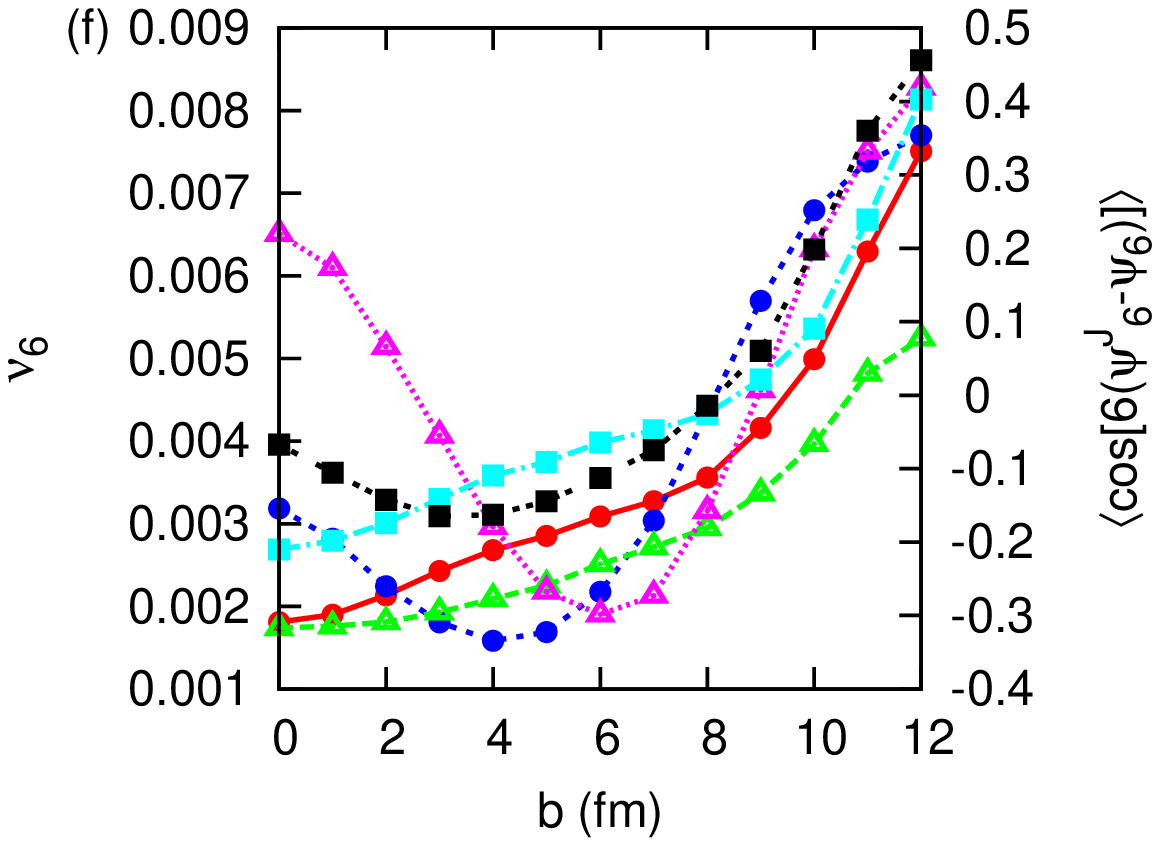}
\caption{(Color online). The $b$ dependence of $\nuh{n}$ and $\ave{\cos[n(\psih{n}-\psi_{n})]}$ for LHC ($\sqrt{s}=2760$ GeV) based on three different models.}
\label{fig:nuvsbLHC2760}
\end{figure}

\begin{figure}
\centering
\includegraphics[scale=0.6, angle=-90]{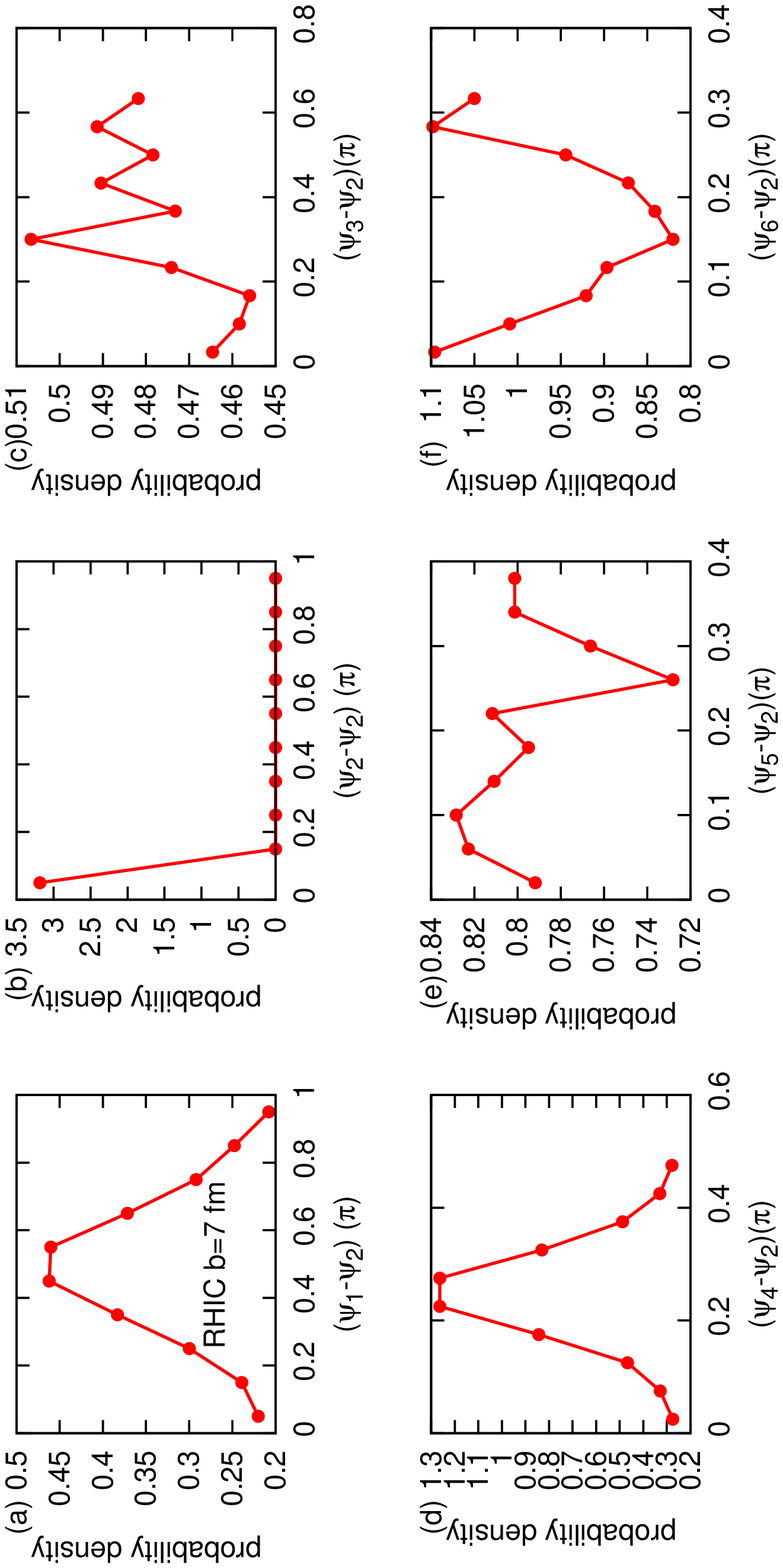}
\includegraphics[scale=0.6, angle=-90]{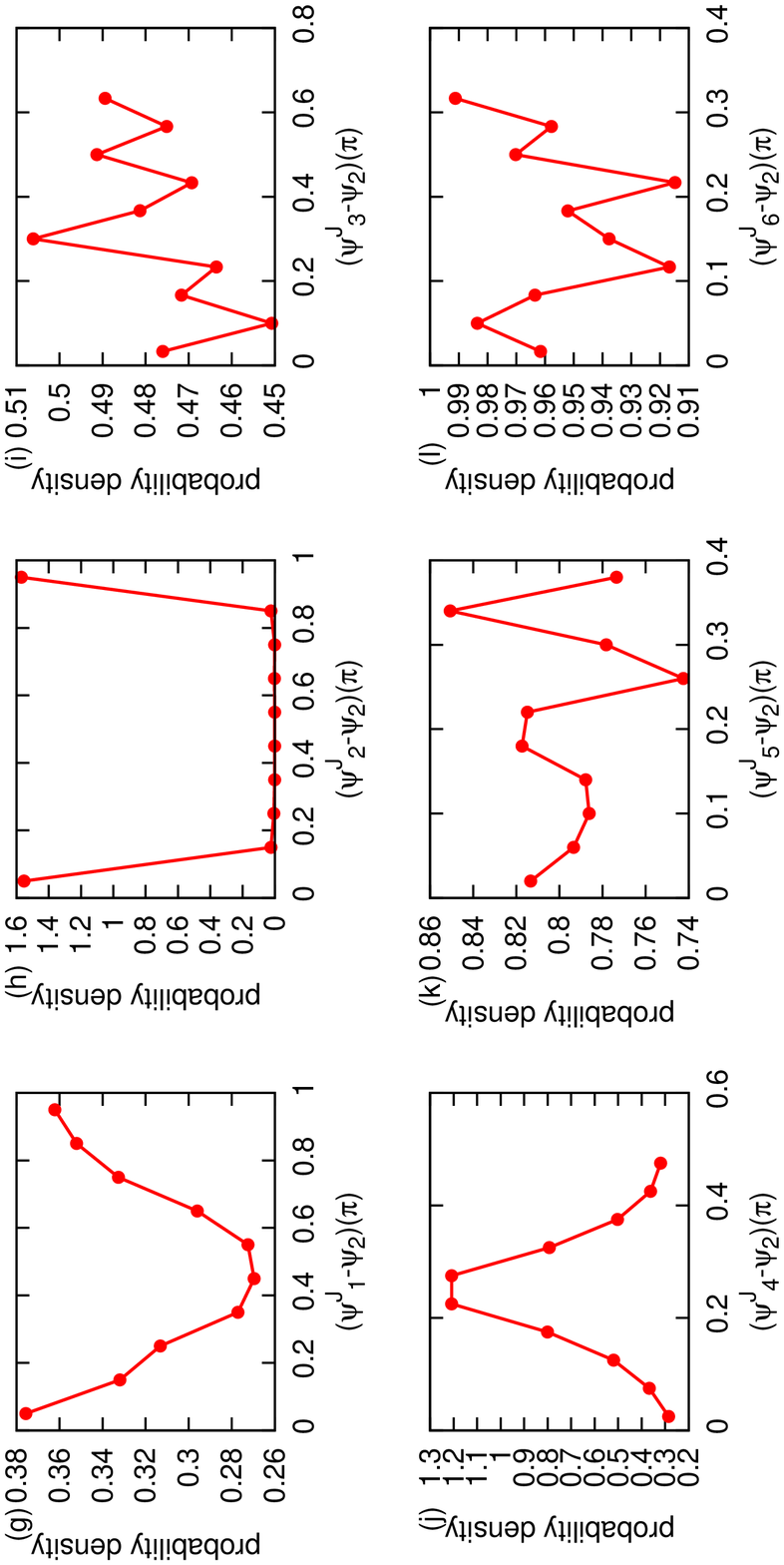}
\caption{(Color online). The distribution of $\psi_{n}-\psi_{2}$ and $\psih{n}-\psi_{2}$ for RHIC ($\sqrt{s}=200$ GeV) calculated by using the NTcE model.}
\label{fig:psi_psij_pp_dis_RHIC_b_7}
\end{figure}

In this section, we focus on results for the jet azimuthal anisotropy. 
Figs.~\ref{fig:nuspectrumRHICLHC} and~\ref{fig:nuspectrumRHICLHCsupp} show the $\nuh{n}$ spectrum [see Eq.~(\ref{eqn:Raadecomp})] in both most central collision ($b=0$ fm) and peripheral collision ($b=7$ fm) as calculated by using different jet energy loss models. We can see that the second harmonics dominate in the peripheral collision reflecting the almond-like geometry in the peripheral collisions, and other harmonics decrease with increasing $n$ in both central and peripheral collisions. More importantly, for the first three harmonics, the NTcE and $\mathrm{L}^{3}$ result for RHIC ($0.2$ TeV) are close, while for LHC ($2.76, \ 5.5$ TeV) the NTcE and $\mathrm{L}^{2}$ result are close. The $b$ dependence of $\nuh{n}$ in the three experiments as predicted by the models is shown in Figs.~\ref{fig:nuvsbRHIC}, \ref{fig:nuvsbLHC2760}, and~\ref{fig:nuvsbLHC5500} (curves labeled as ``model $\nu$''). $< \cos[n(\psih{n}-\psi_{n})] >$, which reflects the correlation between $\psih{n}$ and $\psi_{n}$ \citep{jiajq12}, is also plotted in these figures against different $b$ for different harmonics (curves labeled as ``model $c$'').  
As illustrations, Figs.~\ref{fig:psij-psi_corr_RHIC}, \ref{fig:psij-psi_corr_LHC_2760GeV}, and~\ref{fig:psij-psi_corr_LHC_5500GeV}, at the end of this article, show the distribution of $\psih{n}-\psi_{n}$ computed by using different models for $b=0$ and $7$ fm in the three experiments. Interestingly, the authors in Ref.~\citep{ZhiQiu2011PRC84} presented similar distribution of $\psis{n}-\psi_{n}$ [Here $\psis{n}$ is the angle of $n$th harmonic event plane reconstructed from the final low $p_t$ hadron azimuthal distribution;  see Eq.~(\ref{eqn:psisn})]. By comparing the two, we find that for the second and third harmonics, $\psih{n}$ and $\psis{n}$ have similar distribution relative to the $\psi_{n}$, although $\psih{n}$ distributions computed by three models differ somewhat; for the other higher harmonics, $\psih{n}$ and $\psis{n}$ gains much broader distribution relative to $\psi_{n}$, and $< \cos[n(\psih{n}-\psi_{n})] >$ changes from positive to negative with decreasing $b$ (the value of the transition $b$ depends on individual jet energy loss model). 
On the other hand, the experimental measurement of $\nuh{n}$ is normally not the $\nuh{n}$ shown in Figs.~\ref{fig:nuvsbRHIC}, \ref{fig:nuvsbLHC2760}, and~\ref{fig:nuvsbLHC5500}, but its projection to $\psis{n}$, i.e., $<\nuh{h} \cos[n(\psih{n}-\psis{n})] >$. \footnote{The other measurement of $\nuh{n}$ is to project the $\nuh{n}$ to the second harmonic event plane, i.e., $<\nuh{h} \cos[n(\psih{n}-\psis{2})] >$.} Because of missing soft dynamics in our simulation, we can not calculate $<\cos[n(\psih{n}-\psis{n})]> $ at this stage. However, for the second harmonics $\nuh{2}$, based on the $<\cos[2(\psih{2}-\psi_{2})]> $ shown here and the $\psis{2}-\psi_{2}$ distribution shown in Ref.~\citep{ZhiQiu2011PRC84}, the effect of this angle dispersion should be less than $5\%$ reduction for peripheral collision ($b\geqslant 6$ fm).  This provides solid ground for comparing the data for $\nuh{2}$ with our results to differentiate different jet energy loss models, as carried out in \citep{xzjl2012jqst}. There it was found that the RHIC $\nuh{2}$ data favor the NTcE and $\mathrm{L}^{3}$ models, while the LHC ($2.76$ TeV) $\nuh{2}$ data favor the NTcE and $\mathrm{L}^{2}$ models. This is consistent with $\nuh{2}$ shown in Figs.~\ref{fig:nuvsbRHIC}, \ref{fig:nuvsbLHC2760}, and  \ref{fig:nuvsbLHC5500}. Meanwhile, $\nuh{1}$ and $\nuh{3}$ deliver valuable information about the orders of magnitude of the measured harmonics, although the angle dispersion can bring substantial reduction yet without changing the sign. It is conceivable that the experimental data on $\nuh{n}$ will bring important constraints to the jet quenching models. A tentative try along this direction is pursued in Ref.~\citep{xzjl2012jqst}. 

Let us also take a look at the correlation between $\psih{n}$ and the participant plane $\psi_{2}$. For the central collisions, we can expect $\psih{n}$ to distribute randomly with respect to $\psi_{2}$, which is confirmed by our simulation results for RHIC ($0.2$ TeV). However, in the peripheral collisions, things become tricky. Figure~\ref{fig:psi_psij_pp_dis_RHIC_b_7} shows the $\psi_{n}-\psi_{2}$ and $\psih{n}-\psi_{2}$ distributions at RHIC ($0.2$ TeV) with $b=7$ fm. \footnote{The $\psi_{2}-\psi_{2}$ distribution in principle should be a delta function. The plot (b) of Figure~\ref{fig:psi_psij_pp_dis_RHIC_b_7} serves as a confirmation of our numerics.} For the ISF eccentricities, $\psi_{1}-\psi_{2}$ and $\psi_{4}-\psi_{2}$ are mostly around $\pi/2$ and $\pi/4$, while $\psi_{3}$ distribute randomly. The $\psih{n}-\psi_{2}$ distribution (in Figure~\ref{fig:psi_psij_pp_dis_RHIC_b_7}) seems consistent with the combined information about $\psi_{n}-\psi_{2}$ distribution (in Figure~\ref{fig:psi_psij_pp_dis_RHIC_b_7}) and $\psih{n}-\psi_{n}$ distribution (in Figure~\ref{fig:psij-psi_corr_RHIC}) except for the $1st$ harmonics. From $\psih{1}-\psi_{1}$ and $\psi_{1}-\psi_{2}$ distribution, we expect  $\psih{1}-\psi_{2}$ to be around $\pi/2$, but Figure~\ref{fig:psi_psij_pp_dis_RHIC_b_7} shows $\psih{1}-\psi_{2}$ to be around $0$. This problem is resolved by noticing that the $\psih{1}-\psi_{1}$ distribution is broader when $\psi_{1}-\psi_{2}$ is around $\pi/2$ compared to when $\psi_{1}-\psi_{2}$ is around $0$. Furthermore it implies the  conventional picture that $\psih{1}$ (or $\psis{1}$) randomly distributes is flawed.

\begin{figure}
\centering
\includegraphics[scale=0.6, angle=-90]{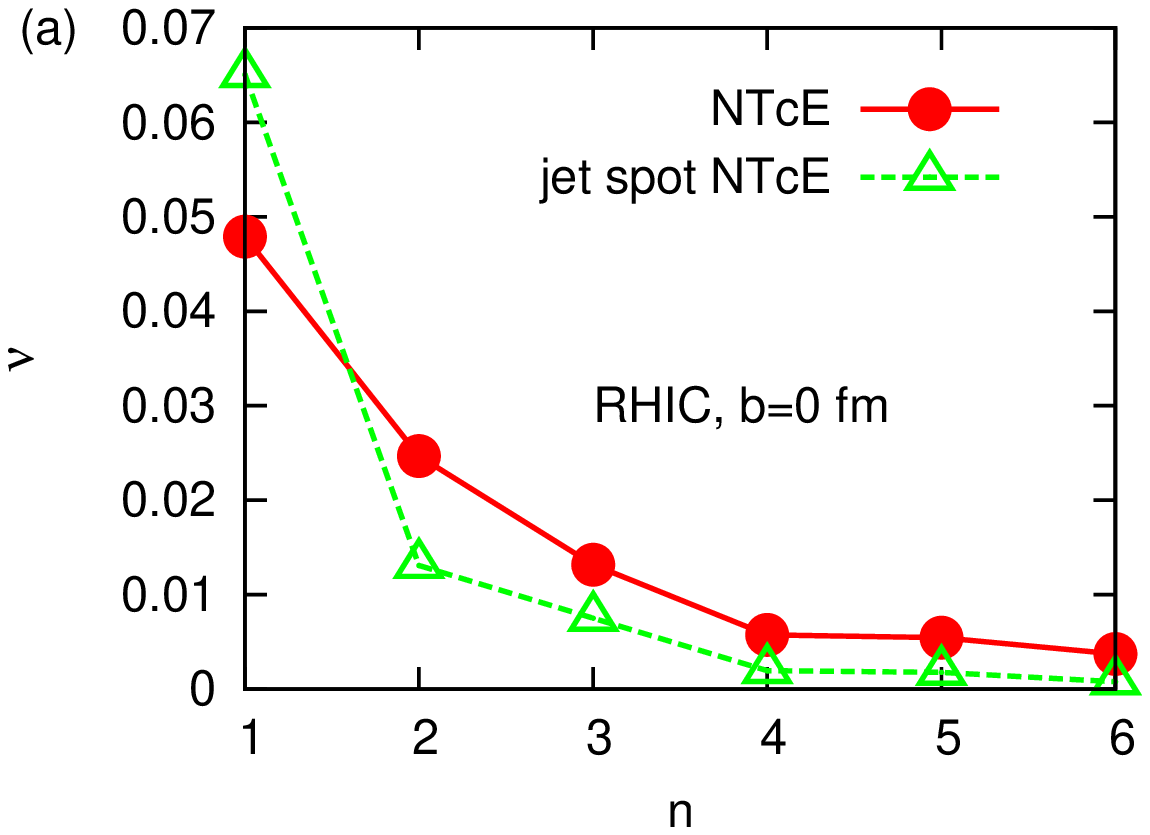}
\includegraphics[scale=0.6, angle=-90]{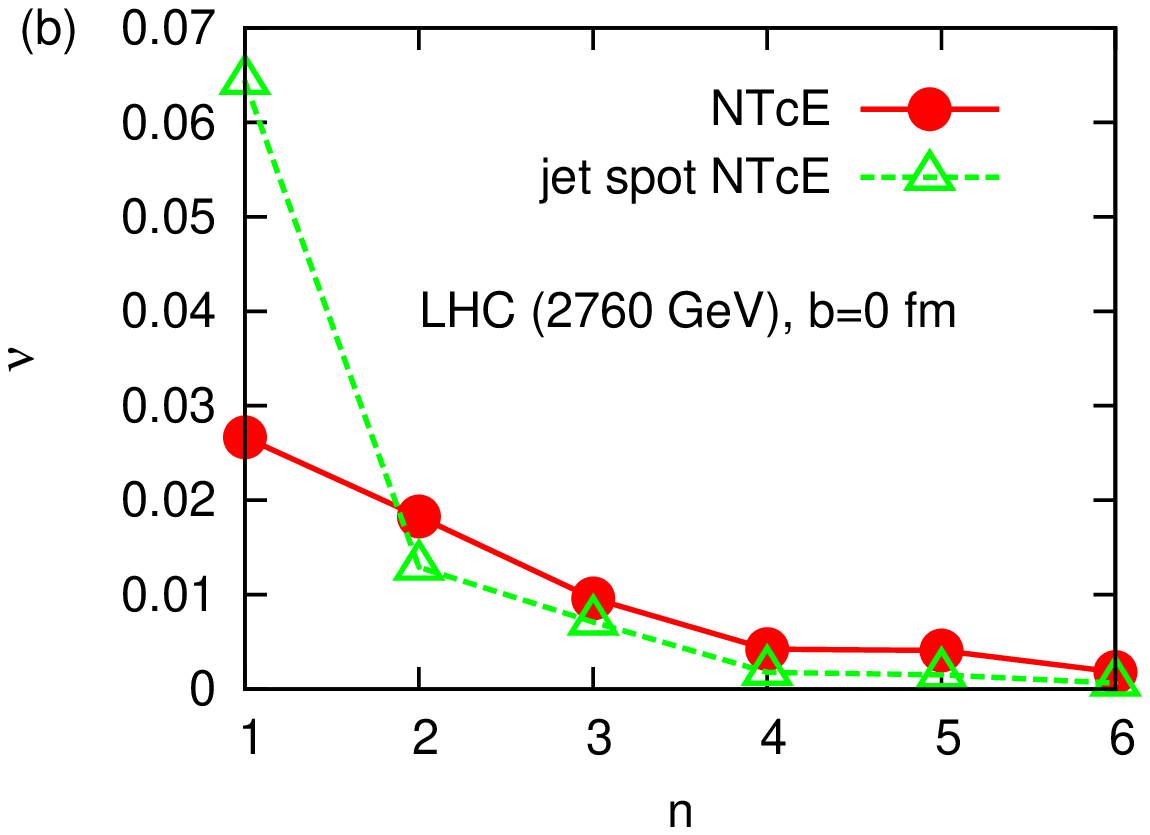}
\caption{(Color online). The spectrum of $\nuh{n}$ for both RHIC ($\sqrt{s}=200$ GeV) and LHC ($\sqrt{s}=2760$ GeV) central collisions due to full calculation of NTcE model (``NTcE'' curve) and the contribution of the jet spot fluctuation in NTcE model (``jet spot NTcE'' curve).}
\label{fig:nuspectrumrhoc}
\end{figure}

\begin{figure}
\centering
\includegraphics[scale=0.6, angle=-90]{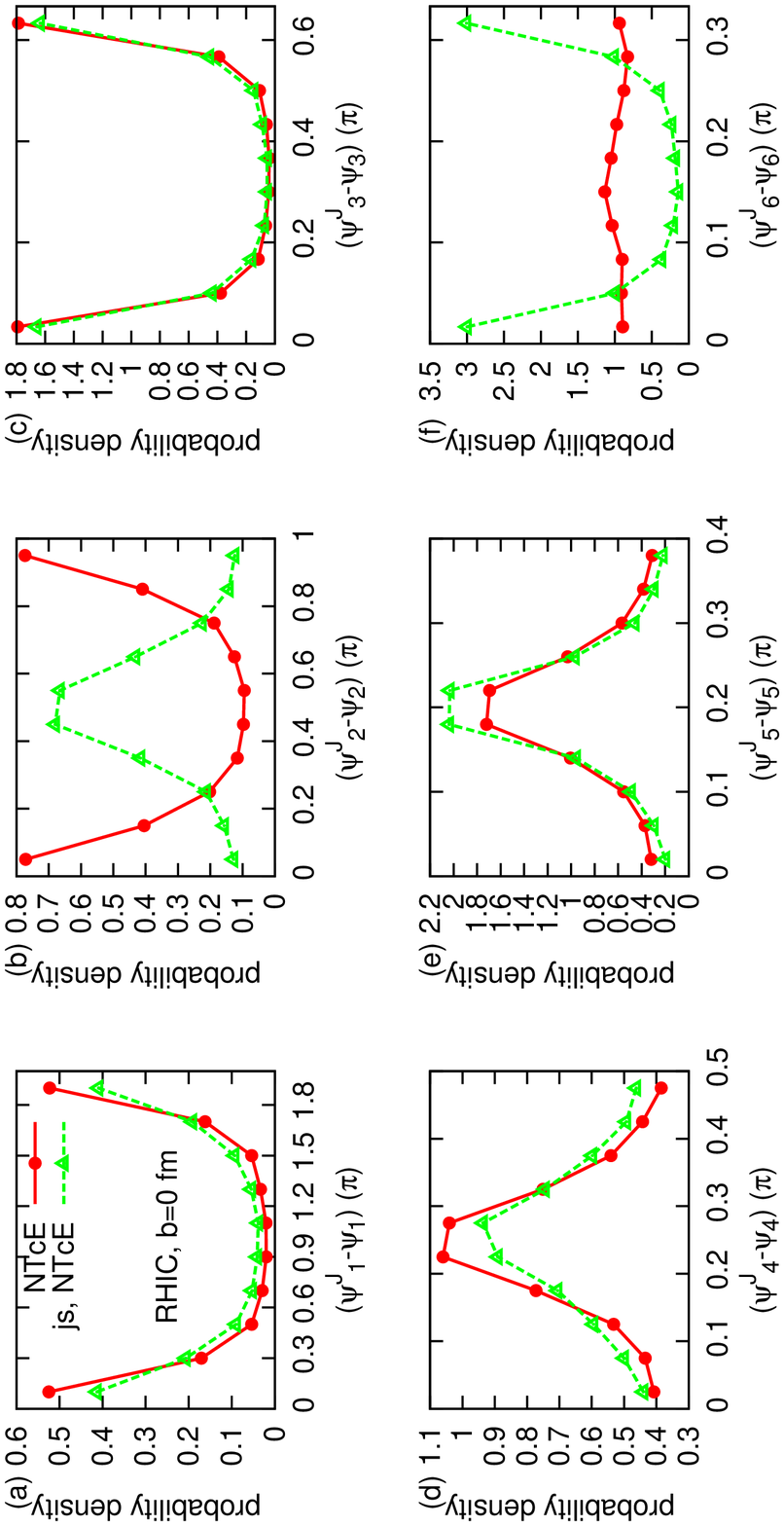}
\includegraphics[scale=0.6, angle=-90]{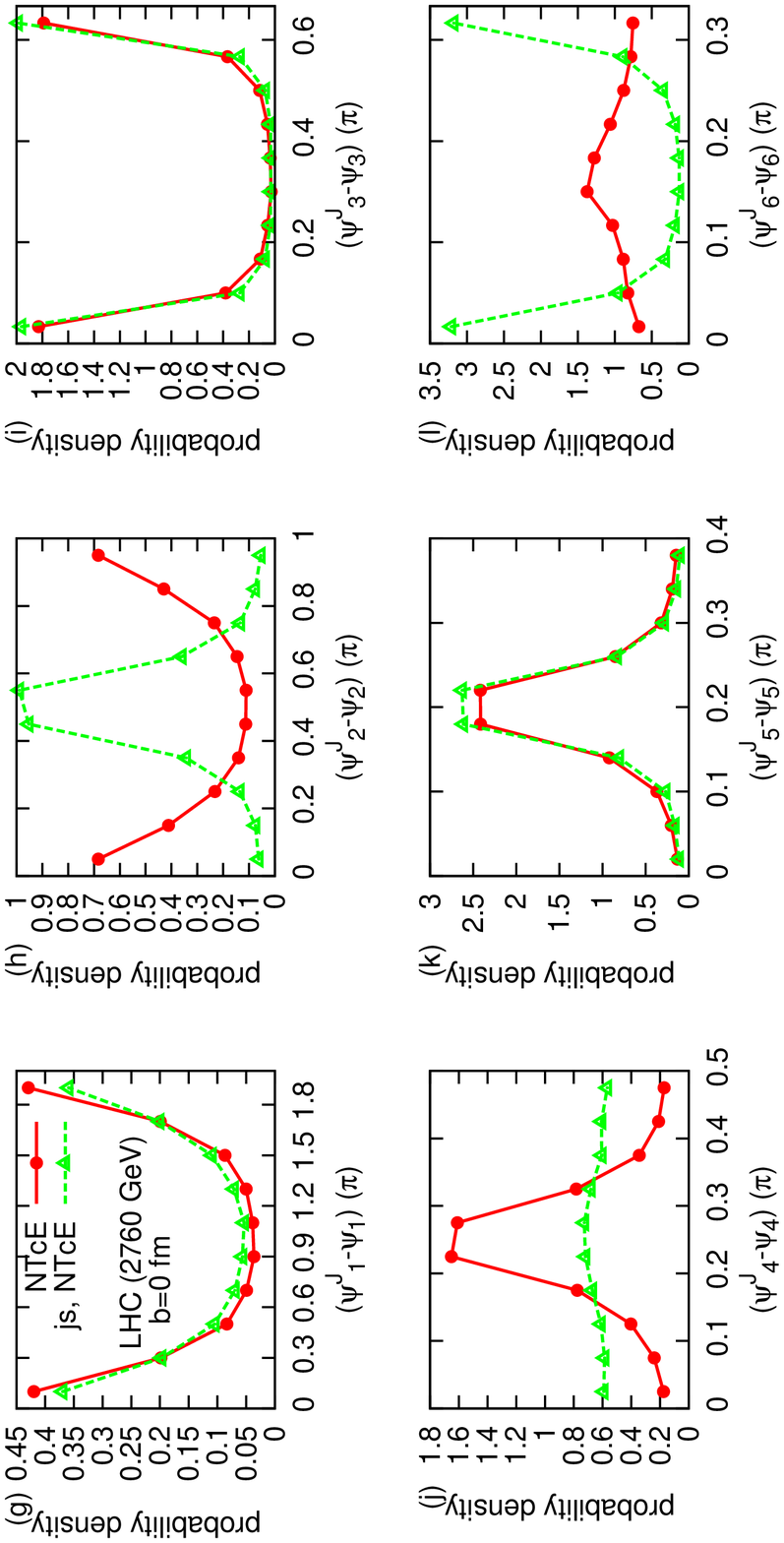}
\caption{(Color online). The distribution of $\psih{n}-\psi_{n}$ for RHIC ($\sqrt{s}=200$ GeV) and LHC ($\sqrt{s}=2760$ GeV) due to full calculation of the NTcE model (``NTcE'' curve) and the contribution of the jet spot fluctuation in the NTcE model (``js NTcE'' curve).}
\label{fig:psij-psi_dis_rhoc_RHIC_LHC}
\end{figure}

\begin{figure}
\centering
\includegraphics[scale=0.6, angle=-90]{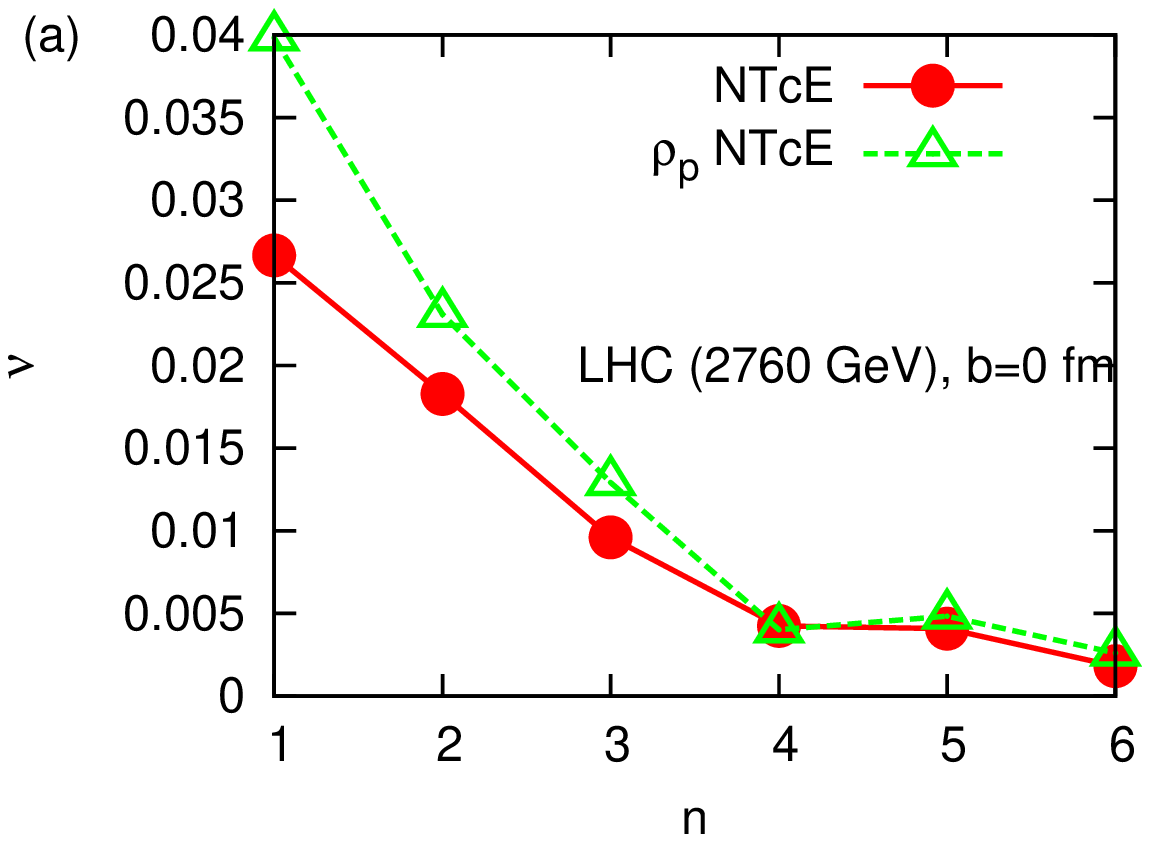}
\includegraphics[scale=0.6, angle=-90]{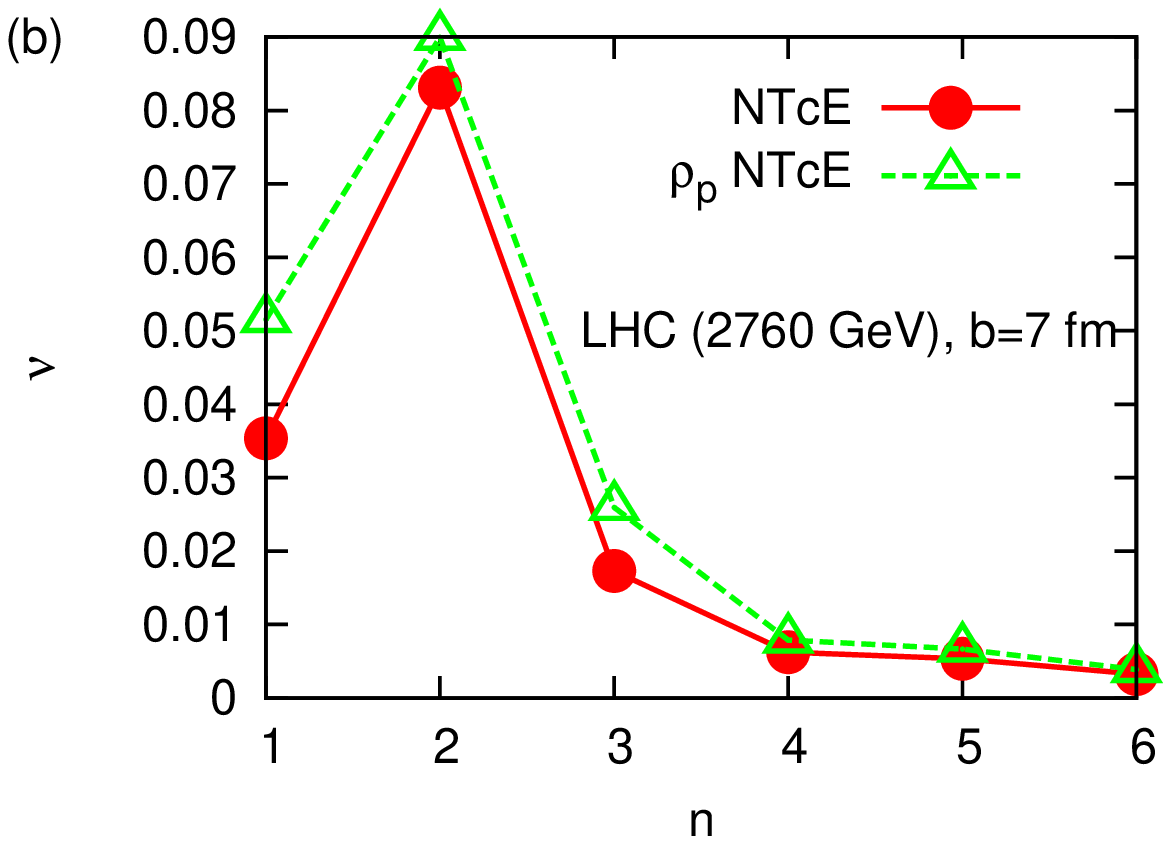}
\caption{(Color online). The spectrum of $\nuh{n}$ for LHC ($\sqrt{s}=2760$ GeV) due to full calculation of the NTcE model (``NTcE'' curve) and the same model with entropy density scaled with $\rhop$ (``$\rhop$ NTcE'' curve).}
\label{fig:nuspectrumnomixing}
\end{figure}  

It is also interesting to separate the contribution to $\nuh{n}$ from jet spot (JS) fluctuation and from the matter density fluctuation. Figure~\ref{fig:nuspectrumrhoc} shows the $\nuh{n}$ due to the full fluctuation (i.e., total response) and only due to the jet spot fluctuation (i.e., JS response) in the central collision at RHIC and LHC ($2.76$ TeV) by using the NTcE model. Here the JS response is computed by removing the fluctuations in the matter density. We can see that, for the first three harmonics, the jet spot contribution is significant. The difference between total response and JS response can be considered as the response due to the matter density fluctuation (i.e., MD response). To compute the difference, we need information about the sign of the responses as shown in  Figure~\ref{fig:psij-psi_dis_rhoc_RHIC_LHC} which plots the corresponding $\psih{n}-\psi_{n}$ distribution. The plots indicate that, for the first and third harmonics in both experiments, the total and JS response are positive, but in the second harmonics, the total response is positive and the JS response is negative. Combining the information about $\nuh{n}$ with the  $\psih{n}-\psi_{n}$ distribution, we find that, for the first harmonics, the total response is dominated by the JS response while the MD response plays a cancellation role; for the second harmonics, the MD response is positive and the JS response plays a cancellation role; for the third harmonics, both MD and JS response are positive. For the other higher order harmonics, the JS response becomes less significant. The result is also consistent with our earlier simple estimate \citep{xzjl2012}. This shows that the jet anisotropy is a probe to both the initial collision density profile at very early stage of $AA$ collision and the matter density profile during the jet traveling through the medium; different $\nuh{n}$ with the two playing different roles put important constraints on modeling both effects. 

Furthermore, is the jet anisotropy sensitive to the composition of the entropy density? In Figure~\ref{fig:nuspectrumnomixing}, we compare the results for LHC ($2.76$ TeV) based on two different media ($b=0$ and $7$ fm); the NTcE calculation is the same as the NTcE for LHC ($2.76$ TeV) in Figure~\ref{fig:nuspectrumRHICLHC} assuming the two-component profile, $S(\tau_{0}) \sim (1-\delta) \times \rhop/2+\delta \times \rhoc$, but the ``$\rhop$ NTcE'' assumes $S(\tau_{0}) \sim \rhop$. This shows that the first three harmonics are sensitive to different entropy densities. Especially, $\nuh{1}$ with $\rhop$ density profile is $60\%$ and $40\%$ bigger than with the mixed profile at $b=0$ and $7$ fm. The $\psih{n}-\psi_{n}$ distributions in the two cases, as we have checked, are quite close. Although the conventional procedure uses the multiplicity vs $b$ to calibrate $\delta$ in the two-component entropy density, our results suggest studying jet anisotropy to be another way to constrain the density profile.

\section{Hard-soft correlation} \label{sec:HScorrelation}

Motivated by the experimental analysis in Ref.~\citep{star10}, we study the (un)triggered azimuthal correlation between hard and soft hadrons at mid-rapidity. Following Refs.~\citep{star10,Bielckiova2004}, the triggered dihadron correlation is the pair distribution relative to angle difference, i.e., $\phih-\phis$, with $\phih$ constrained in specific region, 
$\mathcal{R}$. Here $\mathcal{R}$ is composed of four different pieces: $\phi_{\alpha}-\phi_{\beta}\leqslant\phih-\psie\leqslant\phi_{\alpha}+\phi_{\beta}$ ($\phi_{\alpha}$ is in the first quadrant) and other three with $\phi_{\alpha} \to -\phi_{\alpha}, \pi+\phi_{\alpha}, \pi-\phi_{\alpha}$. 
In  Appendix~\ref{app:correlation}, we derive necessary formula for the computation. Similar calculation has been done in Ref.~\citep{Bielckiova2004}, but we keep track the difference between $\psis{n}$ ($\psih{n}$) and $\psis{2}$ (also called the event plane angle $\psie$). To compute the untriggered correlation, we can simply set $\phi_{\alpha}=\phi_{\beta}=\pi/4$. 
Because the medium evolution in the transverse plane and the subsequent hadronization are not included in our simulation, we simply assume that (1) the event plane angle $\psie$ is the same as $\psi_{2}$;  (2) $\nus{n}=\chi^{s}_{n}\e{n}$ for the first four harmonics \citep{xzjl2012} \footnote{Ref.~\citep{ZhiQiu2011PRC84} shows that $\nus{n} \sim \e{n}$ is valid for the first three harmonics in general but only valid for the fourth in the central collisions.}; and (3) perfect alignment between $\psis{n}$ and $\psi_{n}$. We set $\chi^s_{n=1,2,3,4,5}\approx 0.15,0.26,0.21,0.14,0.086$ respectively (see e.g., \citep{ZhiQiu2011PRC84,Alver:2010dn,Luzum:2011jpg, Teaney:2010vd}), which are for the associated hadrons with $p_{t}$ around $2-4$ GeV at RHIC ($0.2$ TeV). 
\begin{figure}
\centering
\includegraphics[scale=0.7, angle=-90]{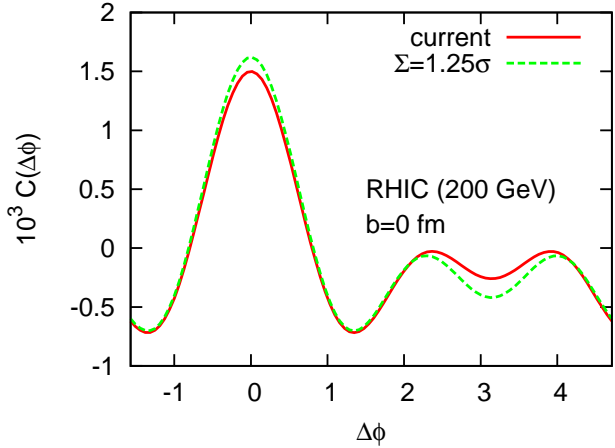}
\caption{(Color online). The untriggered dihadron correlation $C(\dphi)$ in central collisions at RHIC ($\sqrt{s}=200$ GeV) as calculated in our simulation (``current'' curve) and our estimate (``$\Sigma=1.25\sigma$'' curve) in Ref.~\citep{xzjl2012}. The two calculations are all based on the NTcE model. $\Sigma=1.25\sigma$ is a reasonable choice of a parameter in the previous estimate.}
\label{fig:Cdeltaphi_comparision}
\end{figure}  

In Ref.~\citep{xzjl2012}, we made a simple estimate of the untriggered dihadron correlation in the central collisions at RHIC ($0.2$ TeV). Here we can improve the previous calculation by simulating the realistic density profile and including the $\psih{n}-\psi_{n}$ distribution ($\psih{n}=\psi_{n}$ was assumed before). Let us apply the following decomposition: 
\[
C(\dphi)\equiv 2\times\left[\sum_{n} V_{n\Delta}\cos(n\dphi)\right] \ ,
\]
where the definition of $C(\dphi)$ can be found in Appendix~\ref{app:correlation}. 
Here we collect the results of the current calculation (previous estimate) \footnote{The results of the previous estimate, as shown here, are with parameter $\Sigma$ set to $1.25\sigma$. See Ref.~\citep{xzjl2012} for details.}: $V_{1\Delta}=2.3\ (2.8) \times 10^{-4}$, $V_{2\Delta}=3.1\ (3.0) \times 10^{-4}$, 
and $V_{3\Delta}=2.1\ (2.3) \times 10^{-4}$ (higher harmonics are at the $10\%$ level of the first three). Both calculations are based on the same NTcE model. Figure~\ref{fig:Cdeltaphi_comparision} shows the comparison for $C(\dphi)$. We can see that the structures of $C(\dphi)$ are quite close. 

\begin{figure}
\centering
\includegraphics[scale=0.7, angle=-90]{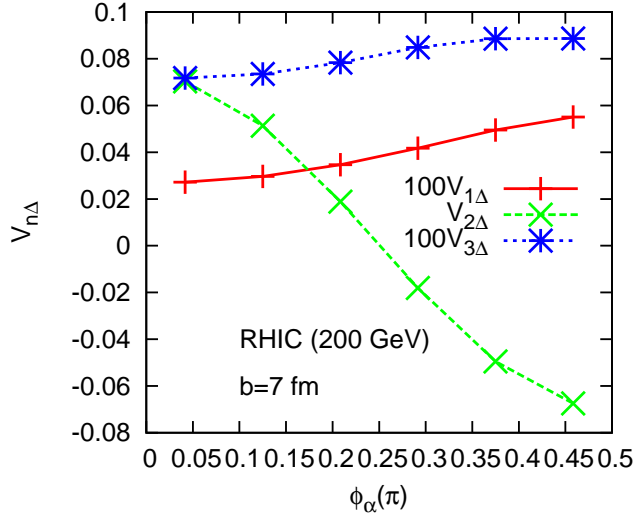}
\caption{(Color online). The triggered dihadron correlation $V_{n\Delta}$ vs trigger direction $\phi_{\alpha}$ with $b=7$ fm at RHIC ($\sqrt{s}=200$ GeV). $\phi_{\beta}$ is set to $\pi/24$, following Ref.~\citep{star10}. The calculation is based on the NTcE model.}
\label{fig:VDelta_triggered}
\end{figure}  

In Figure~\ref{fig:VDelta_triggered} following Ref.~\citep{Luzum11plb}, we show our results for $V_{n\Delta}$ ($n=1,\, 2,\, 3$) vs trigger direction $\phi_{\alpha}$ with $b=7$ fm at RHIC ($0.2$ TeV) based on the NTcE model. The $\phi_{\beta}$ is set as $\pi/24$, as in Ref.~\citep{star10}. The even harmonics $V_{2\Delta}$ and $V_{4\Delta}$ (not shown in the plot) are dominated by $\cos(2\dpsis{2})t_{2}$ and $\cos(4\dpsis{4})t_{4}$ [see Eq.~(\ref{eqn:formula_triggered_correlation})], which lead to $\cos(2\phi_{\alpha})$ and $\cos(4\phi_{\alpha})$ dependence. The odd harmonics $V_{1\Delta}$ and $V_{3\Delta}$ are much smaller (in the plot, they are scaled up by a factor of $100$). The $V_{1\Delta}$ has some $\phi_{\alpha}$ dependence because of the mentioned nonuniform distributions of $\psih{1}-\psi_{2}$ and $\psi_{1}-\psi_{2}$ as shown in Figure~\ref{fig:psi_psij_pp_dis_RHIC_b_7}, while $V_{3\Delta}$ depends much less weakly on $\phi_{\alpha}$. The author in Ref.~\citep{Luzum11plb} extracted $V_{n\Delta}$ directly from the unsubtracted correlation data in Ref.~\citep{star10}.\footnote{Although the trigger $p_t$ is around $2-4$ GeV in the extraction \citep{Luzum11plb}, it was mentioned that, for the higher $p_t$ trigger, $4-6$ GeV, the $\phi_{\alpha}$ and associate $p_t$ dependence of $V_{n\Delta}$ are similar to those with lower $p_t$ trigger.}
Note our $V_{1\Delta}$ is positive, but $V_{1\Delta}$ of Ref.~\citep{Luzum11plb} is negative, which, however, is consistent with the result that $\nus{1}$ of the associated hadron is negative when $p_t$ is below $1-2$ GeV, and positive when $p_t$ becomes bigger. In addition, the extracted $V_{1\Delta}$ can have a contribution from the correlation due to momentum conservation, although a rapidity gap was required in the extraction. 
Our $V_{2\Delta}$ is similar to that in Ref.~\citep{Luzum11plb}. However, the overall sign of our estimated $V_{4\Delta}$ is opposite to the  $V_{4\Delta}$ in Ref.~\citep{Luzum11plb}, because we approximate $\cos(4\dpsis{4})$ in Eq.~(\ref{eqn:formula_triggered_correlation}) by $\cos[4(\psi_{4}-\psi_{2})]$. It turns out that, although $\psi_{4}-\psi_{2}$ is around $\pi/4$ (see Figure~\ref{fig:psi_psij_pp_dis_RHIC_b_7}), the hydrodynamic calculation shows for the final hadron spectrum $\psis{4}-\psi_{2}$ is around $0$ \citep{ZhiQiu2011PRC84}. So we expect that, in the full calculation with medium evolution included, the sign of $V_{4\Delta}$ can be corrected (the magnitude could be changed because the linear response approximation fails in this case).
Because $\nuhre{n}$ has no $p_t$ dependence in our calculation, the $p_{t}$ dependence of $V_{n\Delta}$ is the same that of $\nus{n}$, which is consistent with the analysis of $p_t$ dependence in Ref.~\citep{Luzum11plb}.

\section{Summary} \label{sec:summary}

In summary, we have performed a systematic and event-by-event study of the jet azimuthal anisotropy for both $Au-Au$ collisions at RHIC ($\sqrt{s}=0.2$ TeV) and $Pb-Pb$ collisions at LHC ($\sqrt{s}=2.72$ and $5.5$ TeV). The MC Glauber model is used to generate the initial state density profile that fluctuates strongly from event to event. Three different geometric jet-energy-loss models, including NTcE, $\mathrm{L}^{2}$, and $\mathrm{L}^{3}$ that represent the characteristic path-length and density dependence of the near-Tc enhancement, pQCD, and AdS/CFT models, are explored and compared at both RHIC and LHC ($2.76$ TeV). In each event, we extract $\nuh{n}$, the high $p_t$ event plane angles $\psih{n}$ and the initial participant plane angles $\psi_{n}$. 
For the second harmonics, we see $\psih{2}$ is strongly correlated with $\psi_{2}$ in peripheral collisions as predicted by all the models for different experiments. It is reasonable to assume a strong correlation between $\psis{2}$ and $\psi_{2}$, as demonstrated by the hydrodynamic calculation (e.g., \citep{ZhiQiu2011PRC84}). Based on this, we can directly compare the $\nuh{2}$ with the measured $\n{2}$ of high $p_t$ hadrons, as carried out in Ref.~\citep{xzjl2012jqst}. There it was found that  the  NTcE and $\mathrm{L}^{3}$ models can explain the high $p_t$ $\n{2}$ data at RHIC while  the NTcE and $\mathrm{L}^{2}$ models succeed in describing LHC ($2.76$ TeV) data. One therefore concludes that taking together the geometric data at both RHIC and LHC, only the NTcE model describes all data sets. In addition, as discussed in Appendix D in detail, the NTcE model naturally explains lower opaqueness of the medium created at LHC as compared with that at RHIC; i.e., a reduction of jet-medium interaction at $\sim 30\%$ level as also implied by data.   %
We also analyze the correlation between $\psih{n}$ and $\psi_{n}$ for other harmonics. For the first and third harmonics, $\ave{\cos[n(\psih{n}-\psi_{n})]}$ are around $0.6- 0.8$ except in very peripheral collisions, but for the fourth and higher harmonics, even the sign can change although the specific transition impact parameter depends on models and experiments. This seems consistent with the LHC ($2.76$ TeV) data shown in Ref.~\citep{xzjl2012jqst}, which right now are not converging but do indicate negative responses for higher harmonics. 
We further clarify that, in the middle-centrality collisions (e.g., $b=7$ fm) at RHIC, although $\psi_{3}$ and $\psih{3}$ do have a random distribution, the first harmonic turns out to be tricky: $\psi_{1}$ has slight preference in the out-of-plane direction but $\psih{1}$ prefers to be along the in-plane direction. For the fourth harmonics at $b=7$ fm, both $\psi_{4}$ and $\psih{4}$ fluctuate around $\pi/4$ relative to either participant plane $\psi_{2}$ or the reaction plane.
It is also interesting to separate the contributions to the jet anisotropy. In the central collisions at RHIC, we have analyzed the effect of the initial jet spot fluctuation. The difference between the JS response and the full response should be due to the matter density (shape) fluctuation. We see in the first three harmonics their roles are different, which points out the importance of having a coherent picture for all these harmonics in a valid model. We also test the sensitivity of the jet anisotropy to the composition of the matter density in the Glauber model, which shows at LHC ($2.76$ TeV) that the first three harmonics are affected significantly by changing the entropy density from a combination of participant and collision density to the participant density alone. Although the mixing of the two in the matter density is normally calibrated to the observed multiplicity vs $b$, our test does show another interesting way to calibrate this. 

Based on the information about jet anisotropy and assuming that the response of the collective flow to ISF is linear, we proceed to discuss both untriggered and triggered hard-soft correlation. For the central collisions at RHIC, we have improved our calculation of the untriggered correlation in Ref.~\citep{xzjl2012} and confirmed the results there: a strong peak develops on the near side (``hard ridge'') while a double-hump structure shows up on the away side. This could be a possible explanation for the experimentally observed ``hard ridge'', but the away side structure can be more complicated because of the other sources of correlations (e.g., effects of  global momentum conservation and cluster correlation \citep{Betz:2010qh,Bzdak:2012ia}), which requires more detailed study.
By using our rederived formula for the triggered correlation, we have studied the Fourier component of the correlation, $V_{n\Delta}$. 
We find that the trigger-angle dependence of $V_{n\Delta}$ is consistent with the extracted information in Ref.~\citep{Luzum11plb}. As for the associated $p_t$ dependence of the hard-soft correlation, we expect it to be the same as that of $\nus{n}(p_t)$, because of the weak $p_t$ dependence of $\nuh{n}$ as shown in data at RHIC and implemented in our simulation. This also agrees with the findings in Ref.~\citep{Luzum11plb}. However, a decisive comparison with data again requires a good understanding of other sources of correlations.  

We end with discussions on a number of issues to be improved in future studies. A detailed study of the hard-soft correlation naturally requires including a realistic medium evolution in the transverse plane through hydrodynamics~\citep{Renk:2006sx,Bass:2008rv}. The linear response is a crude approximation for the soft dynamics. Moreover, such transverse expansion dynamics can also play a role in the jet quenching, in particular for higher harmonics. In this simulation, the transverse shape of the medium is frozen (matter density decreases as $1/\tau$ because of the longitudinal expansion), but in reality the eccentricities of the matter will reduce with increasing time because of the pressure gradient driven expansion (and such ``self-quenching'' may be a significant factor for higher harmonics) \cite{Molnar:2012fn}. Although qualitatively the majority of jets may not experience this change very much, because the dramatic shape change happens in a relatively later stage, it is certainly interesting to explore this effect especially for harmonics other than the second harmonic, which are mostly driven by the shape fluctuations. Integration of our current jet azimuthal anisotropy study with hydrodynamics is being pursued. Furthermore, the $p_t$ dependence of jet quenching is not implemented in the present geometric models but efforts are underway to include that. Another important uncertainty is related to possible preequilibrium energy loss, which hopefully will be better understood and estimated with improved descriptions for the preequilibrium evolution \cite{Blaizot:2011xf}. We are focusing on the geometrical information about jet anisotropy, but it is important to have a coherent understanding in terms of different variables, which requires a careful study of the hard parton energy loss for example in the near-Tc enhancement picture. Finally, we hope we have made a case here for using the hard probe as a new and sensitive tool for quantifying the initial fluctuations in heavy ion collisions and for discriminating different models for the initial conditions.  Future studies along this direction will be certainly extended to hard probe of initial conditions generated from e.g., CGC-motivated models \cite{Kharzeev:2000ph,Schenke:2012hg} and their comparison with the MC Glauber model.

\acknowledgments

We thank Larry McLerran, U. Heinz, Z. Qiu, M. Gyulassy, G. Torrieri, B. Betz, A. Buzzatti, J. Jia, R. Lacey, D. Molnar, F. Wang, and S. Mukherjee for helpful communications and discussions. We are also grateful to the Institute for Nuclear Theory and the organizers of the INT Workshop on ``The Ridge Correlation in High-Energy Collisions at RHIC and LHC'' during which the reported research was advanced. JL thanks the RIKEN BNL Research Center for partial support. XZ was supported by the Nuclear Theory Center at Indiana University, and is now supported by the US Department of Energy under Grant No. DE-FG02-93ER-40756. 

\appendix

\section{The structure of the simulation code} \label{app:flowchart}

\begin{figure}
\centering
\includegraphics[scale=0.8, angle=0]{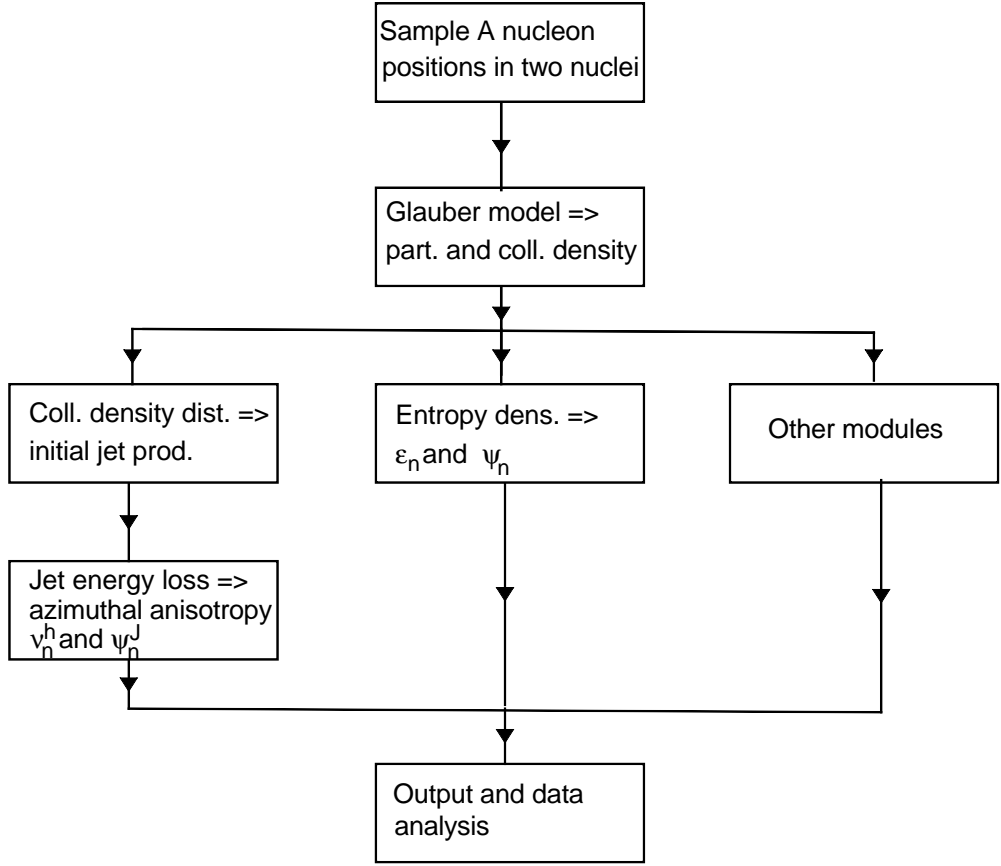}
\caption{The structure of the simulation code.} \label{fig:flowchart}
\end{figure} 

Figure~\ref{fig:flowchart} summarizes the overall structure of the simulation in each event. It starts with the sampling of nucleon positions in the two colliding nuclei. Based on the Glauber model, we can identify the binary collision pairs, which leads to the participant density and collision density by using proper smearing procedure. This provides the basic information for all the following calculation modules. In the jet quenching module, we calculate the azimuthal distribution of the $R_{AA}$ as described in the text and get $\nuh{n}$ and $\psih{n}$. The second module analyzes the initial entropy density, which gives $\e{n}$ and $\psi_{n}$. In principle other modules can be included in the code, for example the EM field calculation. Then the results from different modules are grouped together and recorded in the output file. To get an ensemble of events, the collision geometry, e.g., impact parameter for $AA$ collision, needs to be sampled, which is not shown here. \footnote{In $U-U$ collisions, the collision geometry is more complicated, requiring the sampling of impact parameter and the relative orientation of the two nuclei.} After the initialization of the collision geometry, the whole calculation shown in Fig.~\ref{fig:flowchart} can be carried out.

\section{The formula for the triggered dihadron correlation} \label{app:correlation}

In this section we derive the formula for computing the triggered dihadron correlation mentioned in Sec.~\ref{sec:HScorrelation}, based on the following soft and hard hadron azimuthal distributions:
\begin{align}
\frac{d\Ns}{d\phis} & \sim 1+ 2 \sum_{n} \nus{n} \cos[n(\phis-\psis{n})] \ , \label{eqn:psisn}  \\
\frac{d\Nh}{d\phih} & \sim 1+ 2 \sum_{m} \nuh{m} \cos[m(\phih-\psih{m})] \ . \label{eqn:psihn} 
\end{align}
According to the definition in Sec.~\ref{sec:HScorrelation}, let us calculate the following:	
\begin{align}
\int_{\mathcal{R}}  d\phis d\phih   \frac{d\Ns}{d\phis} \frac{d\Nh}{d\phih} \delta(\phis-\phih-\dphi) \sim 1+C(\dphi)\equiv 1+2\sum_{n}\nus{n}\nuhre{n} \cos(n\dphi) \ . 
\end{align}
Here $\mathcal{R}$ is composed of four different pieces: $\phi_{\alpha}-\phi_{\beta}\leqslant\phih-\psie\leqslant\phi_{\alpha}+\phi_{\beta}$ ($\phi_{\alpha}$ is in the first quadrant) and the other three with $\phi_{\alpha} \to -\phi_{\alpha}, \pi+\phi_{\alpha}, \pi-\phi_{\alpha}$. We keep track of the difference between $\psis{n}$ ($\psih{n}$) and  $\psie$, which is different from  Ref.~\citep{Bielckiova2004}. By using Eqs.~(\ref{eqn:psihn}) and~(\ref{eqn:psisn}), we get 
\begin{align}
&\nuhre{n} \left[1+2\sum_{K=2,4...} \nuh{K} {\cos(K\dpsih{K})} t_{K}\right] \notag  \\
=& \delta_{n,even} {\cos(n\dpsis{n})} t_{n} \notag  \\
&+ \sum_{m,n}^{m+n=even} \nuh{m} t_{m+n} {\cos[m\dpsih{m}+n\dpsis{n}]}  \notag  \\
&+ \sum_{m,n}^{m-n=even} \nuh{m} t_{m-n} {\cos[m\dpsih{m}-n\dpsis{n}]} \notag  \\
=& \nuh{n}{\cos[n(\dpsih{n}-\dpsis{n})]}+\delta_{n,even} {\cos(n\dpsis{n})} t_{n} \notag \\[5pt]
&+\sum_{K=2,4...} \nuh{n+K} t_{K} {\cos[(n+K) \dpsih{n+K} -n\dpsis{n} ]} \notag  \\[5pt]
&+\sum_{K=2,4...}\nuh{|n-K|} t_{K} {\cos[(n-K) \dpsih{|n-K|} -n\dpsis{n}]} \ . \label{eqn:formula_triggered_correlation}
\end{align}
In the above expression, $\dpsih{n}\equiv\psih{n}-\psie$; $\dpsis{n}\equiv \psis{n}-\psie$; $t_{n}\equiv \frac{\sin(n\phi_{\beta})}{n\phi_{\beta}} \cos(n\phi_{\alpha})$. If we assume all the reference angles $\psis{n}$ and  $\psih{n}$ are exactly correlated with $\psie$, the above formula would become the one in Ref.~\citep{Bielckiova2004}. To compare with the data, we need to  average $C(\dphi)$ over many events, i.e., $\ave{C(\dphi)}$. In principle, $C(\dphi)$ should also have $\sin(n\dphi)$ components, but their amplitudes should be very small with many events averaged. [We will simply use $C(\dphi)$ as $\ave{C(\dphi)}$ from now on.]

\section{Discussion on  the near-$T_c$ enhancement and the trace anomaly} \label{app:anomaly}

We discuss here the possible connection between  the near-$T_c$ enhancement of jet-medium interaction and the QCD trace anomaly $(\epsilon-3p)/T^4$ (with $\epsilon,p,T$ the energy density, pressure and temperature) which as an interaction measure also develops a strong near-$T_c$ peak as shown by lattice QCD simulations \citep{Borsanyi:2010cj,Cheng:2009zi}.  One may naturally ask whether these two nonperturbative effects near $T_c$ could share a common underlying picture. We notice that there have been attempts \citep{Ficnar:2012yu} to develop nonconformal holographic models based on gauge/gravity duality that can mimic the QCD trace anomaly well but face tensions in describing jet quenching phenomenon. In this appendix, we will show that the near-$T_c$ enhancement and the trace anomaly may be consistently described together by a plasma of magnetic monopoles based on the ``magnetic scenario'' in \citep{Liao:2006ry,Liao:2008vj,Chernodub,Ratti:2008jz}.   

We consider QCD near-$T_c$ plasma as an ensemble of thermal magnetic monopoles with certain density $n_m(T)$ and effective mass $M_m(T)$. Let us first compute the contribution to trace anomaly from this magnetic component. As a crude approximation let us first neglect the potential energy and express the trace anomaly through the density, i.e.,   
\begin{eqnarray}
\frac{\epsilon-3p}{T^4} = \frac{n_m}{T^3}  \, \chi_M
\end{eqnarray}
with the dimensionless coefficient $\chi_M$ defined as:
\begin{eqnarray}
\chi_M \equiv  {\bigg <} \frac{\tilde{M}^2}{\sqrt{x^2+\tilde{M}^2}}  {\bigg >} = \frac{ \int_0^\infty \frac{x^2 dx}{exp\left(\sqrt{x^2+\tilde{M}^2}\right)-1} \frac{\tilde{M}^2}{\sqrt{x^2+\tilde{M}^2}}}{\int_0^\infty \frac{x^2dx}{exp\left(\sqrt{x^2+\tilde{M}^2}\right)-1} }
\end{eqnarray}
In the above the thermal monopole mass $\tilde{M}= M_m/T$ is about $1.5- 2$ when close to $T_c$ \citep{D'Alessandro:2010xg,Bornyakov:2011dj}, which corresponds to $\chi_M \approx 0.7- 1.1\,\,$. We therefore have roughly the relation  
\begin{eqnarray}
\frac{\epsilon-3p}{T^4} \approx \left( 0.7- 1.1 \right) \times \frac{n_m}{T^3}  
\end{eqnarray}
Now the monopole density close to $T_c$, according to lattice results and model studies \citep{D'Alessandro:2010xg,Bornyakov:2011dj,Chernodub,Liao:2008vj,Ratti:2008jz}, is about $n_m/T^3 \approx 2- 3$, so we get an estimate of the monopole contribution to trace anomaly near $T_c$ of about
\begin{eqnarray}
\frac{\epsilon-3p}{T^4} \approx 1.4 - 3.3
\end{eqnarray}
Note that the peak value of trace anomaly found from lattice simulations is about $\frac{(\epsilon-3p)}{T^4} |_{peak}\approx 4$ \citep{Borsanyi:2010cj,Cheng:2009zi}, so the above contribution is very consistent with lattice data and actually is the dominant part. The mild difference may be made up from the interaction among monopoles: according to the analysis in \citep{Liao:2006ry,D'Alessandro:2010xg} the interaction potential energy is comparable with the kinetic, so could contribute a similar amount to the trace anomaly. We also note that the rapid decrease of trace anomaly with increasing temperature can be understood from the rapidly dropping density $n_m/T^3$ for $T>T_c$. 

Let us now turn to the discussion of the jet-medium interaction within the same magnetic scenario. With  a  picture similar to the GLV model for jet energy loss \citep{Gyulassy:2003mc}, while replacing the Coulomb electric scattering centers by the magnetic ones in the near-$T_c$ plasma, we expect roughly the amount of medium kick received by a penetrating electric jet to scale as
\begin{eqnarray}
\sim \alpha_E(T) \alpha_M(T) n_m(T)
\end{eqnarray}
While the electric coupling $\alpha_E$ and magnetic coupling $\alpha_M$ both depend on temperature scale, their product has to be unity as per the famous Dirac condition, i.e., $\alpha_E(T)\, \alpha_M(T)=1$. Therefore, we see that if we normalize the medium kick by the entropy density at the same temperature, we will get 
\begin{eqnarray}
\kappa \sim \frac{\alpha_E(T) \alpha_M(T) n_m(T)}{s(T)} = \frac{n_m/T^3}{s/T^3}  
\end{eqnarray}
Again the monopole density $n_m/T^3$ strongly peaks at $T_c$  \citep{D'Alessandro:2010xg,Bornyakov:2011dj,Liao:2008vj} while $s/T^3$ relatively slowly decreases from high temperature to low temperature \citep{Borsanyi:2010cj,Cheng:2009zi}. As a result the normalized jet-medium interaction strength $\kappa$ develops a strong near-$T_c$ enhancement. To make a more quantitative statement, let us compare the $\kappa$ at two temperatures: from $1.5T_c$ to $T_c$ the monopole density increases by a factor of $\sim 3$ while the entropy density $s/T^3$ decreases by a factor of $\sim 2$, and therefore the $\kappa$ increases by a factor of $\sim 6$ from $1.5T_c$ down to $T_c$ which is well in accord with the assumed enhancement in the NTcE model. 

To sum up the above qualitative discussions and estimates, we see that the presence of an emergent monopole plasma near-$T_c$ with the density and thermal mass indicated by lattice and model studies, can consistently describe the strong near $T_c$ peaks in both the trace anomaly and the jet-medium interaction. 

\section{An estimate of average opaqueness from RHIC to LHC} \label{app:opaqueness}

In this part we make a simple estimate, in the NTcE model, of the evolution of the average opaqueness of the fireball from  RHIC to LHC. The hot medium created in heavy ion collisions is neither homogeneous nor static, and therefore one can only talk about the opaqueness of the fireball created in the collision on an average sense, i.e., averaging both over the spatial distribution and the time evolution. For simplicity let us consider the perfectly central collision with $b=0$. We will use this case as an indicator of the opaqueness evolution with collision beam energies. The average opaqueness seen by a jet going through the medium along a particular path $P$ can be estimated as follows: 
\begin{eqnarray}
<\kappa[s(l)] >_P = \frac{\int_P \, \kappa[s(l)]\, s(l)\, l\, dl }{\int_P  \, s(l)\, l\, dl } 
\end{eqnarray} 
Note that, even for the same geometry and same jet path in fireballs created at different collision energies, the $s(l)$ will be very different and the convolution with $\kappa[s(l)]$ is rather nontrivial due to the peculiar near-$T_c$ enhancement form in Eq.~(\ref{Eq_kappa}). 
This is of course further subject to averaging over paths, i.e., over all initial jet spots weighed by collision density as well as all possible jet directions. We will use a simple optical Glauber model (with the same parameter sets as in the previous MC Glauber simulations) to evaluate this average. Note that the longitudinal boost-invariant expansion effect is also included as in the previous MC calculations, i.e., the density $s(l)$ on a specific point of the path is really a function of both $l$ and time $t$. 

To give a quick idea, we can simply consider a jet initiating right from the center $\vec r_\perp =0$: with this simplest path we can easily obtain $<\kappa>_{\rm RHIC} \, : \, <\kappa>_{\rm LHC} \,\approx\, 1 : 0.75$. Clearly we see a shift toward less opaqueness as a consequence of two factors:  the near-$T_c$ structure of $\kappa(s)$ and the shift to higher density matter at LHC. In contrast, all models with a constant $\kappa$ will see no change in the opaqueness from RHIC to LHC.  A full evaluation averaging over initial jet spots and paths with fluctuating initial condition gives the following evolution of average opaqueness: 
\begin{eqnarray}
<\kappa>_{\rm RHIC} \, : \, <\kappa>_{\rm LHC} \,\approx\, 1 : 0.72
\end{eqnarray}

We note this is not only qualitatively but quantitatively in agreement with the earlier analysis in \citep{Betz:2012qq} which found that in order to describe the LHC $R_{AA}$ data the jet-medium coupling parameter $\kappa$ (which is a constant in the model of \citep{Betz:2012qq}) has to be reduced by a factor $\sim 30\%$ from the RHIC value in the same model.  What we want to emphasize here is that such a reduction of (average) jet-medium interaction implied by data from RHIC to LHC is naturally borne out from the strong near-$T_c$ enhancement of jet-medium interaction.

\newpage 

%

\begin{figure}
\centering
\includegraphics[scale=0.6, angle=-90]{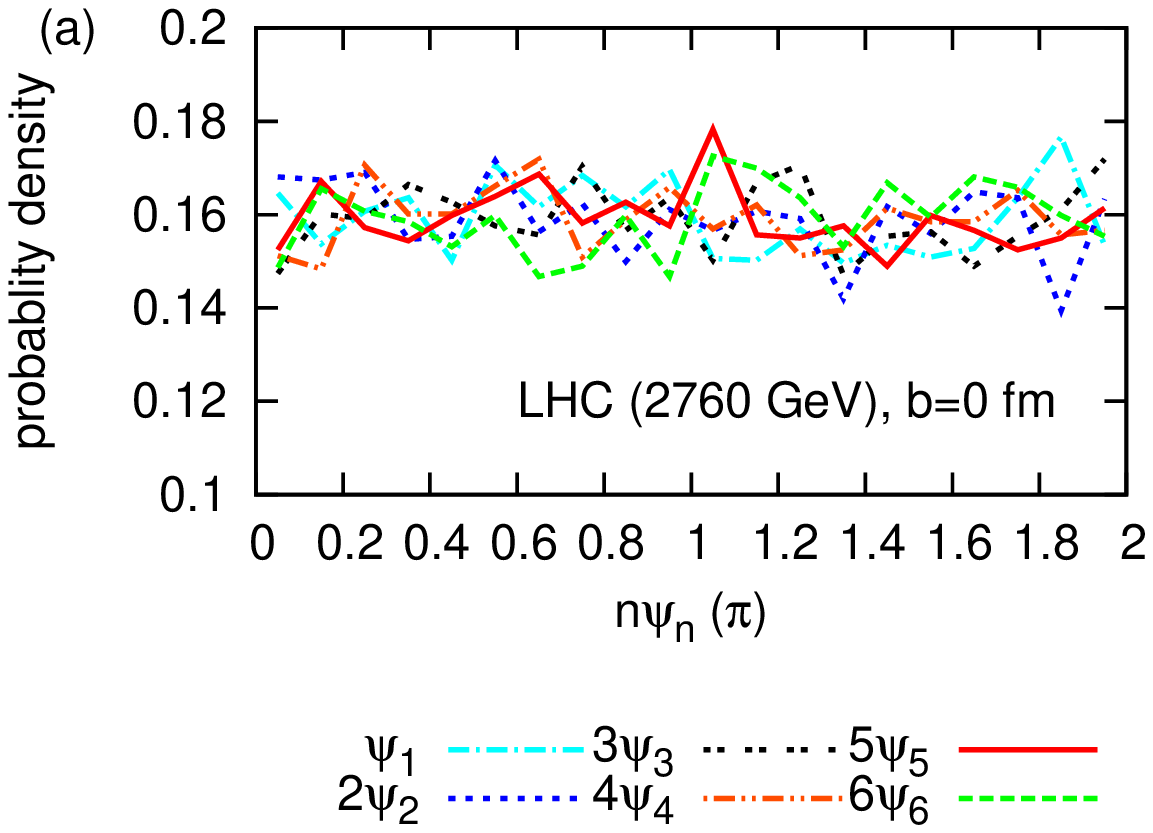}
\includegraphics[scale=0.6, angle=-90]{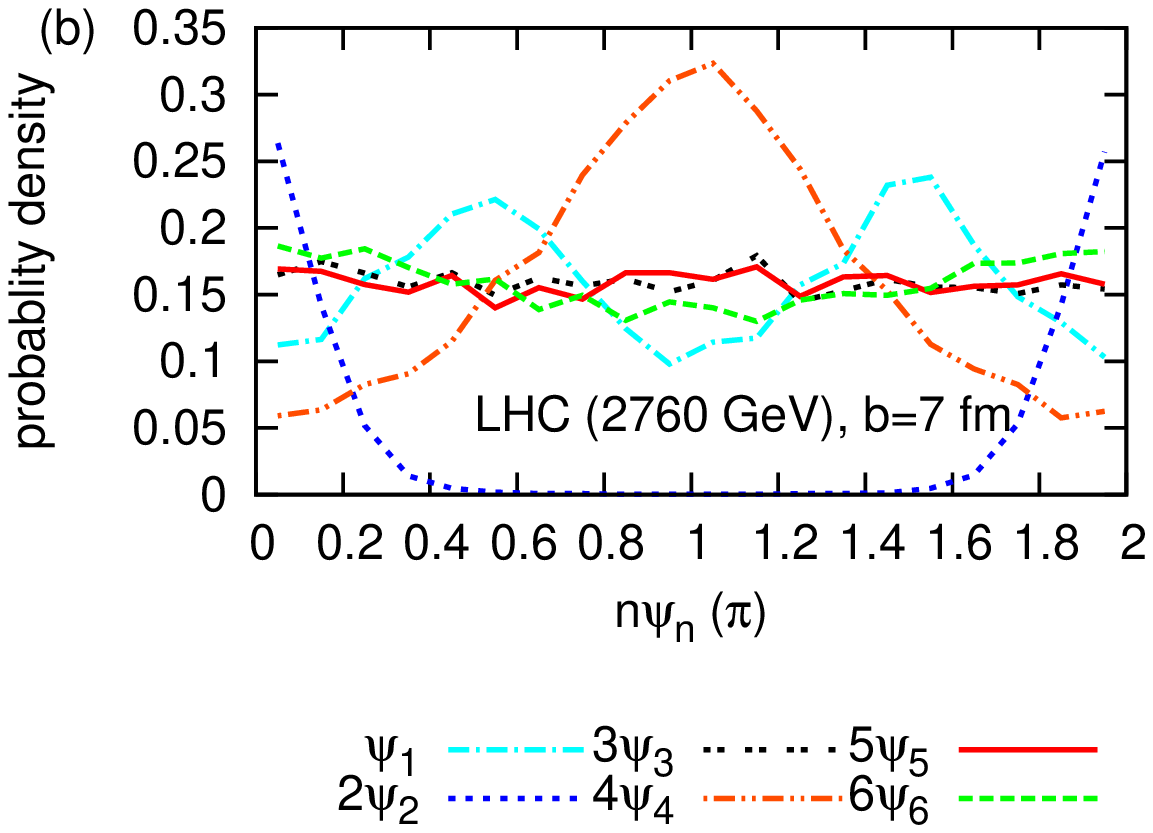}
\includegraphics[scale=0.6, angle=-90]{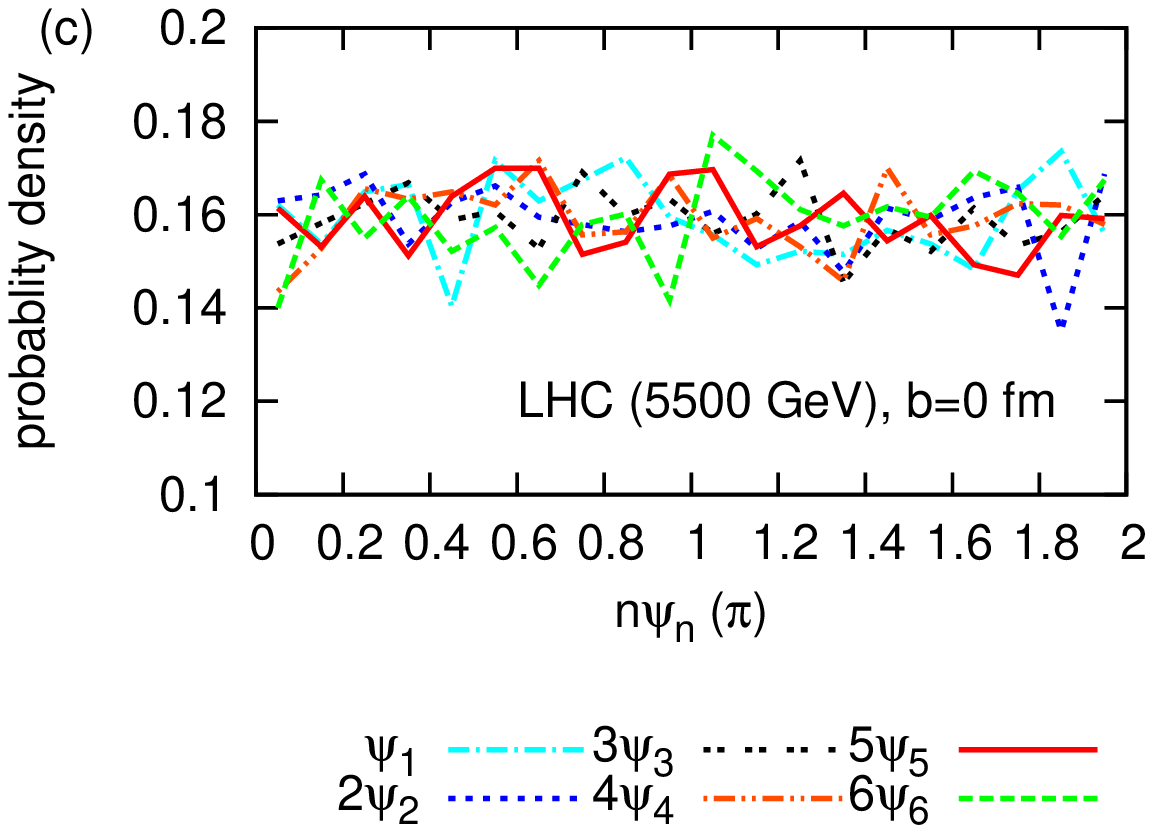}
\includegraphics[scale=0.6, angle=-90]{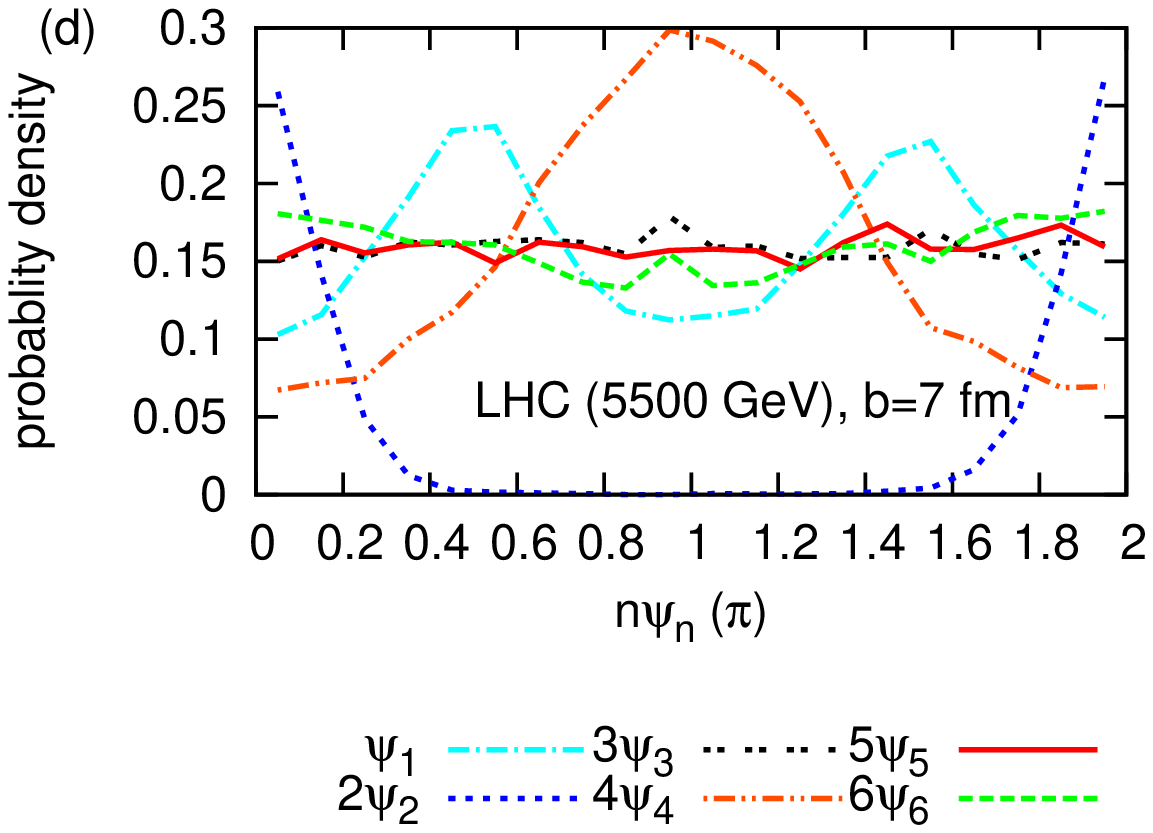}
\caption{(Color online). The distribution of $\psi_{n}$ with $b=0, \ 7$ fm at  LHC ($\sqrt{s}=2760, \ 5500$ GeV). For $b=7$ fm, (2$\psi_{2}$)'s probability density is scaled down by a factor $0.3$.}
\label{fig:psidissupp}
\end{figure} 

\begin{figure}
\centering
\includegraphics[scale=0.6, angle=-90]{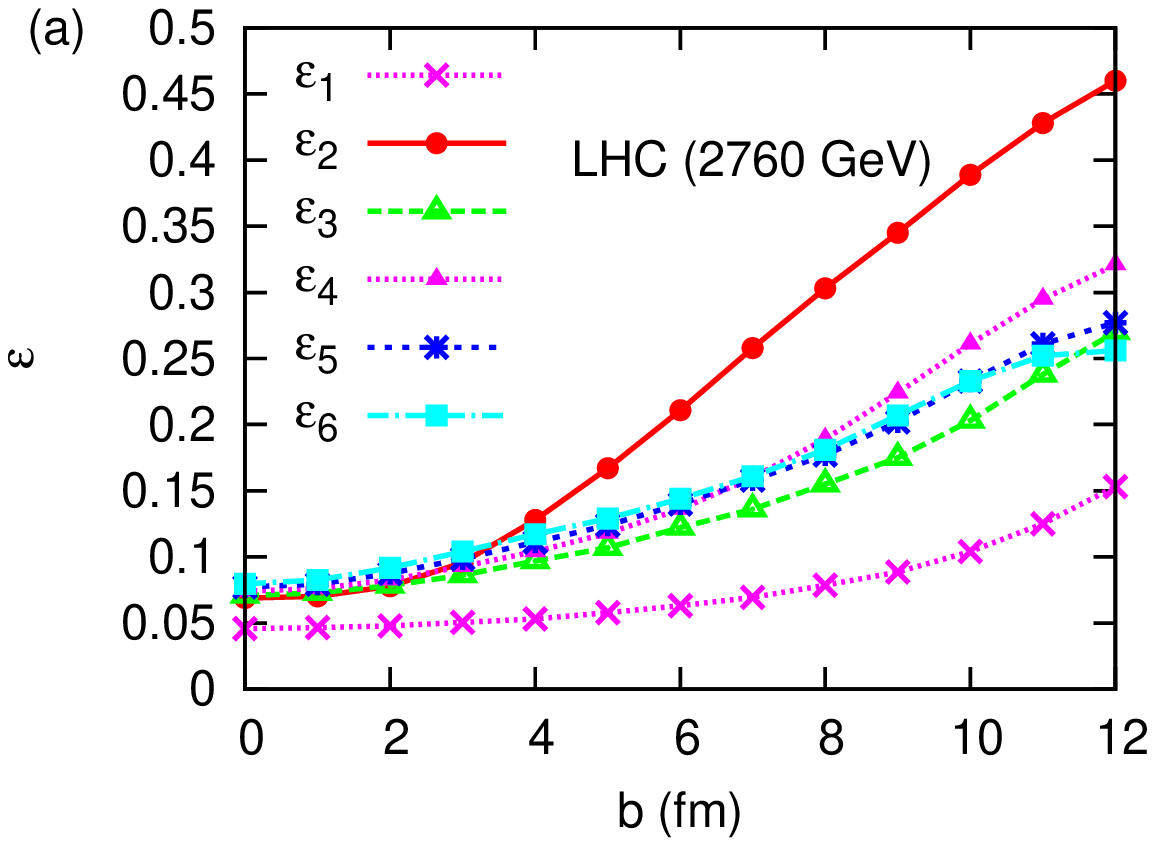}
\includegraphics[scale=0.6, angle=-90]{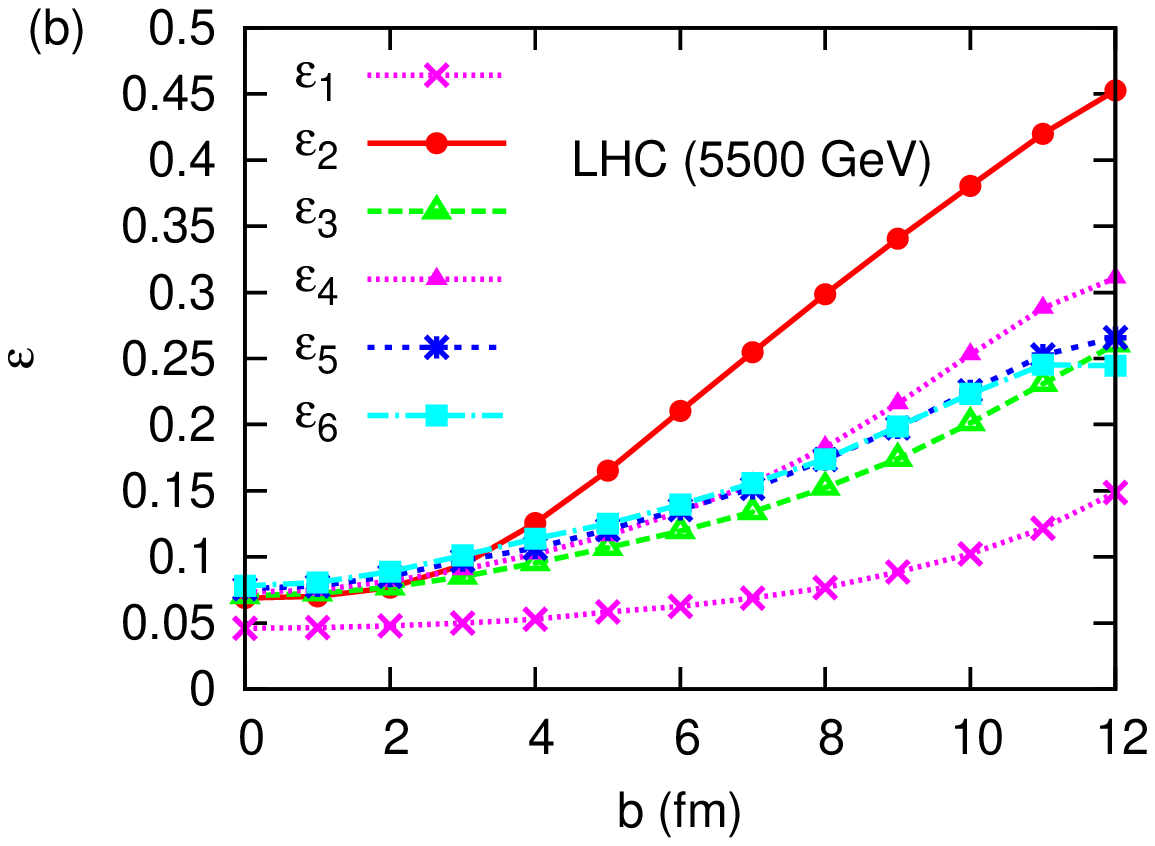}
\caption{(Color online). The $b$ dependence of $\e{n}$ at LHC ($\sqrt{s}=2760, \ 5500$ GeV).}
\label{fig:epsvsbsupp}
\end{figure}

\begin{figure}
\centering
\includegraphics[scale=0.6, angle=-90]{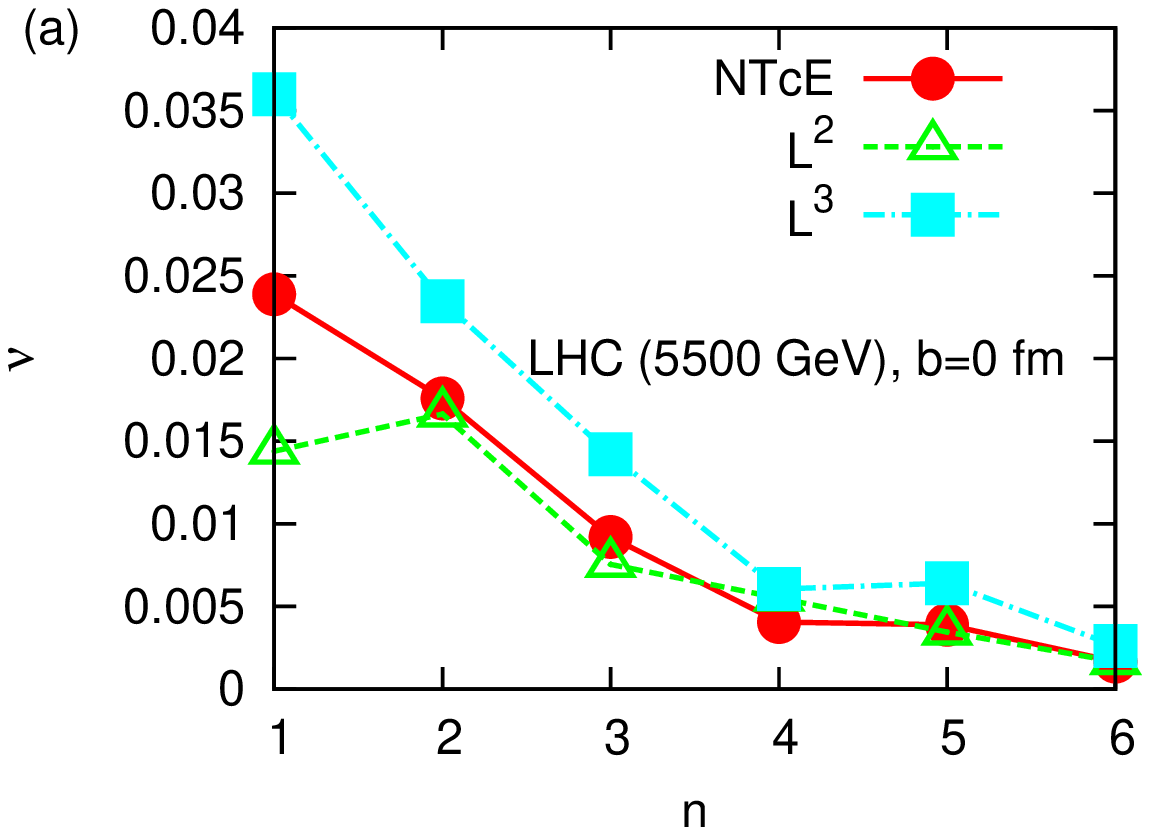}
\includegraphics[scale=0.6, angle=-90]{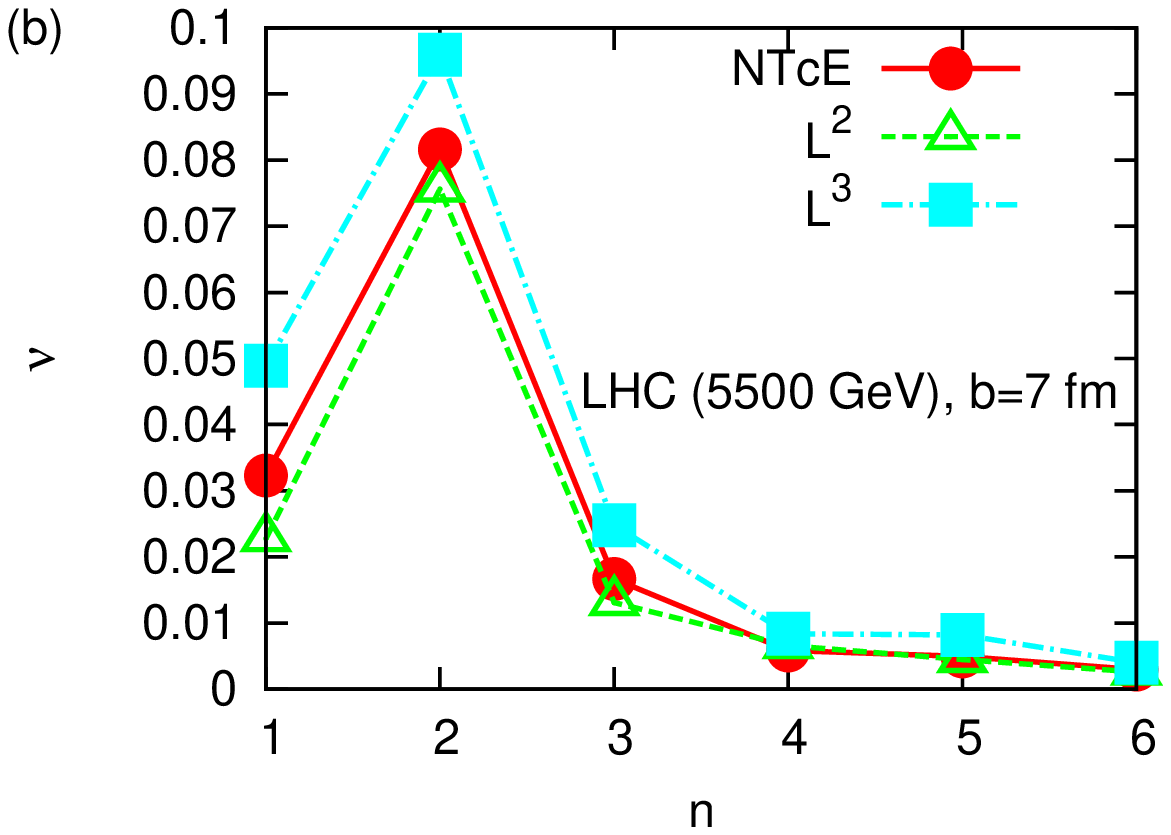}
\caption{(Color online). The spectrum of $\nuh{n}$ for LHC ($\sqrt{s}=5500$ GeV) at $b=0, \ 7$ fm,  based on three different models.}
\label{fig:nuspectrumRHICLHCsupp}
\end{figure}  

\begin{figure}
\centering
\includegraphics[scale=0.6, angle=-90]{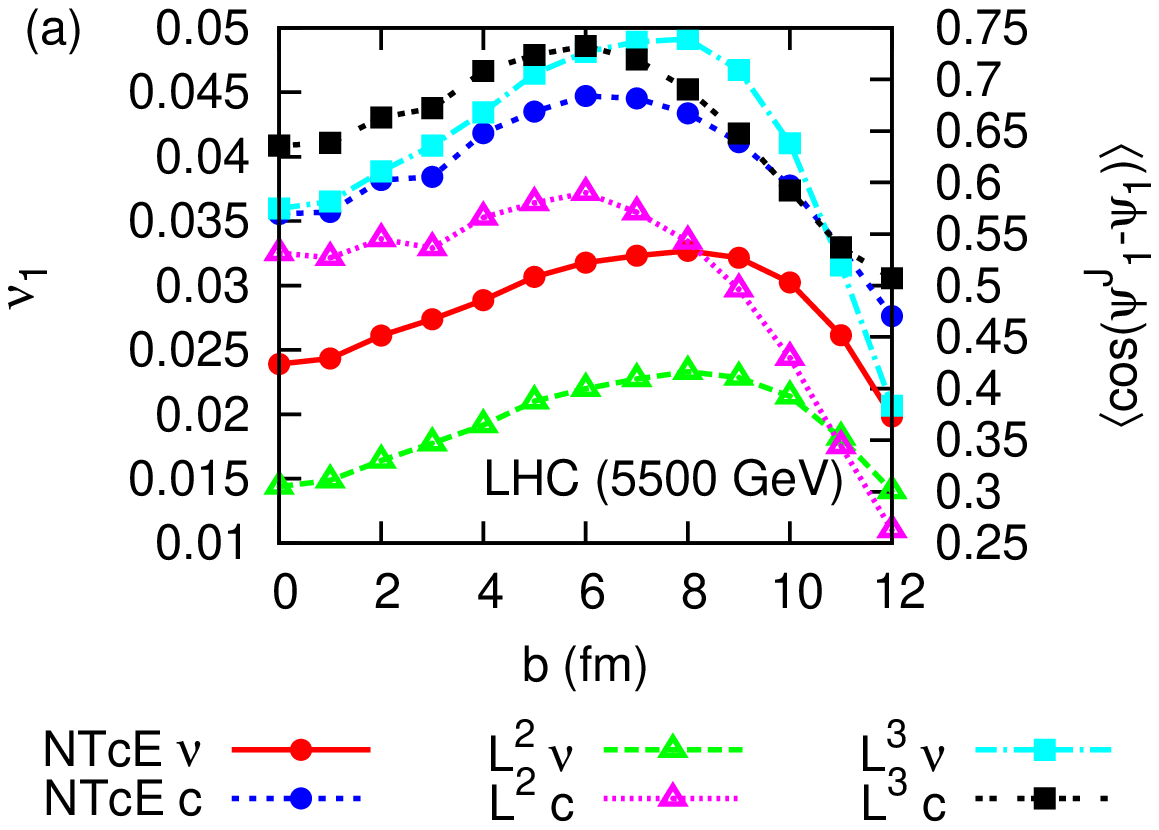}
\includegraphics[scale=0.6, angle=-90]{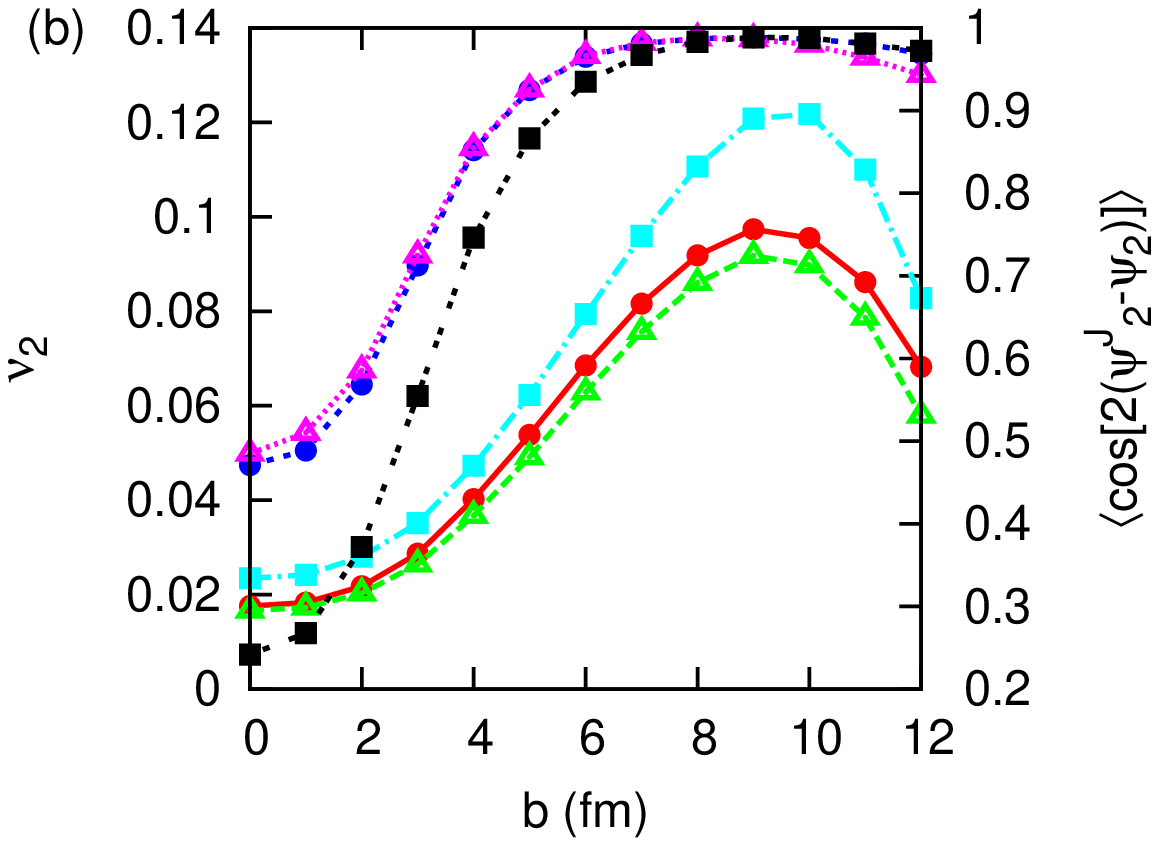}
\includegraphics[scale=0.6, angle=-90]{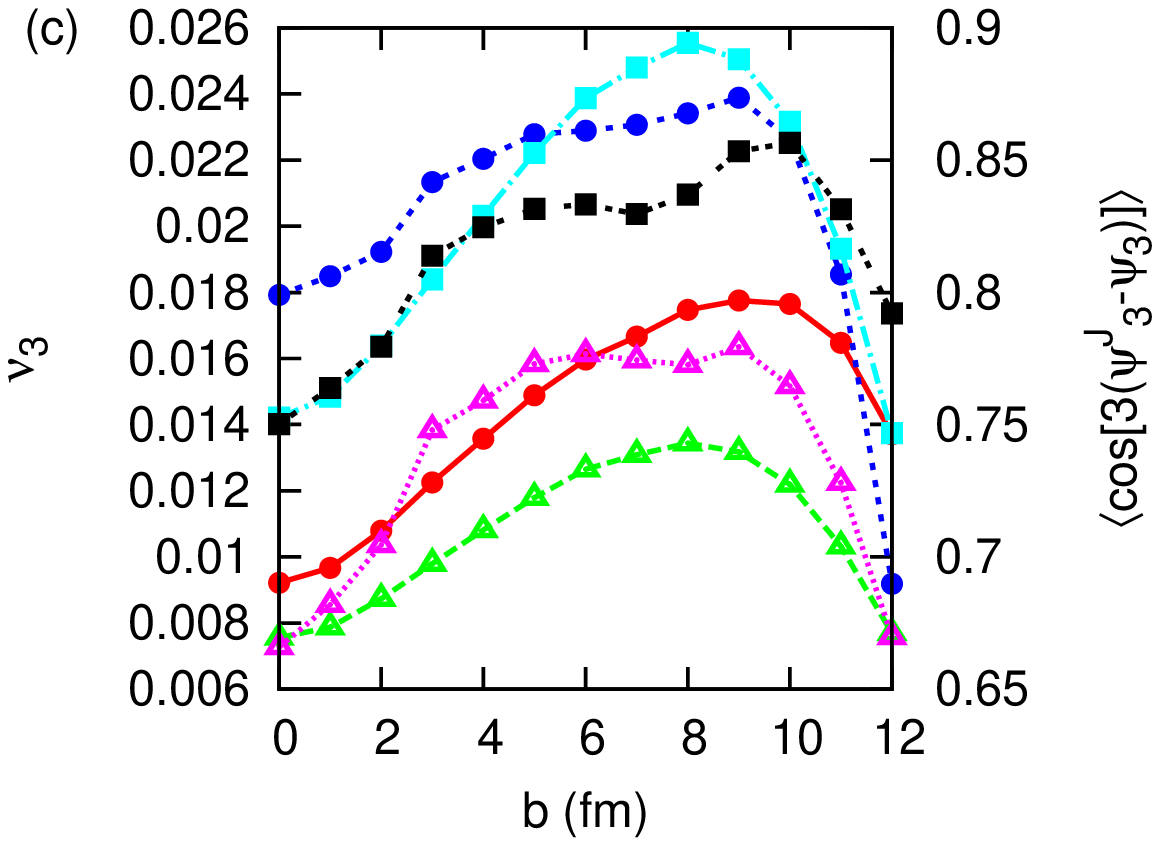}
\includegraphics[scale=0.6, angle=-90]{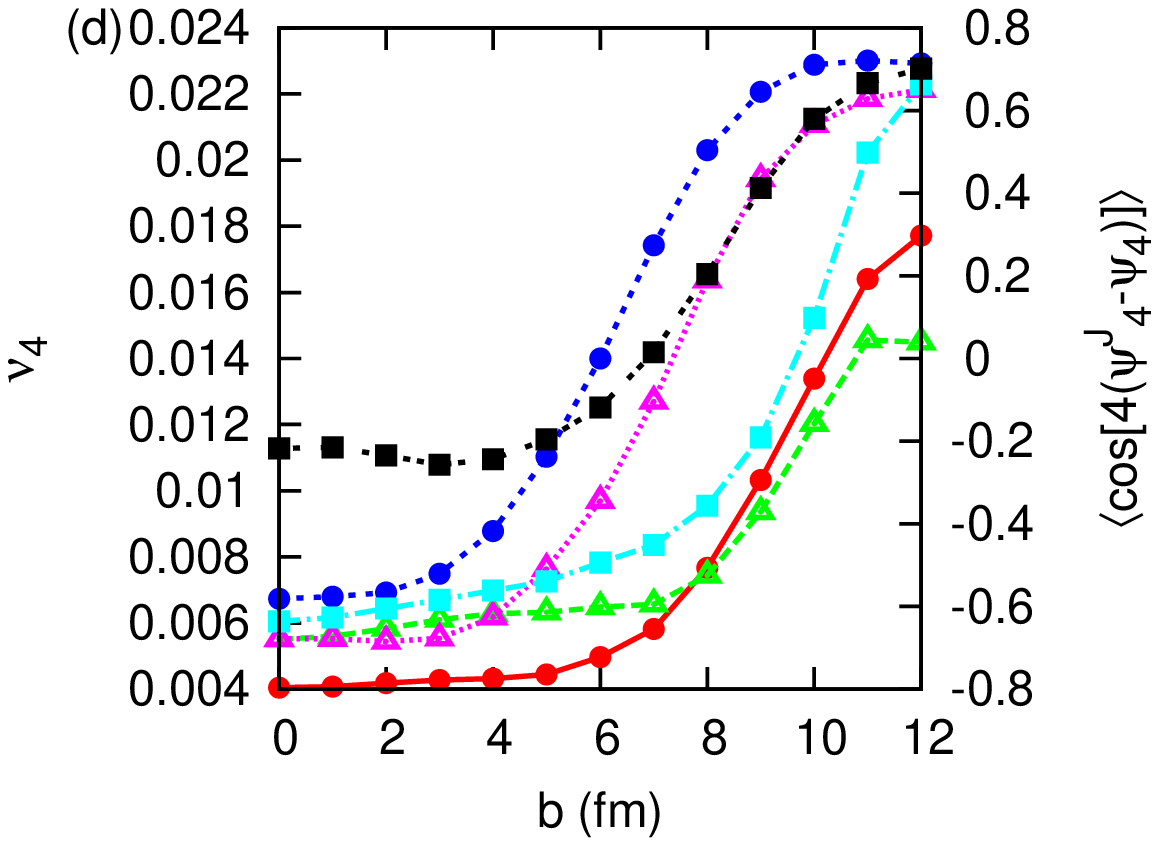}
\includegraphics[scale=0.6, angle=-90]{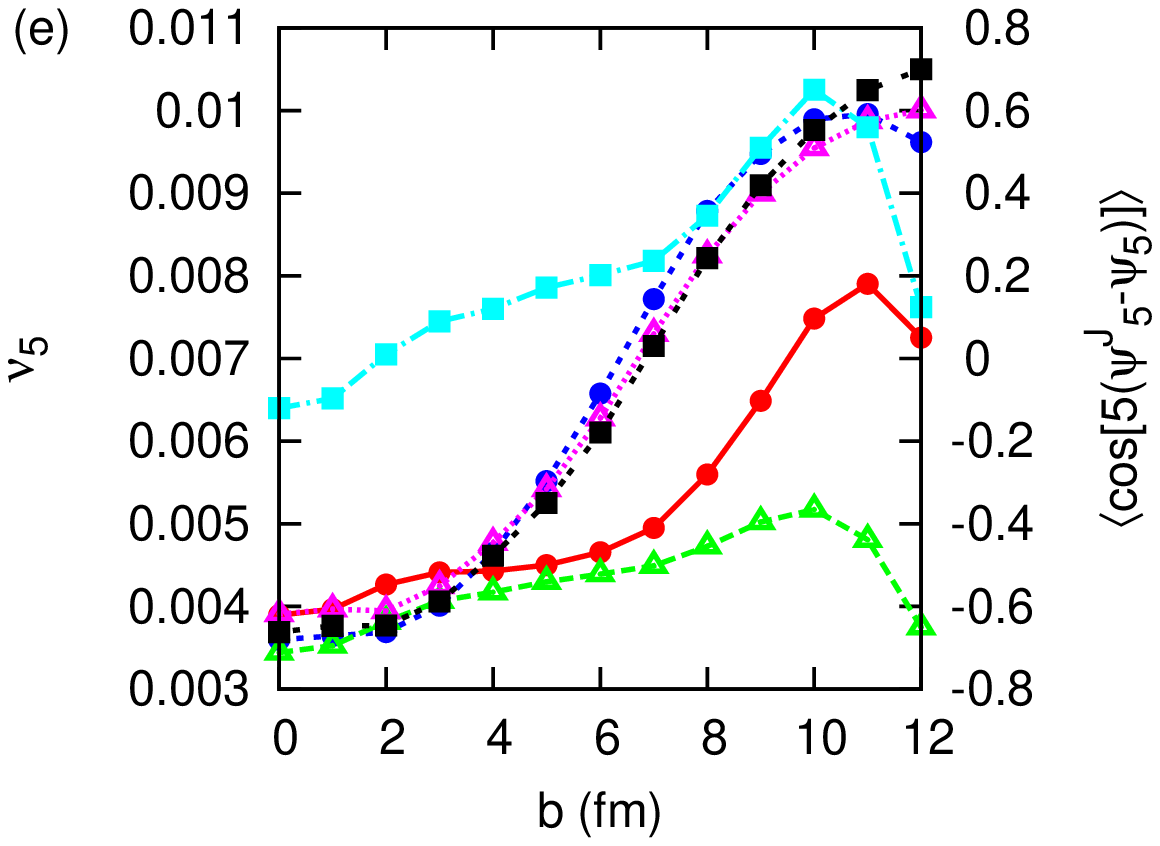}
\includegraphics[scale=0.6, angle=-90]{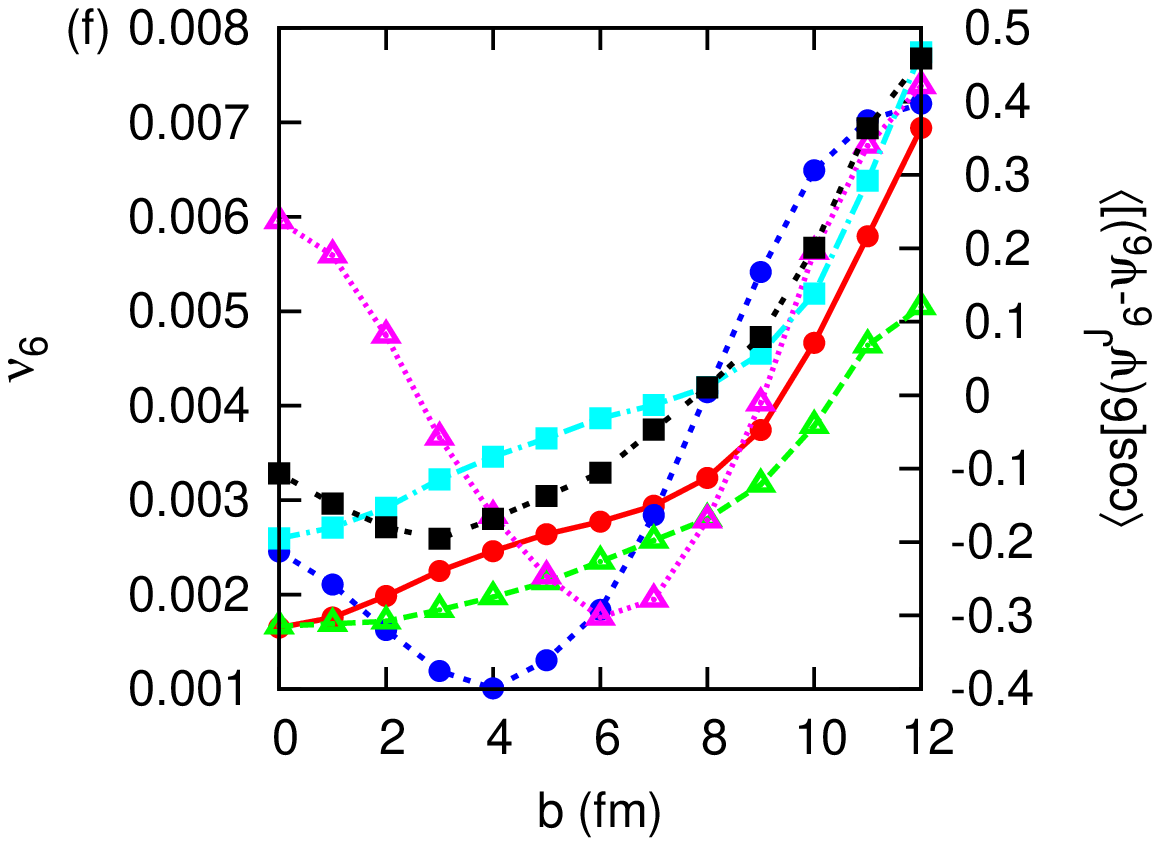}
\caption{(Color online). The $b$ dependence of $\nuh{n}$ and $\ave{\cos[n(\psih{n}-\psi_{n})]}$ for LHC ($\sqrt{s}=5500$ GeV) based on three different models.}
\label{fig:nuvsbLHC5500}
\end{figure} 

\begin{figure}
\centering
\includegraphics[scale=0.6, angle=-90]{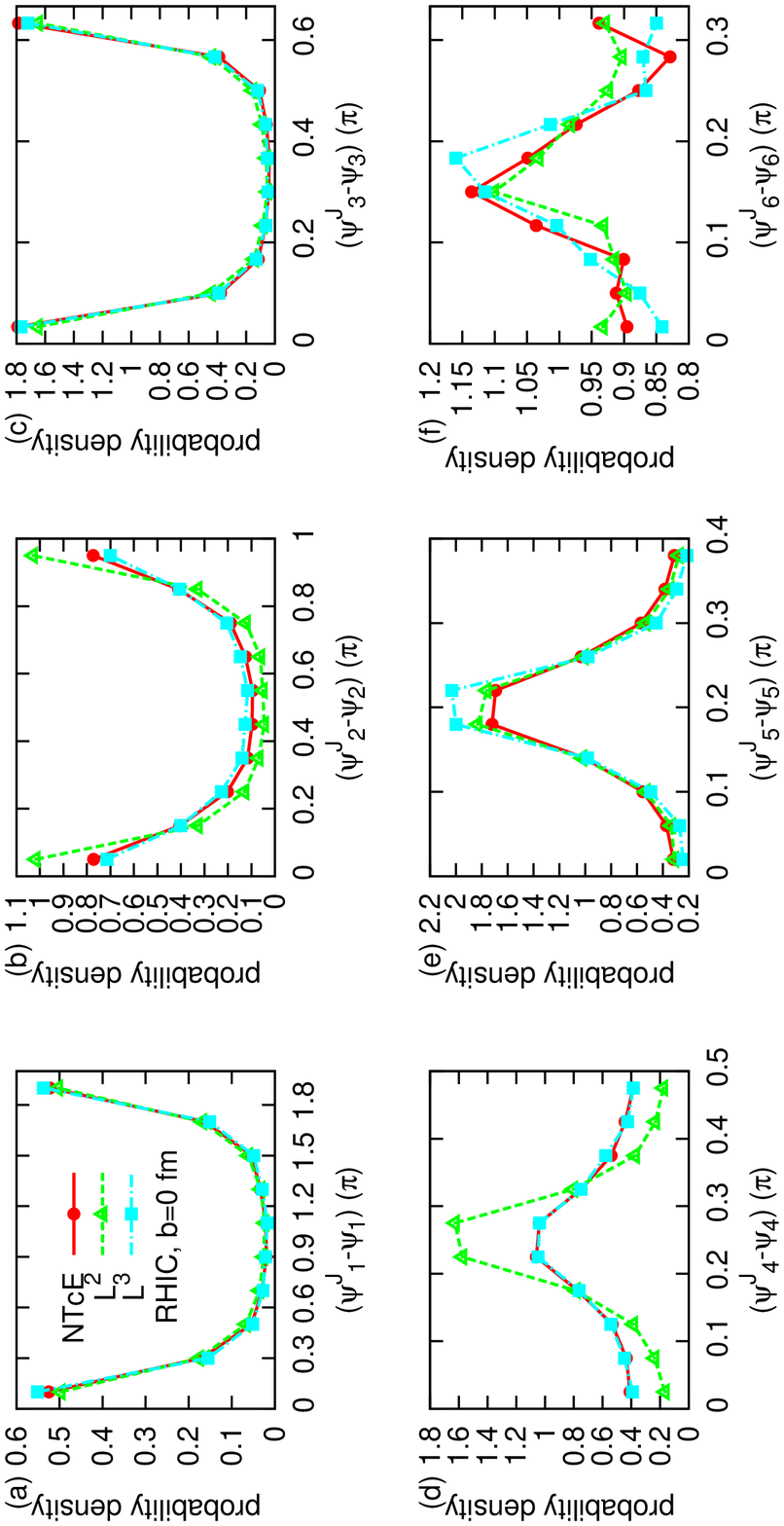}
\includegraphics[scale=0.6, angle=-90]{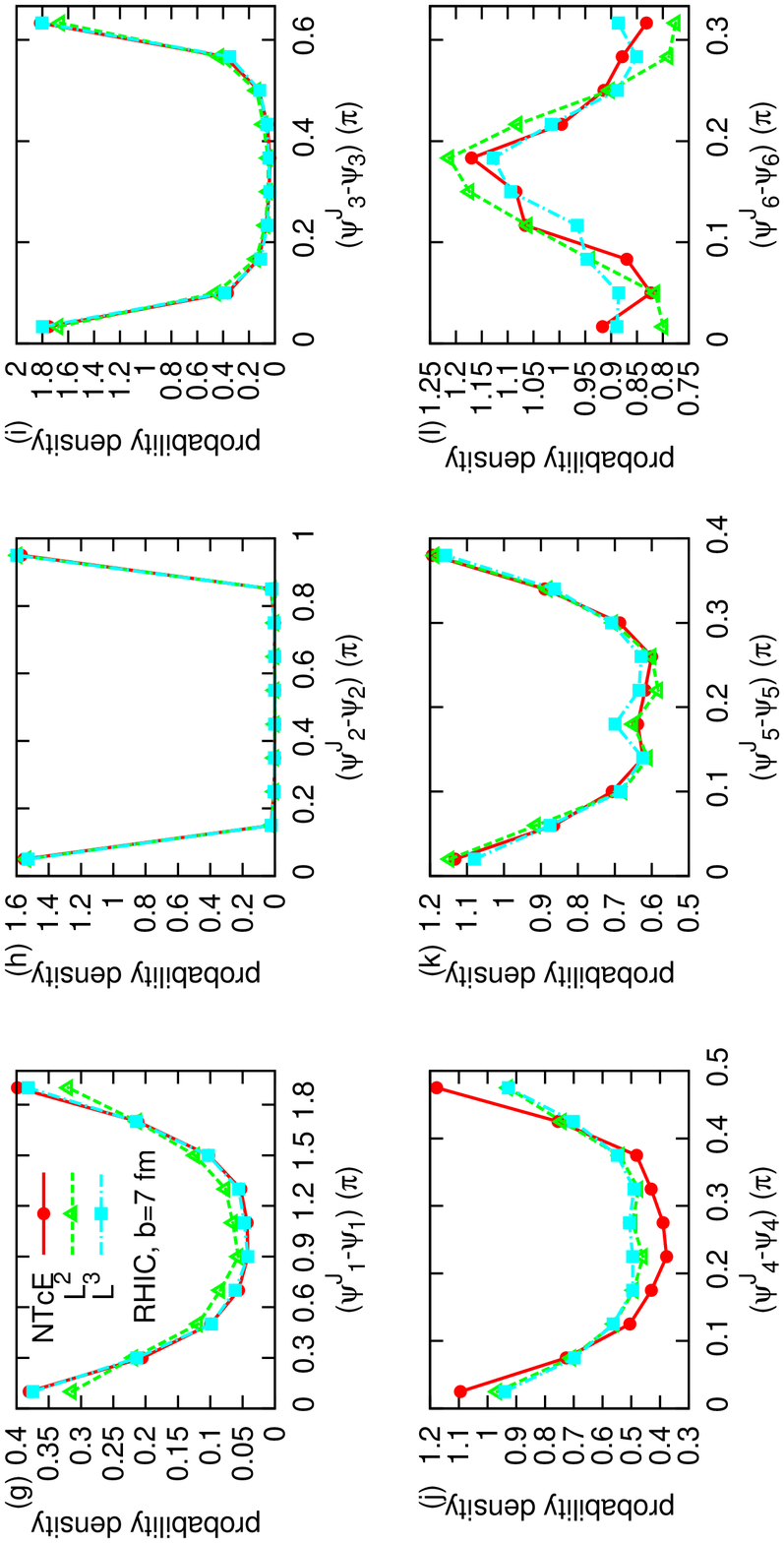}
\caption{(Color online). The distribution of $\psih{n}-\psi_{n}$ for RHIC ($\sqrt{s}=200$ GeV) at $b=0\ , 7$ fm, based on three different models.}
\label{fig:psij-psi_corr_RHIC}
\end{figure}

\begin{figure}
\centering
\includegraphics[scale=0.6, angle=-90]{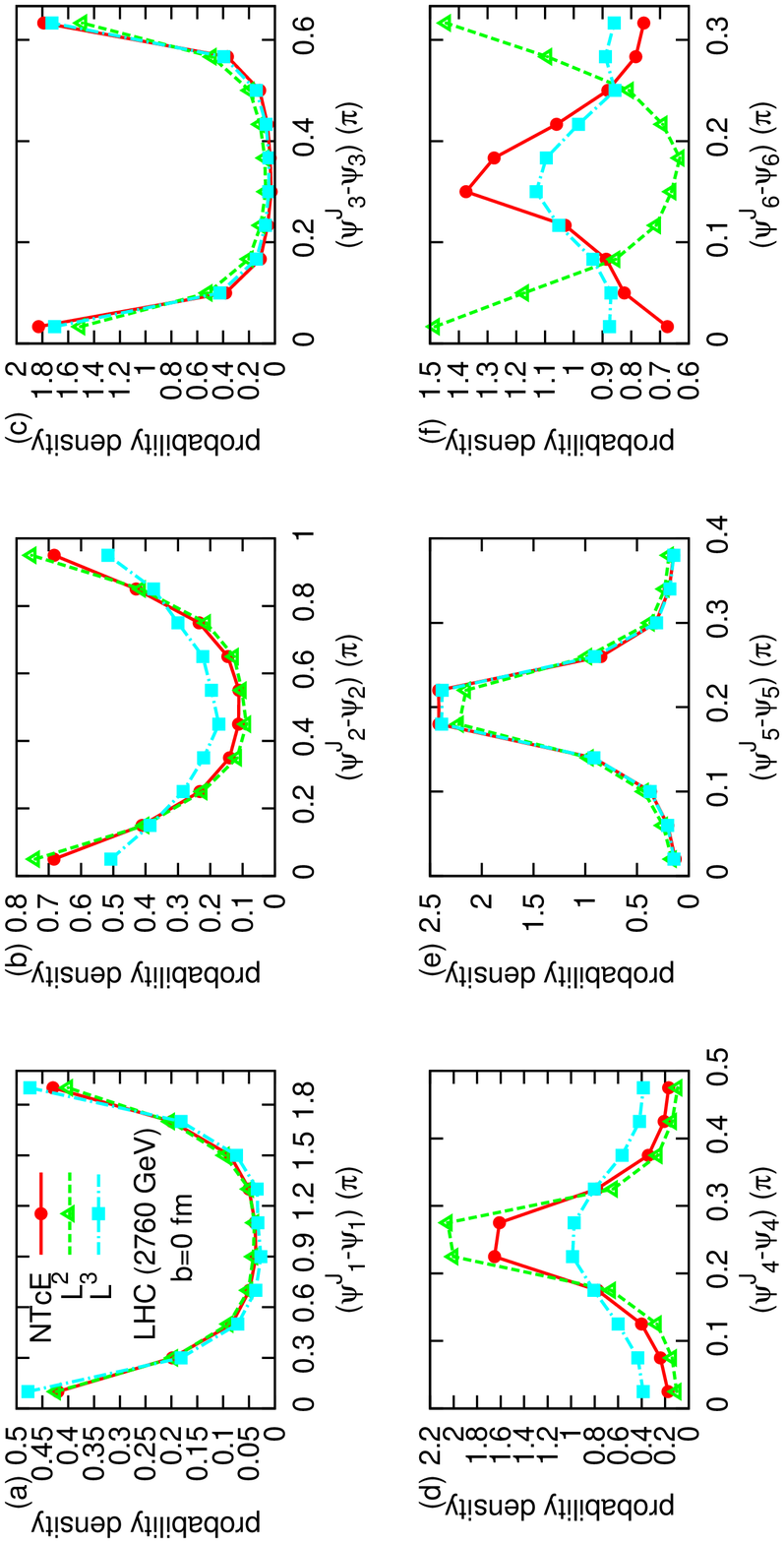}
\includegraphics[scale=0.6, angle=-90]{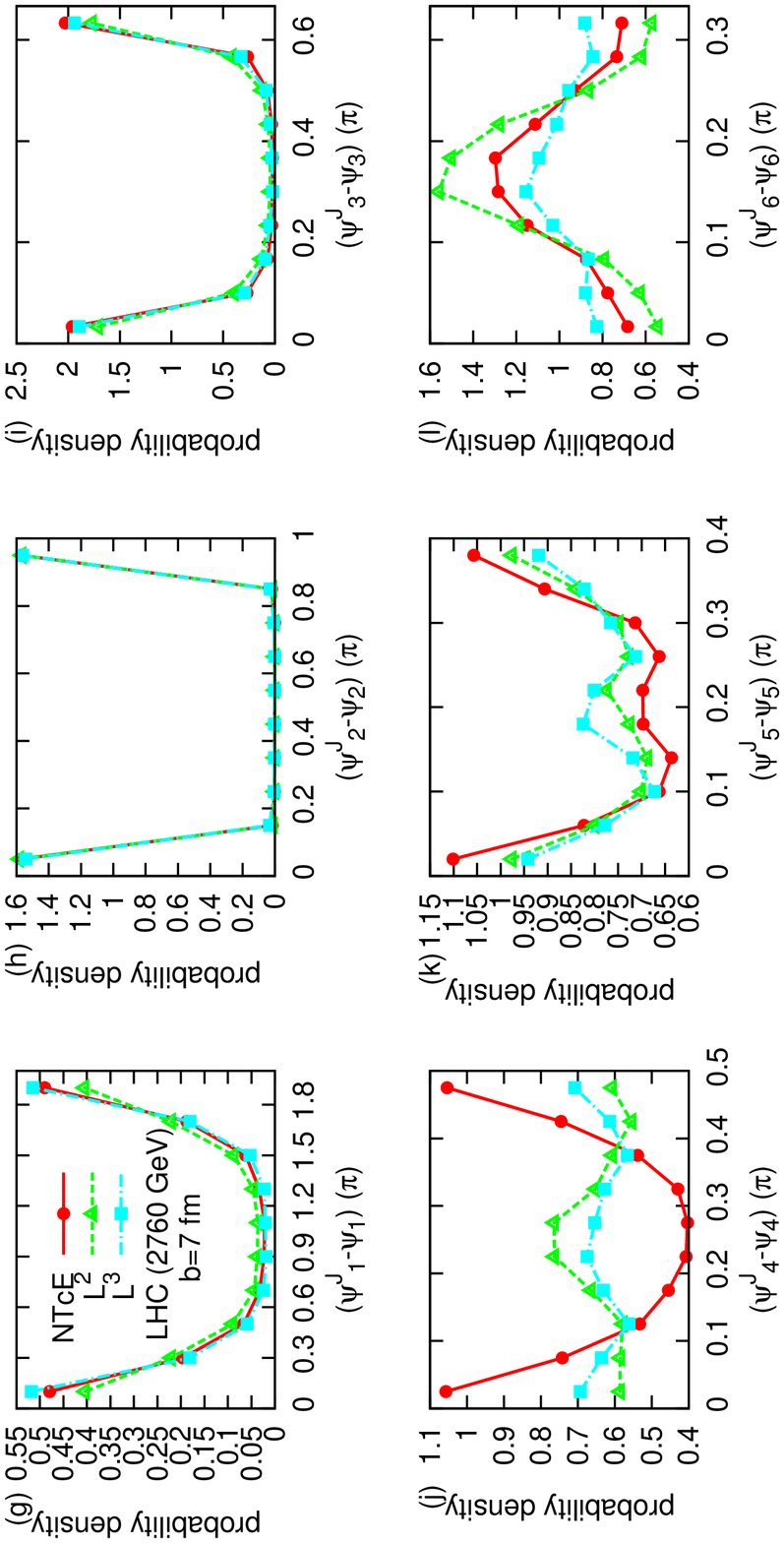}
\caption{(Color online). The distribution of $\psih{n}-\psi_{n}$ for LHC ($\sqrt{s}=2760$ GeV) at $b=0,\ 7$ fm, based on three different models.}
\label{fig:psij-psi_corr_LHC_2760GeV}
\end{figure}

\begin{figure}
\centering
\includegraphics[scale=0.6, angle=-90]{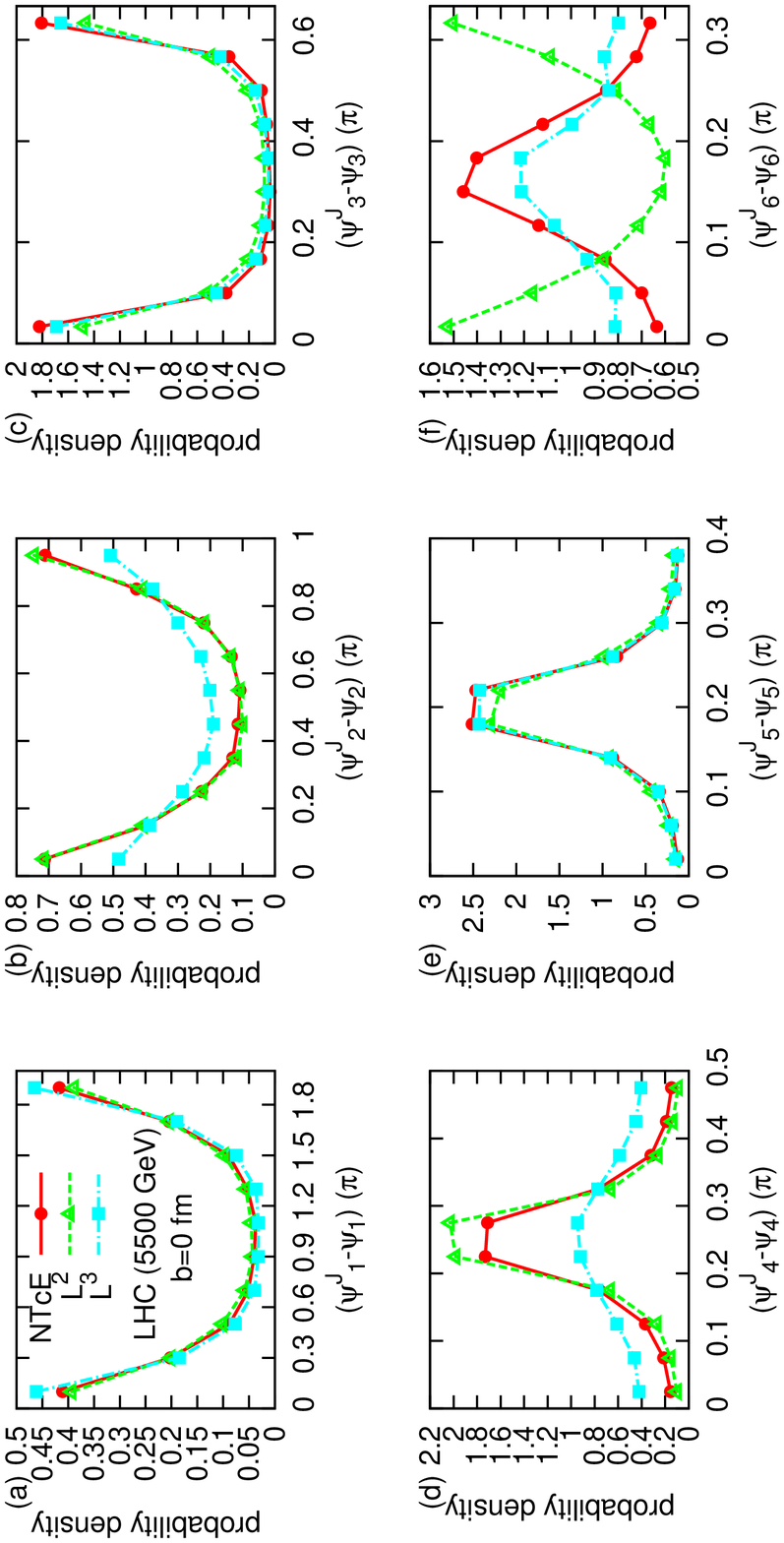}
\includegraphics[scale=0.6, angle=-90]{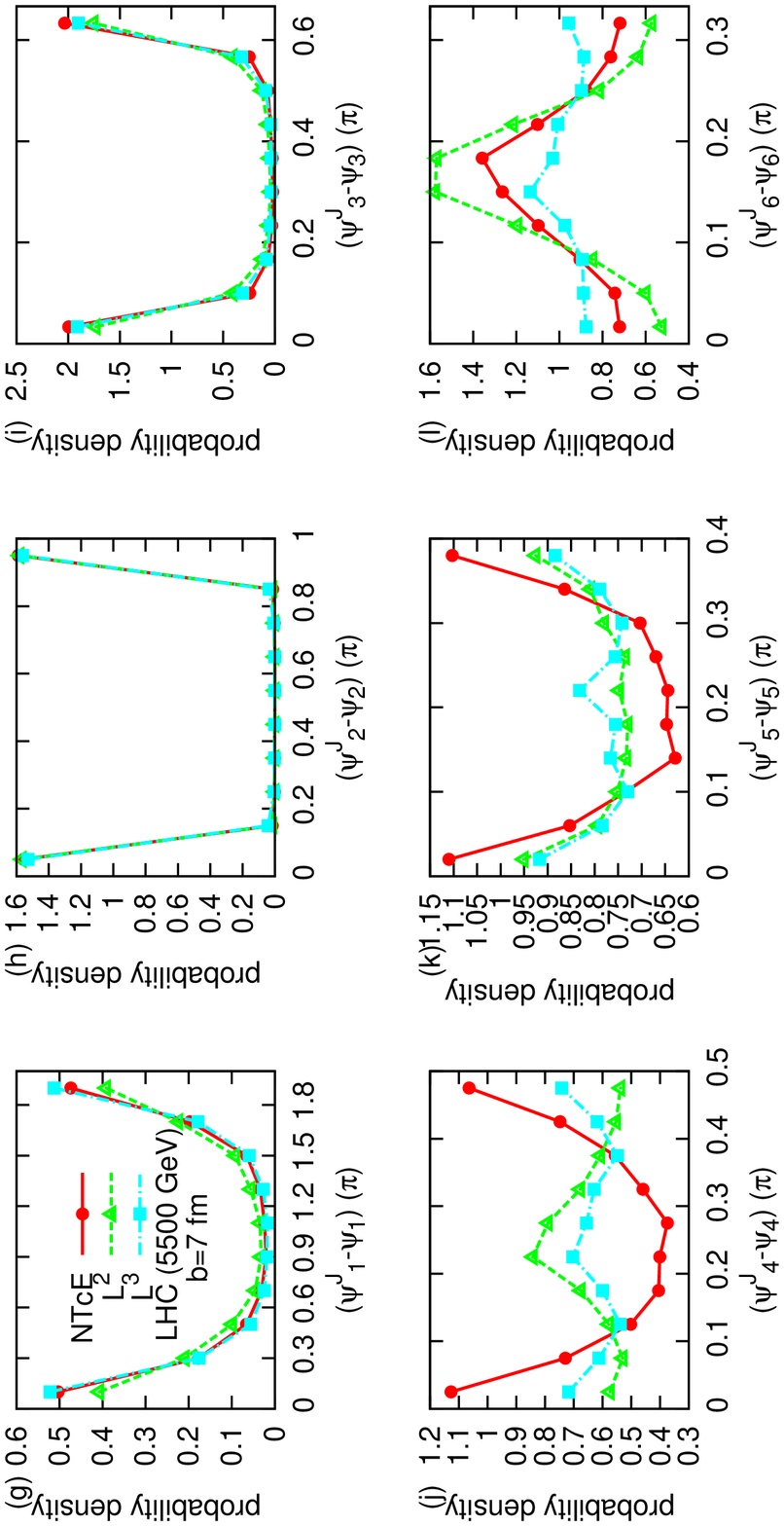}
\caption{(Color online). The distribution of $\psih{n}-\psi_{n}$ for LHC ($\sqrt{s}=5500$ GeV) at $b=0,\ 7$ fm, based on three different models.}
\label{fig:psij-psi_corr_LHC_5500GeV}
\end{figure}

\end{document}